 \documentclass[iop,appendixfloats]{emulateapj}

\newbox\grsign \setbox\grsign=\hbox{$>$}
\newdimen\grdimen \grdimen=\ht\grsign
\newbox\laxbox \newbox\gaxbox
\setbox\gaxbox=\hbox{\raise.5ex\hbox{$>$}\llap
     {\lower.5ex\hbox{$\sim$}}}\ht1=\grdimen\dp1=0pt
\setbox\laxbox=\hbox{\raise.5ex\hbox{$<$}\llap
     {\lower.5ex\hbox{$\sim$}}}\ht2=\grdimen\dp2=0pt

\shorttitle{Bipolar jet launching}
\shortauthors{Fendt \& Sheikhnezami}

\usepackage{amsmath}
\usepackage{float}
\usepackage{ctable} 
\usepackage{color}
\begin{document}
\title{Bipolar jets launched from accretion disks.\\
       II. Formation of symmetric and asymmetric jets and counter jets}
\author{Christian Fendt and Somayeh Sheikhnezami\altaffilmark{1} }
\affil{Max Planck Institute for Astronomy, K\"onigstuhl 17, D-69117 Heidelberg, Germany}
\altaffiltext{1}{Present address: Department of Physics, Faculty of Sciences, Ferdowsi University of Mashhad, Iran}
\email{fendt@mpia.de, nezami@mpia.de}
\begin{abstract}
We investigate the jet launching process from accretion disks extending our recent
study (paper I) to a truly bipolar setup.
We perform axisymmetric MHD simulations of the disk-jet interaction on a computational
domain covering both hemispheres, 
in particular addressing the question of an intrinsically asymmetric origin of jet / counter jet systems.
Treating both hemispheres simultaneously, we overcome the equatorial plane symmetry 
boundary condition used in most previous studies which naturally fosters a symmetric 
evolution.
For the magnetic diffusivity prescription we apply an $\alpha$-parametrisation, 
considering both, {\em globally} models of diffusivity, and {\em local} models.
We first approve the quality of our numerical setup by generating perfectly symmetric jets, 
lasting over a 1000s of dynamical time scales.
We then disturb the hemispheric symmetry {\em in the disk}, and investigate the 
subsequent evolution of the outflow.
The evolution first leads to a substantial disk warping with electric currents 
intersecting the equatorial plane. 
We investigate two models, 
i) a disk with (initially) different thermal scale height in both hemispheres, and 
ii) a symmetric disk into which a local disturbance is injected in one hemisphere.
In both cases the disk asymmetry results in asymmetric outflows with mass fluxes 
differing by 10-20\%.
For case i) the jets are launched asymmetric right from the beginning.
However, due to the symmetric diffusivity profile and restoring gravitational forces, 
the outflow asymmetry dies out after several hundred dynamical time scales.
For case ii) the onset of asymmetry in the outflows is delayed as the disturbance injected 
at a larger radius first has to propagate to smaller radii where the bulk of the outflow 
is launched. 
%
For a {\em local} prescription of magnetic diffusivity (following mainly the local sound speed)
the outflow asymmetry is persistent for long time,
implying that diffusive effects play a major role when re-establishing hemispheric
symmetric by gravitational forces.
We find up to 30\% difference in mass flux between jet and counter jet for this setup,
lasting over 1000s of dynamical time scales (i.e. lasting for the whole simulation).
For comparison we have run simulations where jets are launched from a symmetric disk
into an asymmetric coronal hemispheres of different density. 
As expected, the initial jet and counter jet are launched asymmetric, however, 
as soon as the jets have left the coronal asymmetry (within a few hundred dynamical 
time steps) a symmetric flow structure is established.
In summary, our results suggest that the jet asymmetries in protostellar and extragalactic
jets can indeed be generated intrinsically and maintained over long time by disk asymmetries 
and the standard jet launching mechanism.
\end{abstract}
\section{Introduction}
Astrophysical jets as highly collimated beams of high velocity material and outflows of 
less degree of collimation and lower speed are an ubiquitous phenomenon.
The jets are observed over a wide range of luminosity and spatial scale, originating 
from young stars, micro-quasars, or active galactic nuclei.
The common understanding is that magnetohydrodynamics (MHD) are responsible for launching, accelerating, 
and collimating these jets.
Furthermore, it is clear that accretion and ejection is related to each other - one efficient way 
to remove angular momentum from a disk is to eject it vertically into a jet or outflow. 

Observational data have confirmed the co-existence of bipolar jets in most jet forming regions.
Jet and counter jet appear, however, typically asymmetric in shape with very few exceptions
(see e.g. \citealt{1990A&A...232...37M}, \citealt{1996ApJ...468L.103R}, \citealt{2007prpl.conf..231R}).
\citet{2013A&A...551A...5E} find similar mean velocities in both lobes of the jets observed, 
while they report asymmetric variations in mass outflow rates and velocities, and suggest that 
the jet launching mechanism on either side of the disk is not synchronized.

One exception is the jet source HH\,212 which ejects an almost perfectly symmetric bipolar structure
\citep{1998Natur.394..862Z}, suggesting that the causal origin of jet knots is located close to
the central engine.
On the other hand, if jets form  naturally asymmetric, and if thus asymmetric jets would
need special conditions to be formed, we may ask what is this kind of "natural" ejection process and
what are the additional conditions for symmetry?

A small number of papers have a addressed this topic.
\citet{1996A&A...306..329C} discuss the MHD origin of the one-sidedness as probably caused by a superposition
of a quadrupolar disk dynamo and a stellar dipole.
The first bipolar jet simulation has been published by \citet{2003A&A...398..825V} and 
follow-up papers.
Recent simulations consider asymmetric ejections of stellar wind components from an offset
multipole stellar magnetosphere \citep{2010MNRAS.408.2083L}.
Velocity asymmetries in protostellar jets have been investigated by \citet{2012A&A...545A..53M}, 
in particular comparing intrinsic and extrinsic mechanisms, 
such as e.g. the field alignment between stellar and disk field or the pressure distribution in 
the ambient cloud. 
Both effects seem to play a role and work on different time scales.
\citet{2012ApJ...759L...1S} discuss semi-analytical models indicating a counter-rotation of the 
inner jet and compare them to apply MHD simulations.

To our knowledge, only very few numerical simulations investigating the bipolar launching of disk jets 
have been performed. 
It is therefore interesting to investigate the evolution of both hemispheres of a jet-disk system
in order to see whether and how a global asymmetry in the large-scale outflow can be governed
by the disk evolution.
In a previous paper (paper I, \citealt{2012ApJ...757...65S}) we have presented detailed 
model simulations for the jet launching process, investigating a comparative set of various launching 
parameters such as magnetization (or plasma-beta), disk diffusivity and diffusivity scale height.
In this paper (paper II) we continue our investigation concentrating on the {\em bipolar}
character of jets and outflows, in particular on the fact that both jet components are mostly 
observed as asymmetric.
We investigate the launching of bipolar outflows from initially symmetric conditions, and
also from a disk structure with slight asymmetries.

With {\em launching}, we denote the  process which conveys material from radial accretion into vertical  ejection, 
lifting it from the disk  plane into the corona, thus establishing a disk wind.
Referring to our notation in paper I we like clarify again that with {\em formation} we denote 
the process of accelerating and collimating an already existing slow disk wind or stellar wind in to a jet beam.
This paper deals with both processes - launching and  formation.

The present paper is organized as follows. 
Section 2 describes the numerical setup, the initial and boundary conditions of our simulations
with particular emphasis to the approach of simulating both hemispheres.
The general evolution of jet launching is presented in section 3,
where we also present the simulation of a perfectly symmetric jet.
Section 4 is then devoted to asymmetric jets and a parameter study comparing jets from different 
disk setups. We will discuss their symmetry properties and how symmetry can be broken by the 
intrinsic disk evolution.

\begin{table*}
\begin{center}
\caption{Grid resolution of various parameter runs.
The equidistant grid usually covers the highly diffusive regions of disk and disk wind,
while the stretched grid covers the weakly diffusive, almost ideal MHD regions further distant
from the launching region.
Simulations discussed in this paper are labeled with a star '*'.
}
\label{tbl:resolution}
 \begin{tabular}{lrclcl}
\hline
\hline
\noalign{\smallskip}  
run         & \multicolumn{2}{l}{equidistant subgrid} & \multicolumn{2}{l}{stretched subgrids} &  \\
            & $n_r$ & grid $r$-extension & $n_r$ & grids $r$-extension &  \\
            & $n_z$ & grid $z$-extension & $n_z$ & grids $z$-extension &  \\
\noalign{\smallskip}
\hline
\noalign{\smallskip} 
sb1*-sb4*      & 800 & $ 0.0 \leq r \leq 60.0 $ &            &                & \\
               & 1214 & $-40.0 \leq z \leq  40 $ &  800, 800  & $-100.0  \leq z \leq -40.$, $40.0 \leq z \leq 100. $ & \\
sb5-sb7        & 800 & $ 0.0 \leq r \leq 60.0 $ &            &                & \\
               & 1214 & $-40.0 \leq z \leq  40 $ &  800, 800  & $-100.0  \leq z \leq -40.$, $40.0 \leq z \leq 100. $ & \\
\noalign{\smallskip} 
\hline
\noalign{\smallskip} 
cf1-cf5         & 800 & $ 0.0 \leq r \leq 60.0 $  &  &  &\\
                & 1214 & $-40.0 \leq z \leq  40 $ &  800, 800  & $-100.0  \leq z \leq -40$, $40.0 \leq z \leq 100 $ & \\
cf6             & 1600 & $ 0.0 \leq r \leq 60.0 $  &  &  & \\
                & 2428 & $-40.0 \leq z \leq  40 $ & 1200, 1200 & $-100.0  \leq z \leq -40$, $40.0 \leq z \leq 100 $ & \\
cf8             & 2428 & $ 0.0 \leq r \leq 40.0 $  &  &  & \\
                & 2428 & $-10.0 \leq z \leq  10 $ & 1000, 1000 & $ -50.0  \leq z \leq -10$, $10.0 \leq z \leq  50 $ & \\
cf14*           & 1200 & $ 0.0 \leq r \leq 50.0 $  &  &  & \\
                & 3600 & $-60.0 \leq z \leq  60 $  &  &  & \\
cf16*           & 2428 & $ 0.0 \leq r \leq 40.0 $  &  &  & \\
                & 2428 & $-20.0 \leq z \leq  20 $ & 1000, 1000 & $ -50.0  \leq z \leq -10$, $10.0 \leq z \leq  50 $ & \\
\noalign{\smallskip} 
\hline
\hline  
\end{tabular}
\end{center}
\end{table*}

%
%
%

\section{Model setup}
As discussed in detail in paper I, we model the launching of MHD jets from slightly sub-Keplerian disks, 
which are initially in pressure equilibrium with a non-rotating corona.
The main goals of paper I were
   i) to determine the mass ejection to accretion rate fraction for a variety of disk physical
      characteristics such as plasma-beta, magnetic diffusivity, or disk scale height, and
  ii) similarly for the angular momentum flux, and
 iii) to determine the main geometry for these jets, such as the asymptotic jet radius and opening 
      angle, or the size of the main jet launching area of the disk.

In the present paper we extend our approach to simulations on a computational domain including
{\em both hemispheres}.
This allows us to investigate the {\em truly bipolar} launching and how the launching symmetry can
affect the symmetry characteristics of jet and counter-jet.

We apply the MHD code PLUTO \citep{2007ApJS..170..228M, 2012ApJS..198....7M}, 
solving the time-dependent resistive MHD equations as described in paper I.
Again, the simulations are performed in a 2.5-dimensional setup in cylindrical coordinates 
(thus applying 3D axisymmetry).

\section{Numerical setup - initial \& boundary conditions}
The major extension from paper I is that we now treat the truly {\em bipolar} launching
of outflows.
In general we apply the same initial disk structure and boundary conditions as before, 
however we extend the disk-outflow system across the equatorial plane into both 
hemispheres.
Above and below the disk, respectively, a hydrostatic corona is prescribed in pressure balance 
with the disk gas pressure (thus, implying a density jump and entropy jump from disk to corona, 
see paper I).

We apply a uniform grid with across the midplane and attached to that scaled grids,
We run simulations applying a different grid resolution (see Tab.\ref{tbl:resolution}).
With the highest-resolution grid we resolve the disk scale height at the innermost radius
with up to $\simeq 10$ grid cells.

\subsection{Boundary conditions}
The main goal of this paper is to investigate the symmetry of bipolar jets launched from a 
diffusive accretion disk.
It is therefore essential to carefully check our numerical setup, in particular the internal 
boundary conditions describing the sink, in order to prevent numerical artifacts generating 
asymmetry.

Compared to paper I, obviously the equatorial-plane boundary condition is now omitted.
The disk itself may now evolve into a warped structure, breaking the intrinsic hemispheric 
symmetry of disk and outflow. 
A disk mid-plane, if it exists, will not be necessarily along the equatorial plane.
Further consequences are e.g. that the electric currents can now flow across the disk 
midplane.

As discussed in paper I, for the outer disk radius we apply n outflow boundary condition.
We feed the inner jet launching area by accretion from the inner disk areas.
Since compared to paper I the physical extent of the computational domain is 
somewhat reduced, also the mass reservoir for disk accretion is smaller, which limits
the disk mass evolution to comparatively shorter time scales.

The remaining boundary conditions are equivalent the conditions similar we have 
applied in paper I, i.e. the outflow boundary conditions (modified from the original code,
see \citealt{2010ApJ...709.1100P}) 
for the radial and vertical outer boundaries, and the axisymmetry boundary condition
along the rotation axis.
The sink boundary conditions allow to absorb the mass and angular momentum of the 
accreting material.

\subsection{Initial conditions}
We apply the same basic initial conditions as in paper I. 
In addition, a few extensions to this setup were made in order to break the symmetry
and to govern an asymmetric evolution.

As in paper I, the initial magnetic field is prescribed by the magnetic flux function 
$\psi$ following \cite{2007A&A...469..811Z},
\begin{equation}
\displaystyle
\psi(r,z) = \frac{3}{4} B_{z,i} r_{\rm i}^2 
            \left(\frac{r}{r_{\rm i}}\right)^{3/4}
            \frac{m^{5/4}}{\left( m^2 + \left(z/r\right)^2\right)^{5/8} },
\label{eq:magini}
\end{equation}
where $B_{z,0}$ measures the vertical field strength at $(r=r_{\rm i},z=0)$.
The (initial) field tension is determined by the parameter $m$. 
We apply $m = 0.4$ as in paper I.

For the disk density and pressure we apply the same distribution as Eq.~(6) of paper I.
However, we also run models with an initially asymmetric disk structure, considering a 
pressure scale height in the upper disk hemisphere $\epsilon = \epsilon_{\rm up} = 0.15$ 
different from the 
pressure scale height in the lower disk hemisphere $\epsilon = \epsilon_{\rm down} = 0.10$. 

The initial disk corona follows the same distribution as in paper I.
In order to compare intrinsic effects of asymmetric bipolar launching with 
external effects, we have, however, also run comparison simulations with an
initially symmetric disk, but with a disk corona of different density/pressure in 
the upper and lower hemisphere, respectively.

\begin{table*}
\begin{center}
\caption{Overview on our simulation runs, 
concerning the character of asymmetry imposed and the diffusivity distribution applied.
Simulation runs discussed in this paper are labeled with a star '*'.
}
\label{tbl:bicases}
\begin{tabular}{lllll}
\tableline \tableline
\noalign{\smallskip} 
run      & character                 & symmetry breaking                                                            & $\eta$ profile\tablenotemark{a} & \\
\noalign{\smallskip} 
\tableline
\noalign{\smallskip} 
sb1* & reference run             & none (symmetric)                                                               & $f_1$   & $\alpha_{m,1} = 3.0$ \\
sb2* & global disk asymmetry     & initial scale height, $(\epsilon_{\rm up},\epsilon_{\rm  down}) = (0.15, 0.1)$ &  $f_1$  &  \\
sb3* & local disk asymmetry      & over pressure injected  at $r=12$                                             & $f_1$    & $\alpha_{m,1} = 3.0$ \\
sb4* & global disk asymmetry  & initial density contrast, $(\delta_{\rm up},\delta_{\rm down})=(10^{-3}, 10^{-4})$ & $f_2$  & $\alpha_{m,2} = 3.0$ \\
sb5  & local disk asymmetry      & over pressure injected  at $r=18$                                             & $f_2$    & $\alpha_{m,2} = 3.0$ \\
sb6  & local disk asymmetry      & over pressure injected  at $r=18$                                             & $f_2$    & $\alpha_{m,2} = 1.0$ \\
sb7  & local disk asymmetry      & over pressure injected  at $r=18$                                             & $f_2$    & $\alpha_{m,2} = 2.0$ \\
\noalign{\smallskip} 
\tableline
\noalign{\smallskip} 
cb1 & global disk asymmetry  & initial scale height, $\epsilon_{\rm up},\epsilon_{\rm down} = 0.15, 0.1 $    & $f_1$ &  \\
cb2-3 & global disk asymmetry  & initial scale height, $ \epsilon_{\rm up},\epsilon_{\rm down} = 0.15, 0.1 $  & $f_2$, $\Gamma = 1/3$ & $\alpha_{m,2} = 0.01$ \\ 
cb4 & global disk asymmetry  & initial scale height,   $ \epsilon_{\rm up},\epsilon_{\rm down} = 0.15, 0.1 $  & $f_2$, $\Gamma = 1/3$ & $\alpha_{m,2} = 0.05$ \\ 
cb5 & global disk asymmetry  & initial scale height,   $ \epsilon_{\rm up},\epsilon_{\rm down} = 0.15, 0.1 $  & $f_3$                 & $\alpha_{m,3} = 0.05$ \\ 
cb6 & global disk asymmetry, high res  & initial scale height, $\epsilon_{\rm up},\epsilon_{\rm down} = 0.15, 0.1 $    & $f_2$, $\Gamma = 1/3$ & $\alpha_{m,2} = 0.05$ \\ 
cb8 & global disk asymmetry, highest res & initial scale height, $\epsilon_{\rm up},\epsilon_{\rm down} = 0.15, 0.1 $  & $f_2$, $\Gamma = 2/3$ & $\alpha_{m,2} = 0.01$ \\ 
cb14* & global disk asymmetry & initial scale height, $\epsilon_{\rm up},\epsilon_{\rm down} = 0.15, 0.1 $  & $f_3$ & $\alpha_{m,3} = 0.1$ \\
cb16* & global disk asymmetry & initial scale height, $\epsilon_{\rm up},\epsilon_{\rm down} = 0.15, 0.1 $  & $f_3$ & $\alpha_{m,3} = 0.1$ \\
\noalign{\smallskip} 
\tableline \tableline
\end{tabular}
\end{center}
\end{table*}

\subsection{Units and normalization}
We apply the same code units and normalization as in  paper I.
Throughout this paper  distances are expressed in units of the inner disk radius $r_{\rm i}$,
while $p_{\rm d,i}$ and $\rho_{\rm d,i}$ are the disk pressure and density at this radius, 
respectively\footnote{The index ''i'' refers to the value at the inner disk radius 
at the equatorial plane at time $t=0$}.
For the sake of comparison to previous papers we may assume $r_{\rm i} = 0.1\,$AU for a
protostellar jet and  $r_{\rm i} = 10\,$Schwarzschild radii for an AGN jet.
Velocities are measured in units of the Keplerian velocity $v_{\rm K,0}$ at the inner disk 
radius. Thus, by assuming smaller inner disk radii, the outflow speed will become larger.
Time is measured in units of $t_{\rm i} = r_{\rm i} / v_{\rm K,i}$, which can be related to 
the Keplerian orbital period $\tau_{\rm K,i} = 2\pi t_{\rm i}$.

The (initial) disk aspect ratio $\epsilon$ is the ratio of the isothermal sound speed to the 
Keplerian speed, both evaluated at disk mid plane, $\epsilon \equiv c_{\rm s} / v_{\rm K}$.
Pressure is  given in units of $p_{\rm d,i} = \epsilon^2 \rho_{\rm d,i} v_{\rm K,i}^2$.
The magnetic field is measured in units of $B_{\rm i} = B_{z,\rm i}$.
We adopt $v_{\rm  K,i} = 1$, $\rho_{\rm d,i} = 1$ in code units.

\begin{figure*}
\centering
 \includegraphics[width=4.4cm]{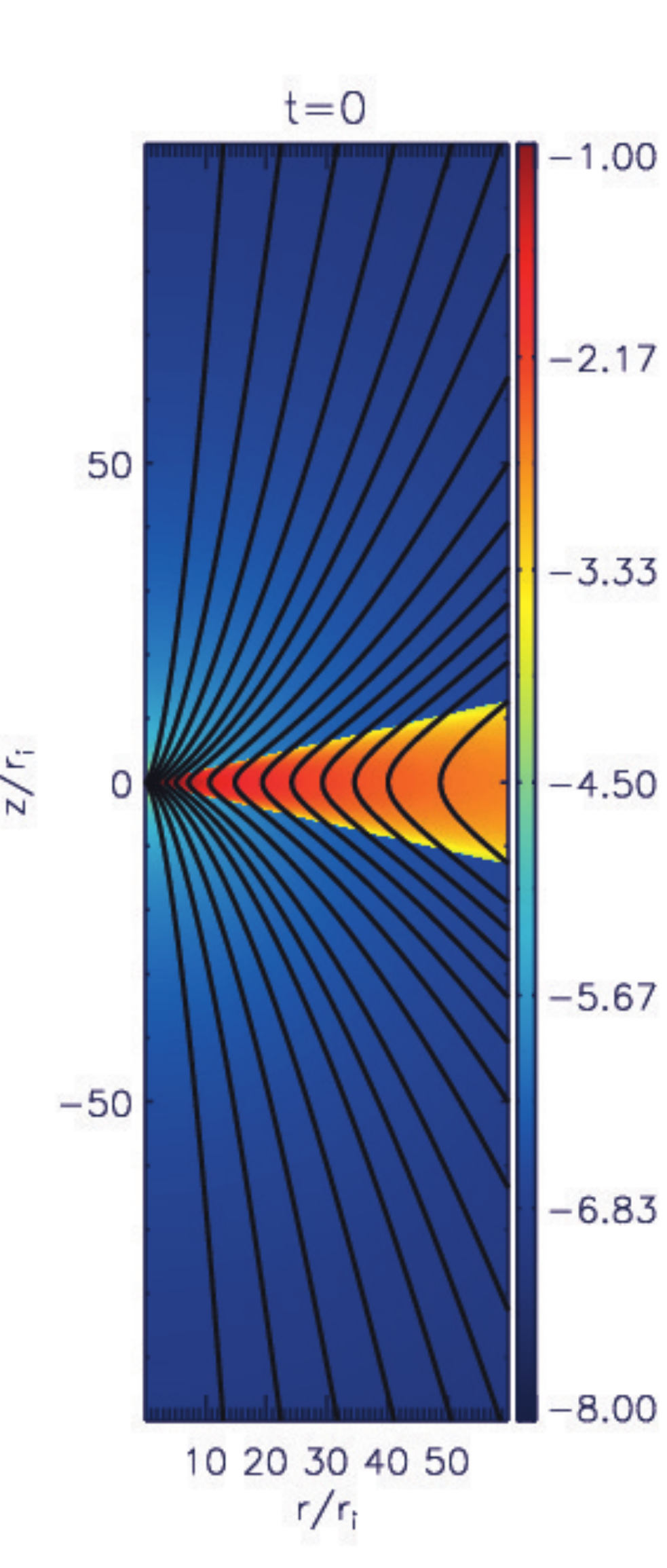}
 \includegraphics[width=4.4cm]{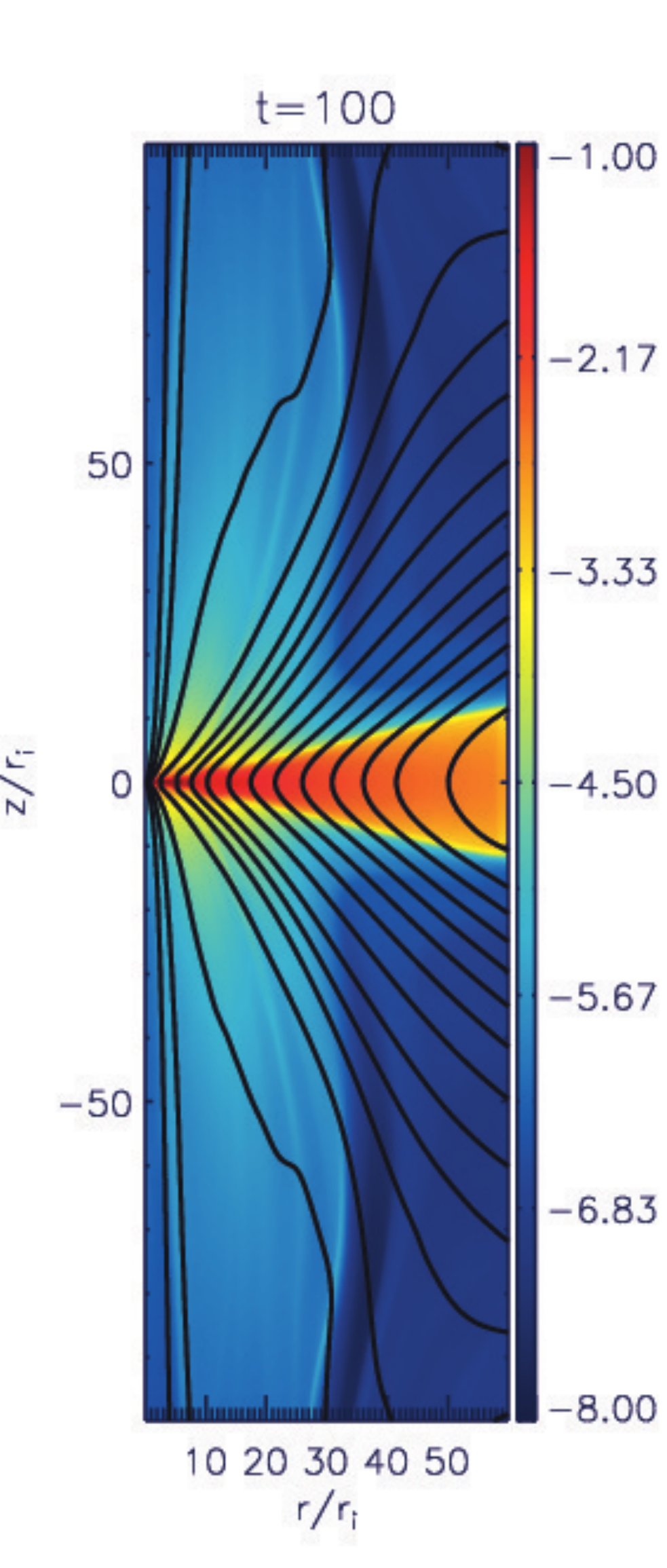}
 \includegraphics[width=4.4cm]{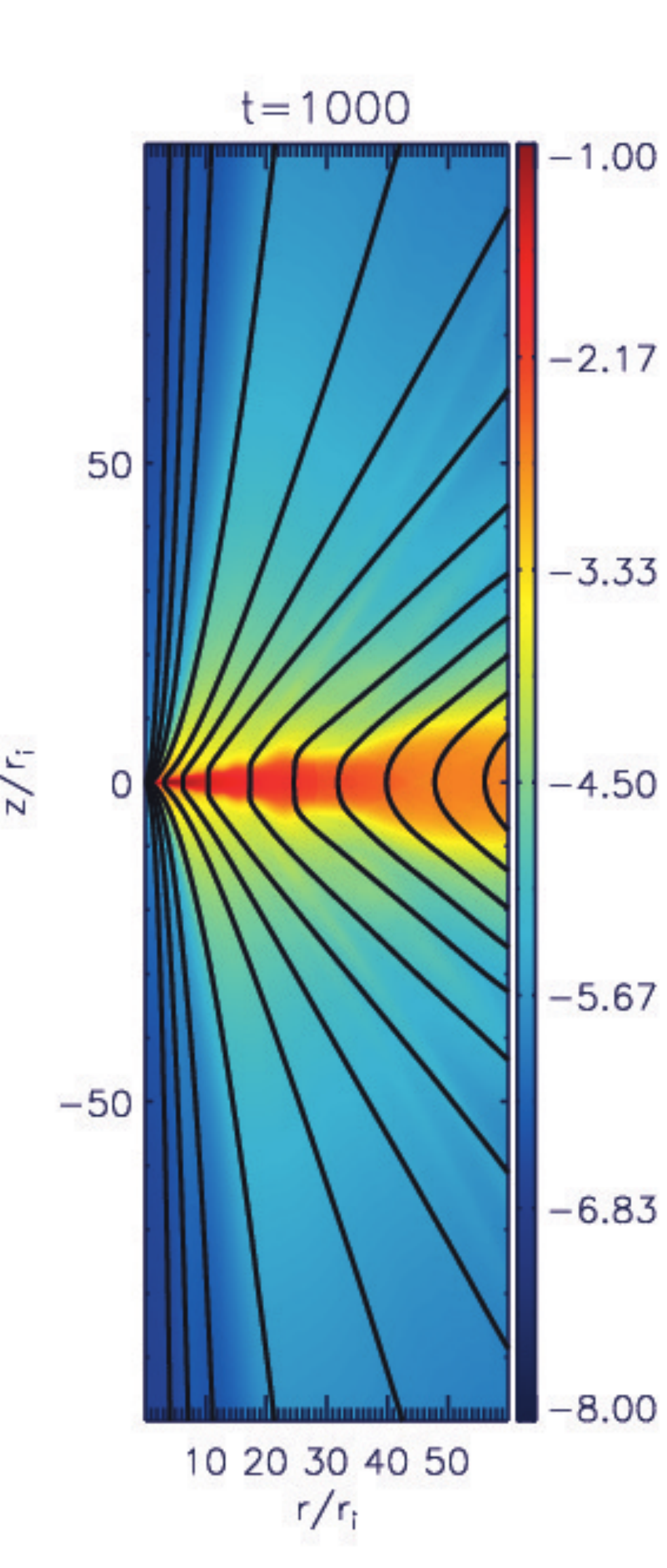}
 \includegraphics[width=4.4cm]{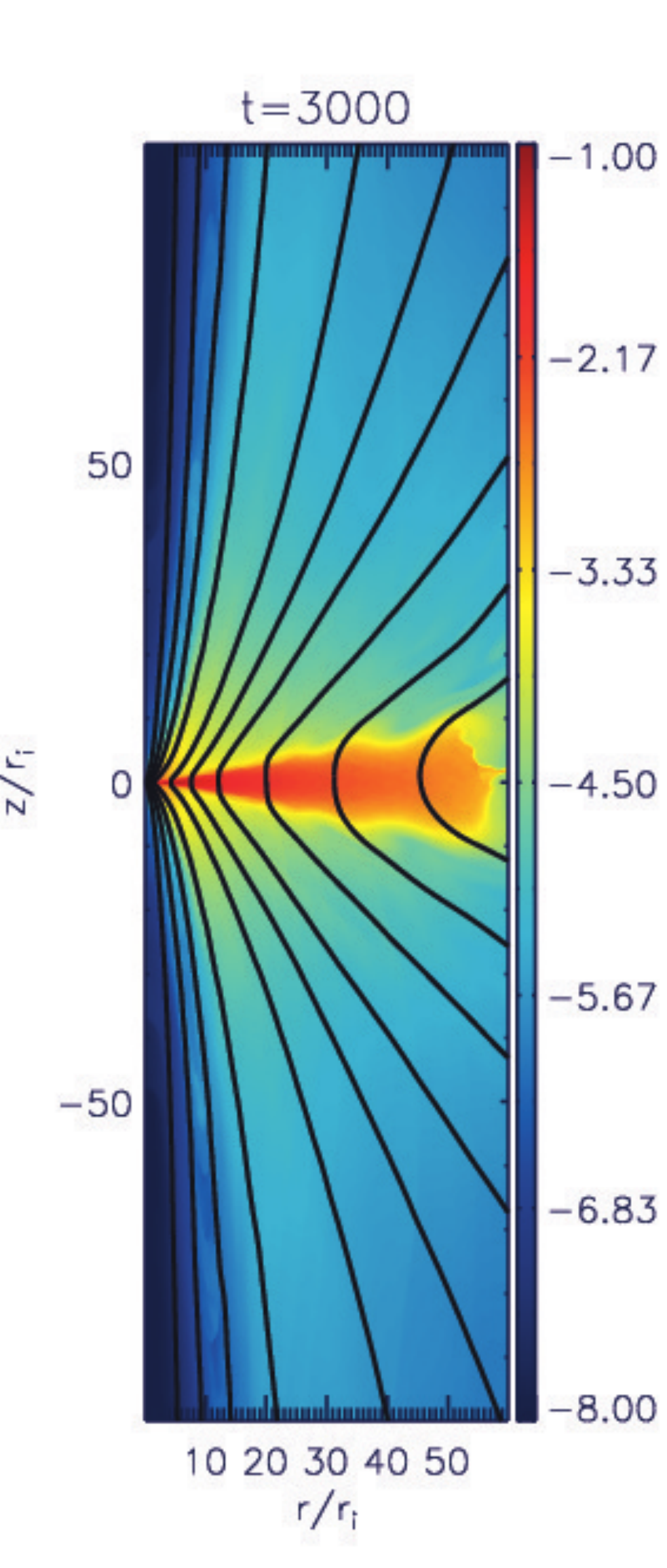}
\caption{Time evolution of the bipolar jet-disk structure for reference simulation sb1 applying 
a fixed-in-time and fixed-in-space magnetic diffusivity distribution Eq.~\ref{eq:magdiff_global}.
Shown is the evolution of mass density (color) and the poloidal magnetic field 
(contours of poloidal magnetic flux $\Psi(r,z)$) for the flux levels 
$\Psi = 0.01, 0.03, 0.06, 0.1, 0.15, 0.2, 0.26, 0.35, 0.45, 0.55,  0.65, 0.75, 0.85, 0.95, 1.1, 1.3, 1.5, 1.7$, 
for the dynamical times steps $t = 0, 100, 1000, 2000, 3000$.
}
\label{fig:bipo1_fix_case1}
\end{figure*}

\begin{figure}
\centering
 \includegraphics[width=8cm]{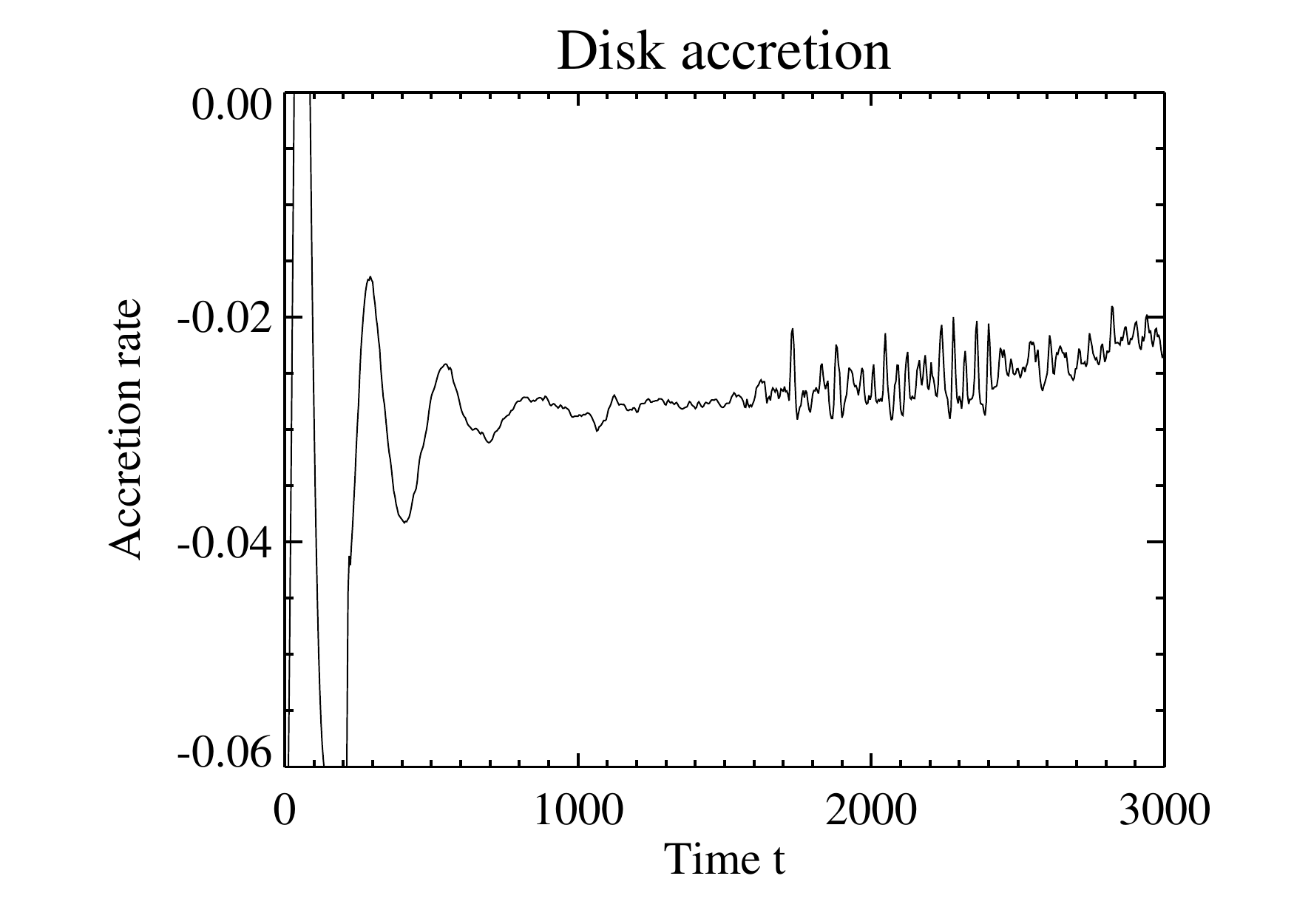}
 \includegraphics[width=8cm]{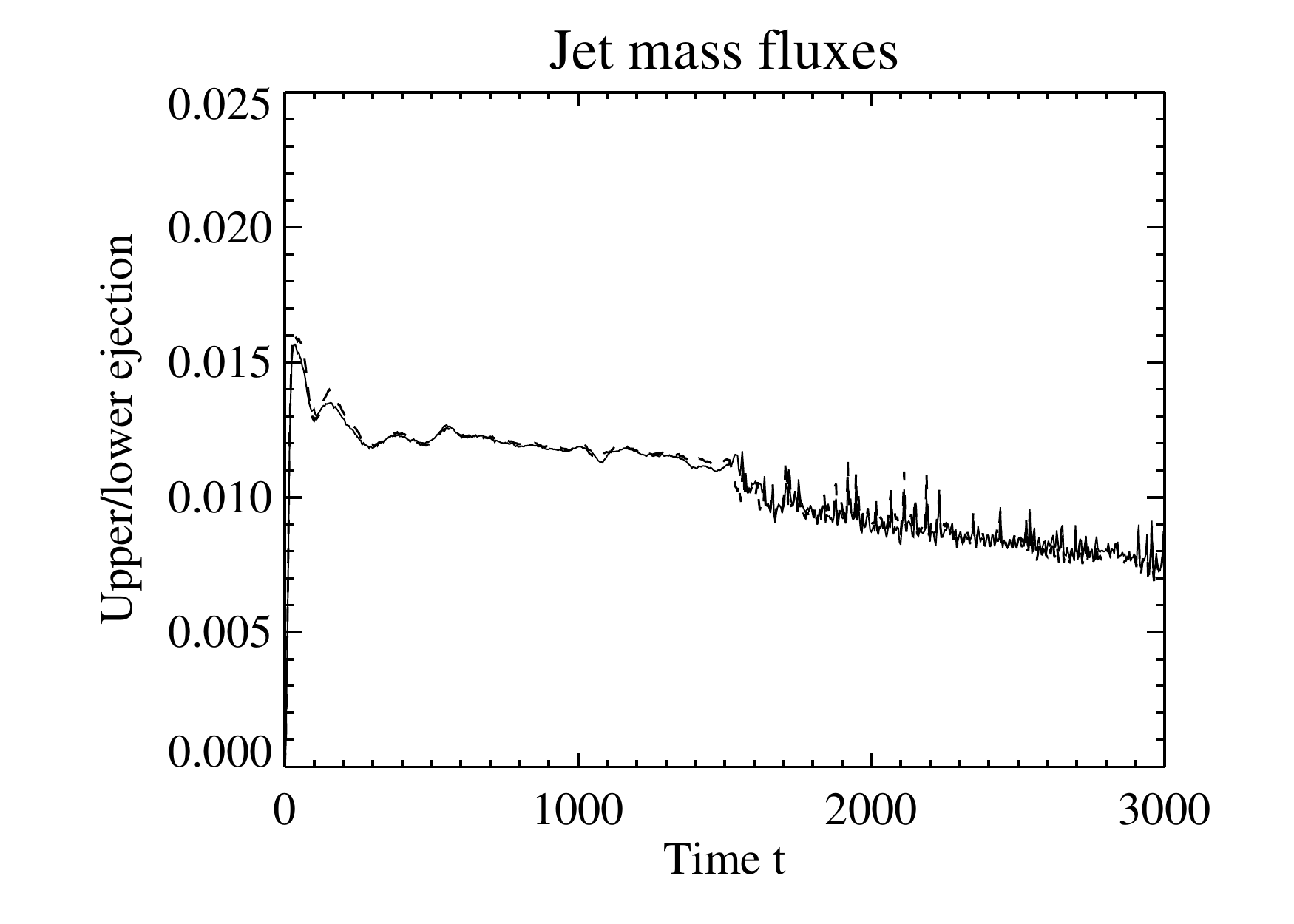}
\caption{Time evolution of the mass fluxes for reference simulation sb1.
Shown is evolution of accretion rate and the mass ejection rates from the upper (solid) and lower (dashed)
disk surfaces (all in code units). 
Ejection rates are measured in the control volumes with $r_1=1.5$ and $r_2 = 10.0$, while the accretion
rate is vertically integrated at $r=10$.}
\label{fig:bipo1_fix_case1_mass}
\end{figure}

\section{Magnetic diffusivity}
In order to extend of the setup of paper I into a truly bipolar configuration we need to re-consider
our model for the magnetic diffusivity $\eta(r,z)$.
An obvious constraint is that the prescription of diffusivity should not {\em a priori} influence 
the symmetry of the system.
Since the disk structure may now evolve asymmetric in both hemispheres, the magnetic diffusivity 
must be able to follow such a disk evolution. 
Naturally, in asymmetric disks the disk midplane does not follow the equatorial plane.

In general, we assume the diffusivity to be anomalous and of turbulent origin.
Since we do not resolve the disk turbulence from first principles (e.g. by resolving the MRI), we 
apply a parameterized diffusivity distribution effectively following in principle an 
$\alpha$-prescription.
We neglect geometrically more complex distributions such as e.g. MRI-inactive dead zones 
of turbulent diffusivity or a more detailed local treatment of MRI turbulence
\citep{1996ApJ...457..355G, 2007ApJ...668L..51P, 2010MNRAS.405...41G, 2012arXiv1210.6664F,
2013A&A...550A..61L, 2013ApJ...767...30B}.

As detailed in paper I, we assume a diagonal diffusivity tensor with the non-zero components
$\eta_{\rm \phi\phi} \equiv  \eta_{\rm p}$, and $\eta_{\rm rr}=\eta_{\rm zz} \equiv  \eta_{\phi}$,
where we denote $\eta_{\rm p}$ as the {\em poloidal magnetic diffusivity}, 
and $\eta_{\phi}$ as the toroidal magnetic diffusivity, respectively.
The anisotropy parameter $\chi = \eta_{\phi} /\eta_{\rm p}$ quantifies the different strength 
of diffusivity in poloidal and toroidal direction.
Here, we apply $\chi = 3.0$ for all simulations.

For the diffusivity profile, we have investigated several options which we will discuss in the 
following.
Table \ref{tbl:bicases} compares the parameter setups for the different magnetic diffusivity 
distributions applied in our simulations.

\subsection{A global magnetic diffusivity prescription}
A first option for the diffusivity distribution is to mirror the profile applied in paper I along 
the equatorial plane,
\begin{equation}
\eta_{\rm p}(r,z) = \alpha_{\rm m,1} f_1(r,z) \equiv 
               \alpha_{\rm m,1} v_{\rm A,0} H_0 \exp{\left(-\frac{2 z^2}{H_{\eta}^2}\right)},
\label{eq:magdiff_global}
\end{equation}
with the Alfv\'en speed $v_{\rm A,0} \equiv v_{\rm A}(r, z=0)$ and the disk thermal scale height 
$H_0 \equiv H(r) = \epsilon r = c_{\rm S}(r,z=0) / v_{\rm K}(r,z=0)$, 
measured at the midplane and at time zero.
Essentially, the diffusivity profile Eq.~\ref{eq:magdiff_global} is geometrically tied to the equatorial 
plane, potentially inducing hemispheric symmetry of the disk and the outflow.

As in paper I, we allow for a diffusivity scale height $H_{\eta}$ larger than the thermal scale height $H_0$ 
(see discussion in paper I).
In fact, \citet{2010MNRAS.405...41G} who investigated the MRI-induced turbulence of accretions disks by 
high-resolution box simulations, finds an {\em increasing} level of turbulence with disk height. 
His simulations attest a maximum level of turbulence at about 2-3 disk pressure scale heights, 
which are in nice agreement with our model approach.
More recent simulations by \citet{2011MNRAS.416..361B, 2013ApJ...764...66S} indicate similar scale heights 
for the turbulent stresses.

In Eq.~\ref{eq:magdiff_global}, both $v_{\rm A}$ and $c_{\rm S}$ can be chosen time-independent
(as in paper I), or evolving in time (as e.g. in \citealt{2010A&A...512A..82M}).
For the sake of comparison, in the present bipolar reference simulation (denoted by sb1) we 
apply a magnetic diffusivity profile $f_1(r,z)$ fixed in time.

\begin{figure*}
\centering
 \includegraphics[width=4.4cm]{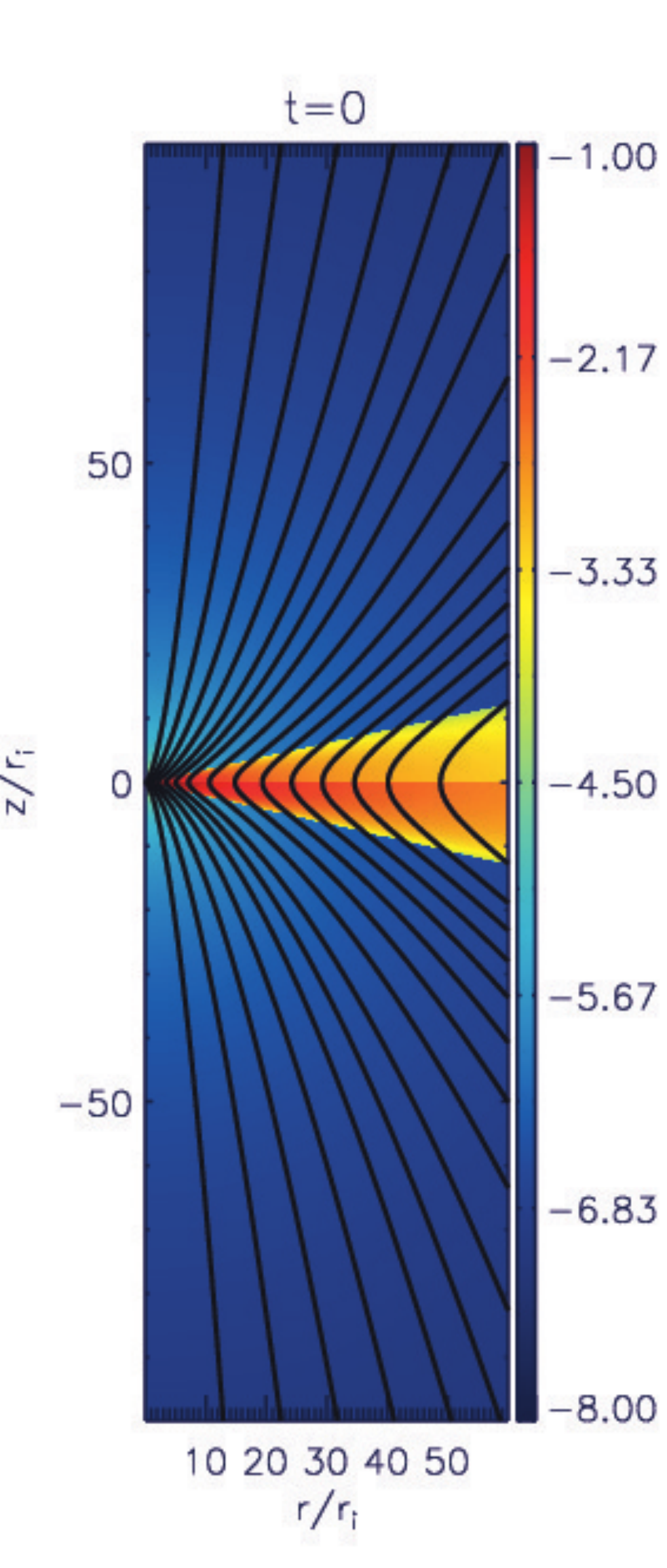}
 \includegraphics[width=4.4cm]{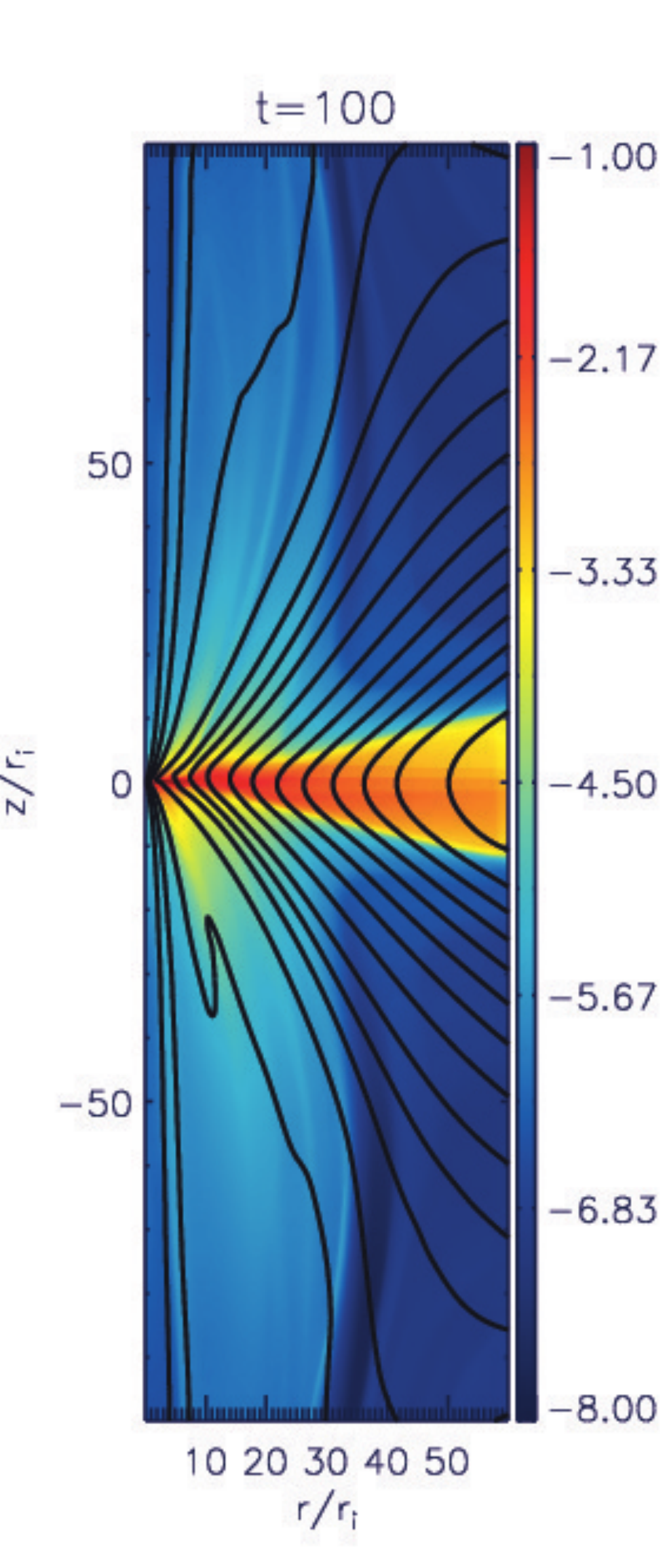}
 \includegraphics[width=4.4cm]{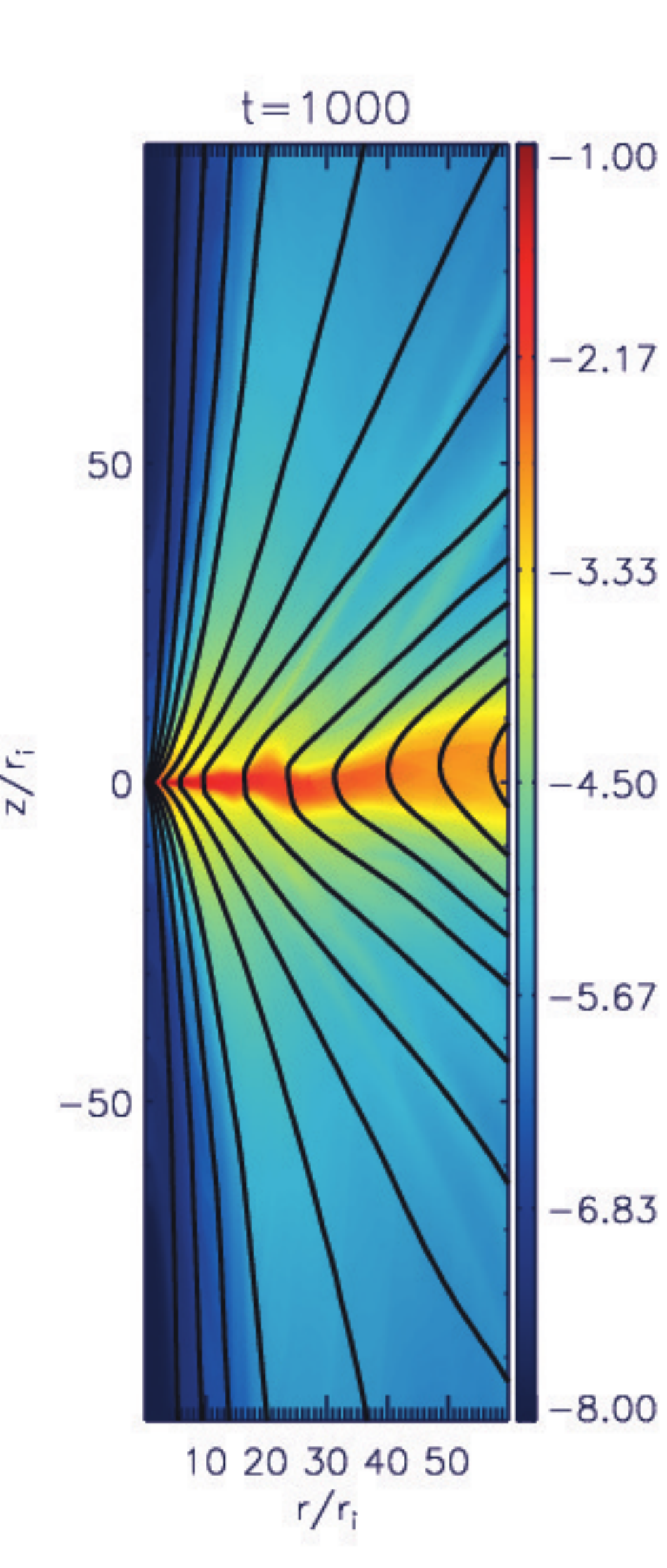}
 \includegraphics[width=4.4cm]{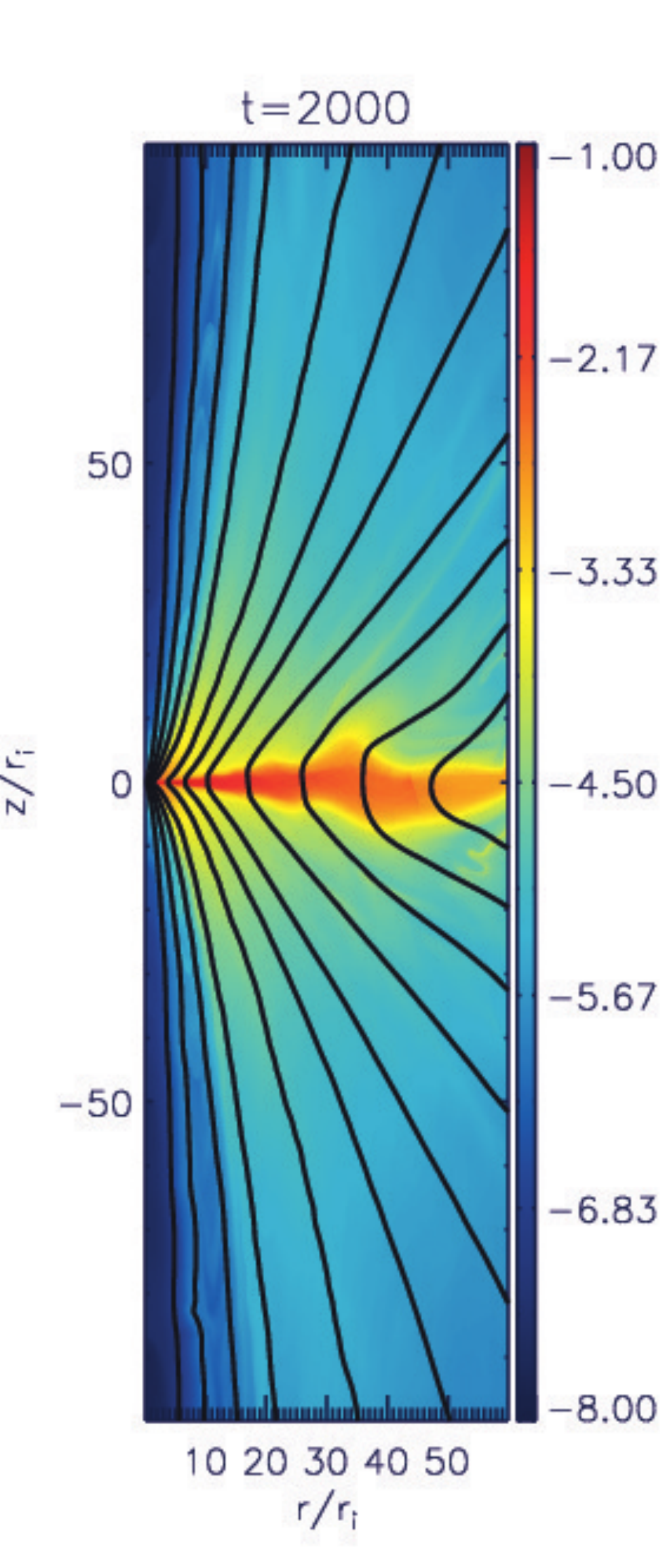}
\caption{Time evolution if the bipolar jet-disk structure for simulation sb2 
applying a fixed-in-time diffusivity profile Eq.~\ref{eq:magdiff_global}, and evolving
from an asymmetric initial state with different thermal scale heights for the upper and lower disk
hemisphere, $\epsilon_{up}=0.15$ and $\epsilon_{down}=0.1$.
We show the evolution for time $t = 0, 100, 1000, 2000$ of the mass density (colors) 
and the poloidal magnetic field (lines), i.e. contours of poloidal magnetic flux $\Psi$ with
flux contours
$\Psi = 0.01, 0.03, 0.06, 0.1$, $0.15, 0.2, 0.26, 0.35, 0.45$,
         $0.55, 0.65, 0.75, 0.85$, $0.95, 1.1, 1.3, 1.5, 1.7$. 
}
\label{fig:bipo1_fix_case2}
\end{figure*}

\begin{figure*}
\centering
 \includegraphics[width=5.8cm]{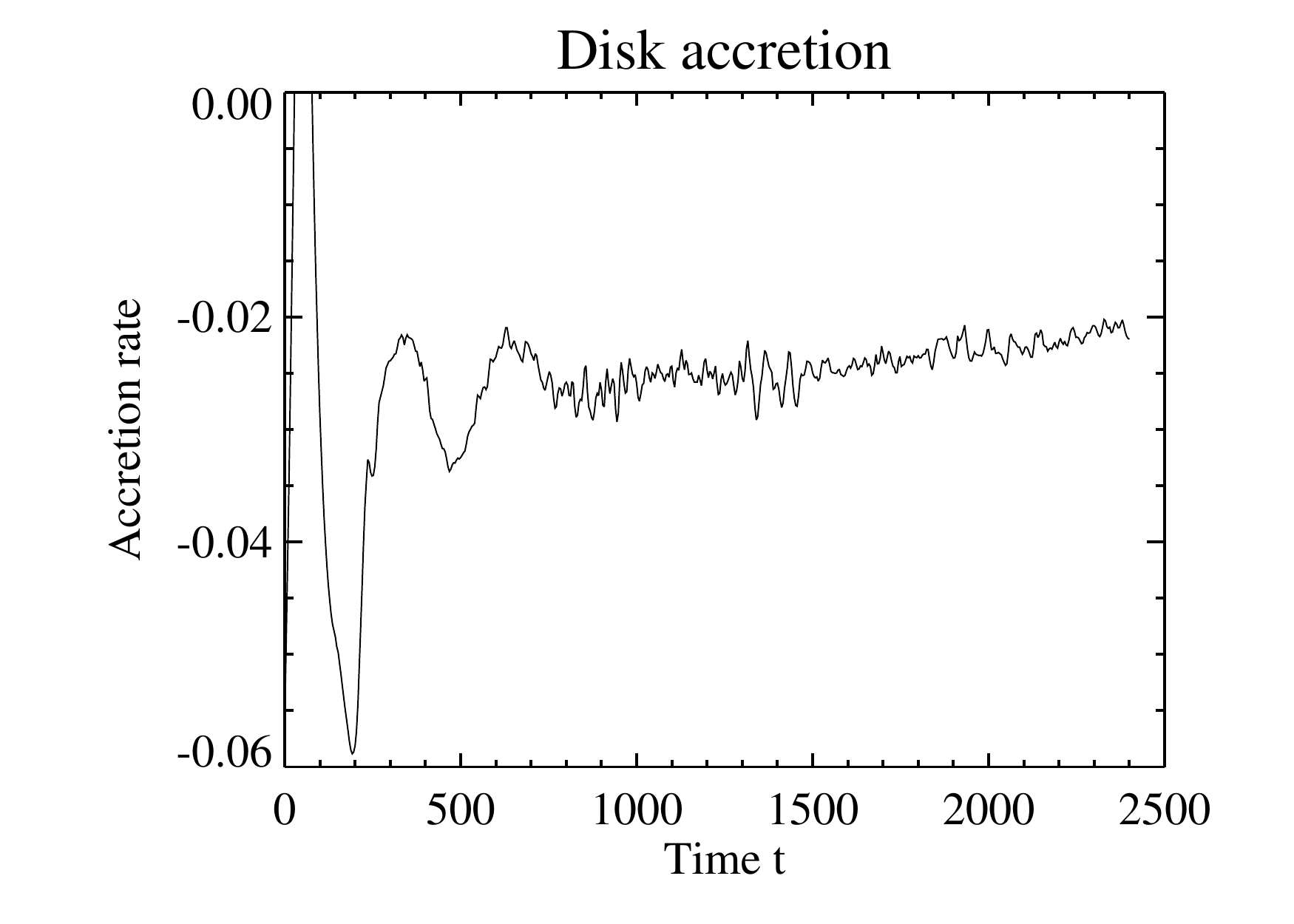}
 \includegraphics[width=5.8cm]{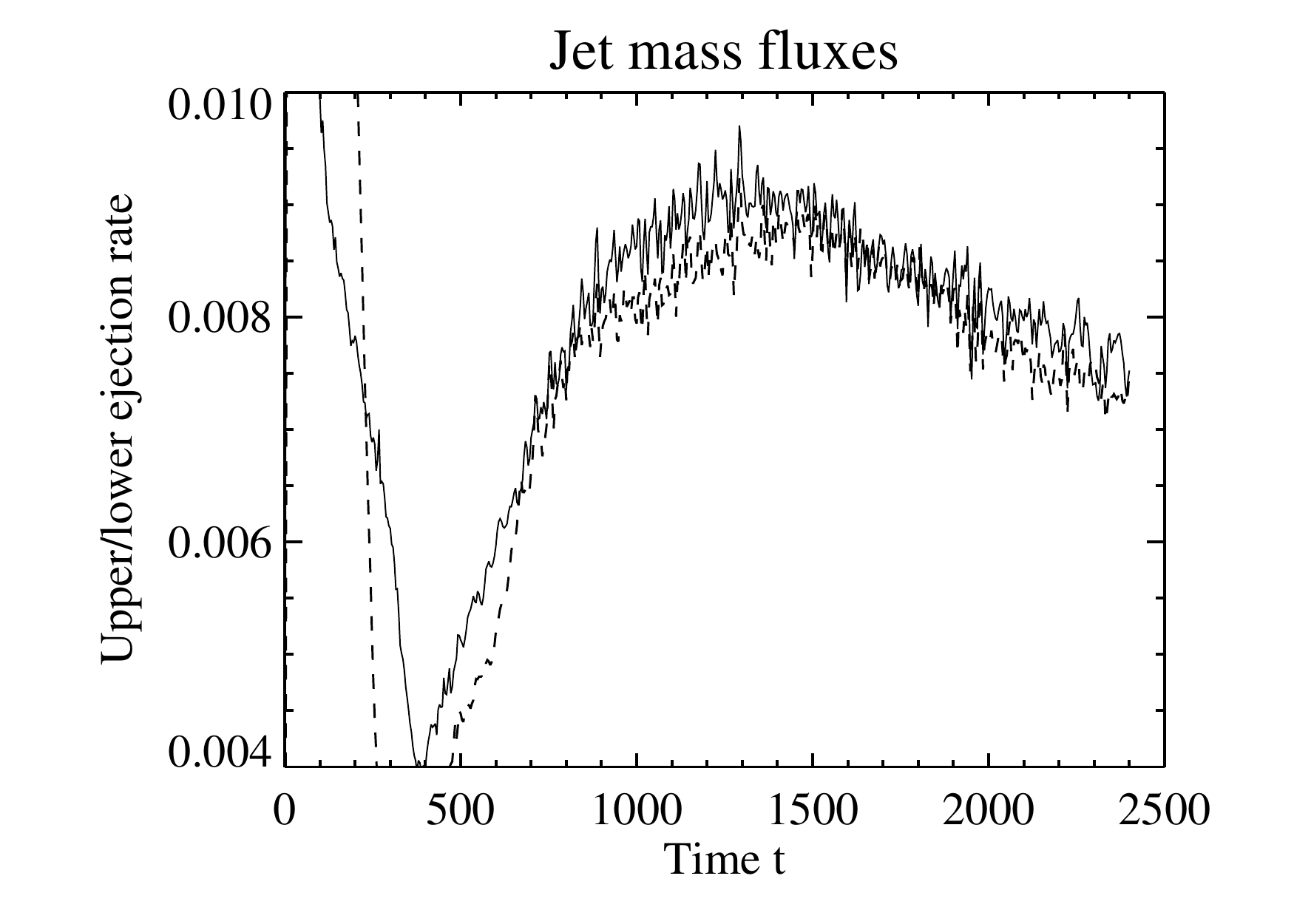}
 \includegraphics[width=5.8cm]{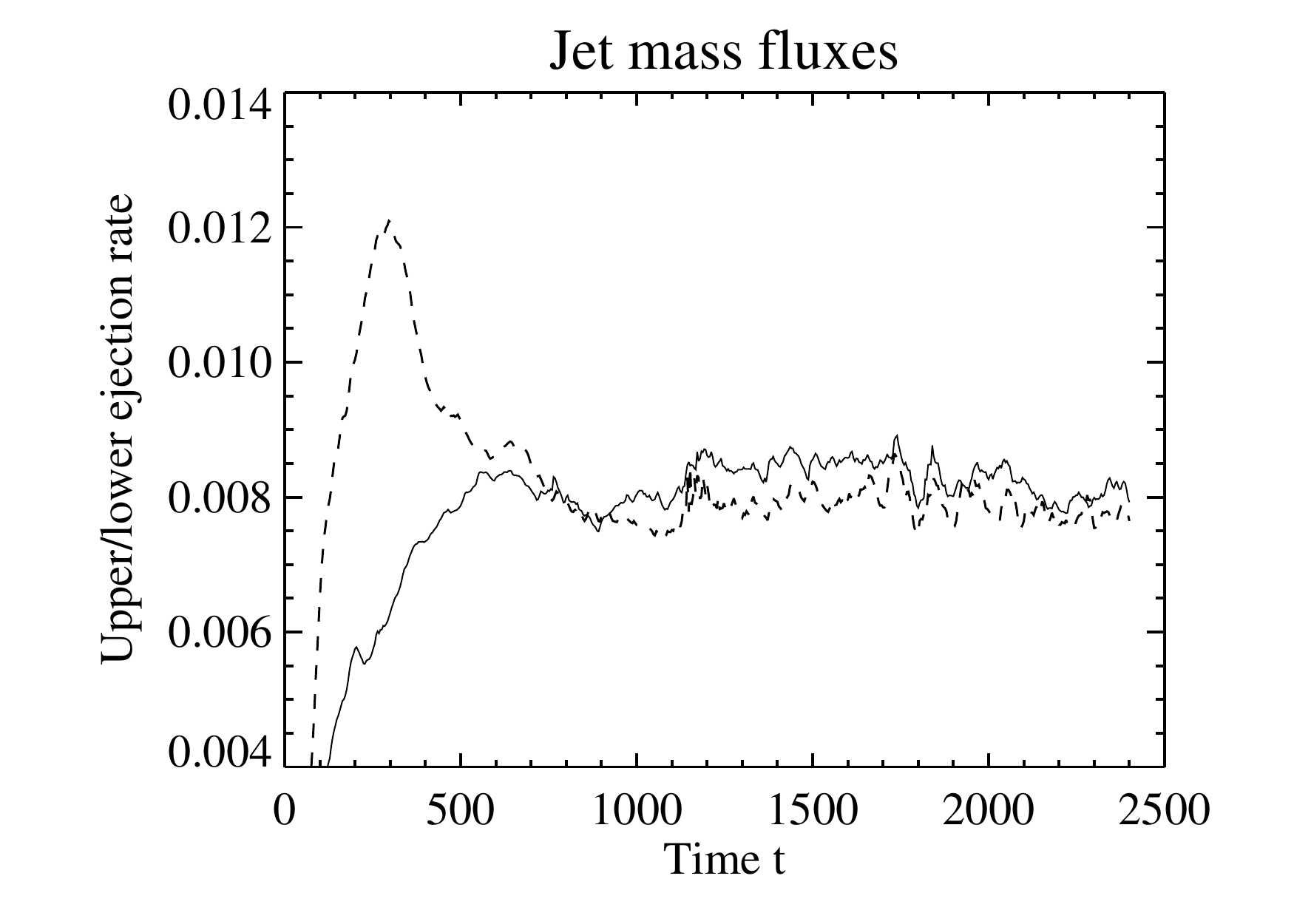}
\caption{Time evolution of the mass fluxes for simulation sb2.
Shown is evolution of accretion rate and the mass ejection rates from the upper (solid) and lower (dashed)
disk surfaces (all in code units). 
Ejection rates (middle) are measured in the control volumes with $r_1=1.5$ and $r_2 = 10.0$, 
while the accretion rate (top) is integrated at $r=10$.
For comparison we show the asymptotic (vertical) mass fluxes through the jet, integrated from 
$r_1=2.0$ to $r_2 = 50.0$ at $z= -75, 75$.}
\label{fig:bicase2_massfluxes}
\end{figure*}

\begin{figure}
 \centering
\includegraphics[width=6cm]{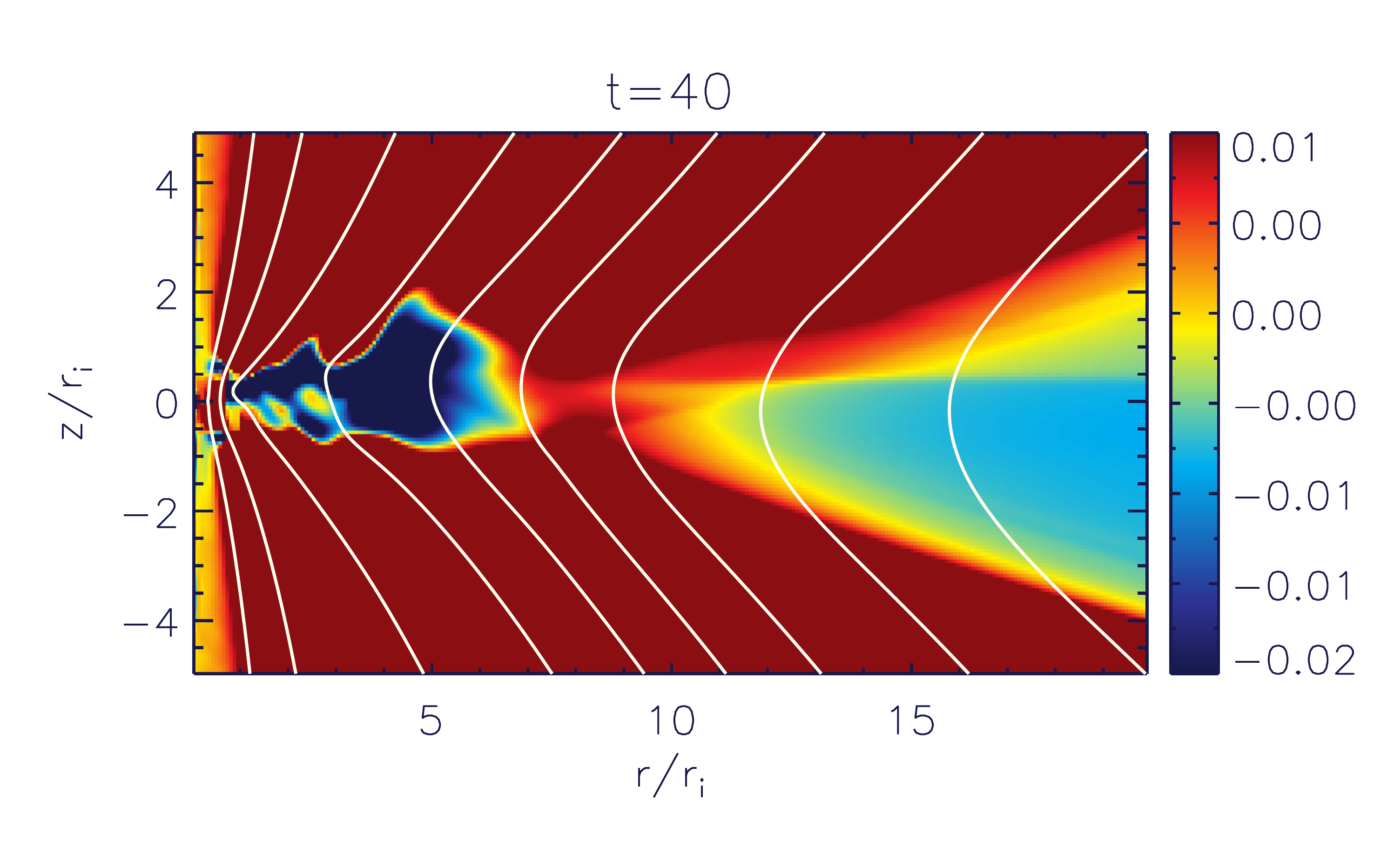}
\includegraphics[width=6cm]{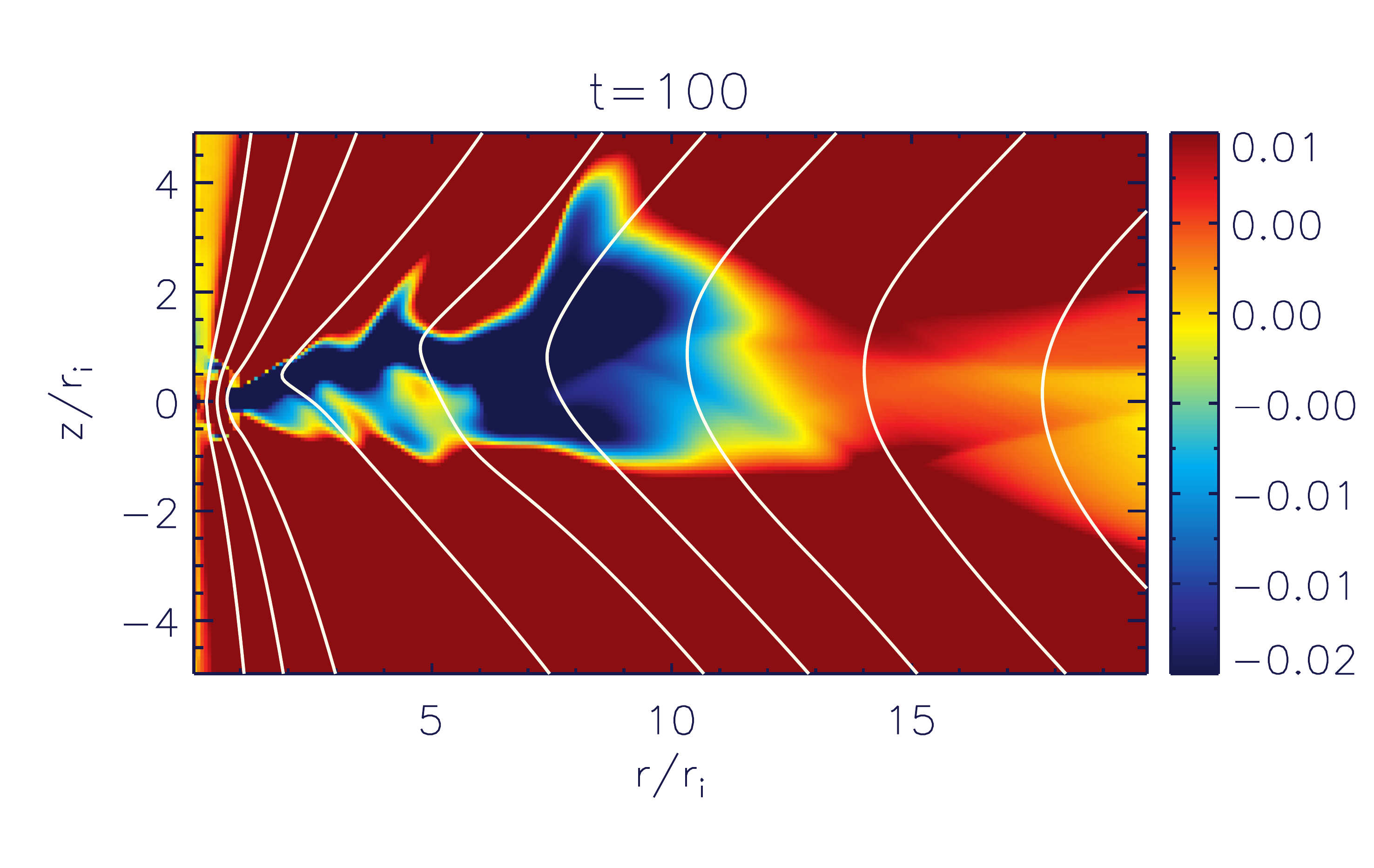}
\includegraphics[width=6cm]{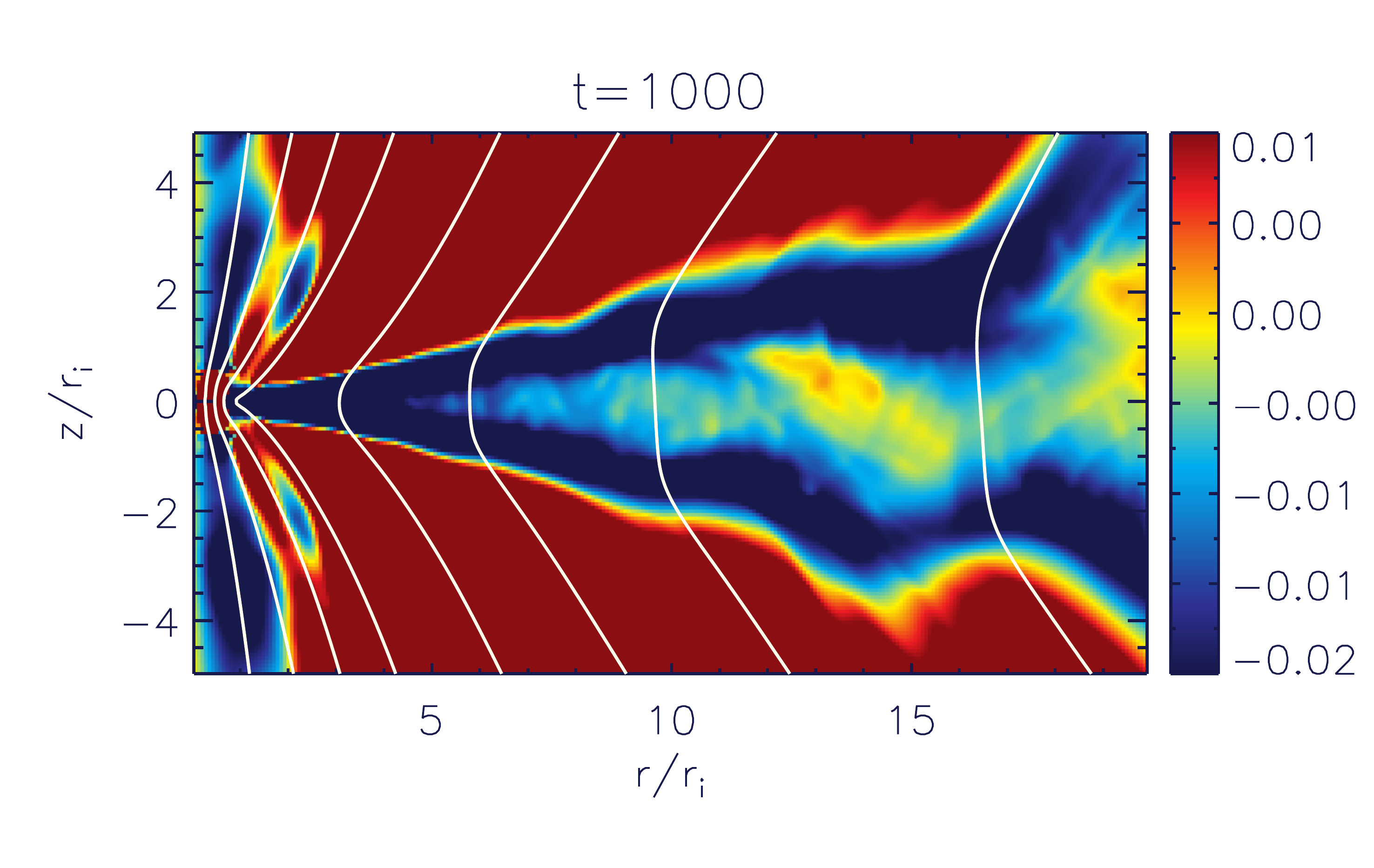}
\includegraphics[width=6cm]{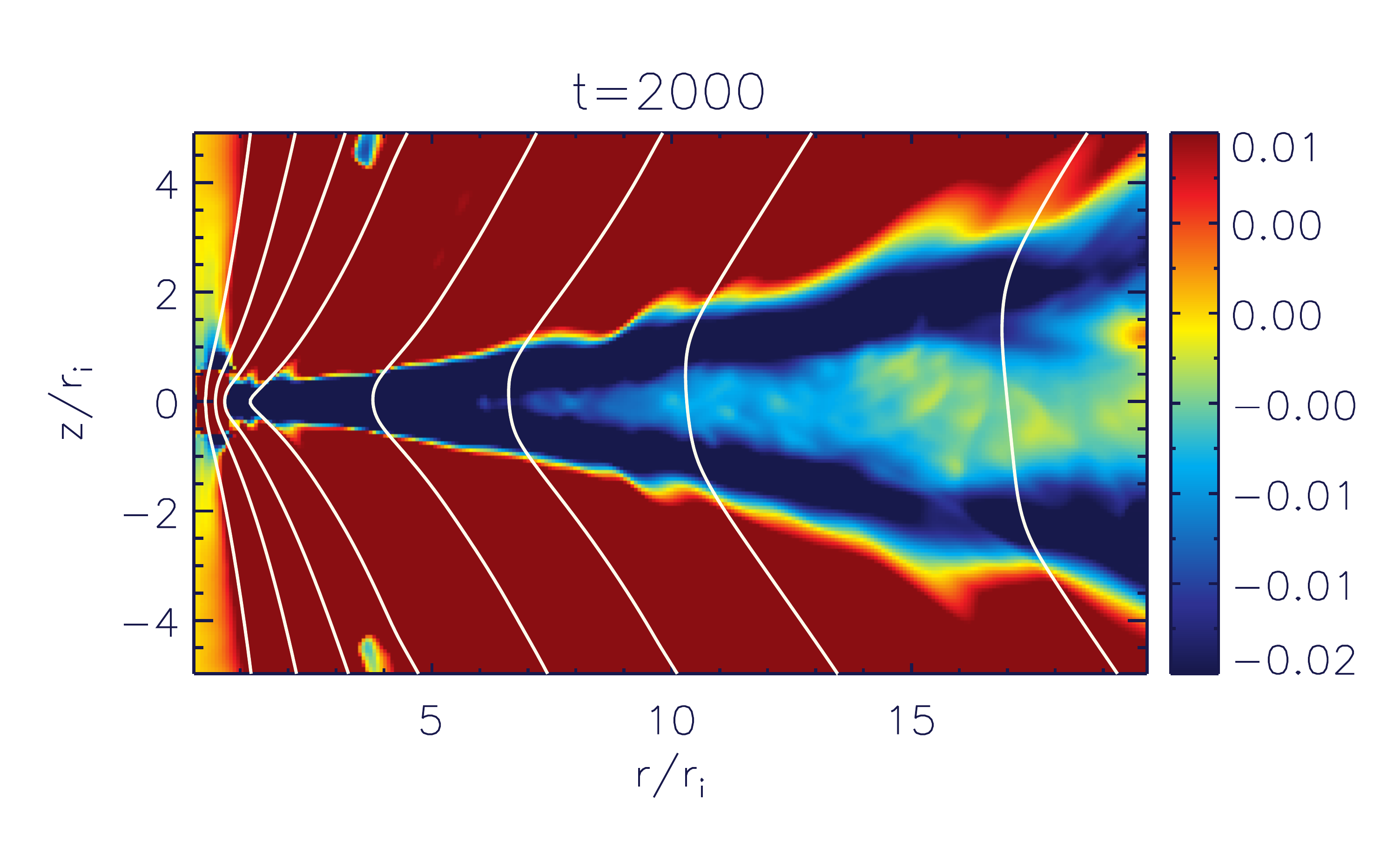}
\caption{Disk evolution for the jet-disk system sb2. 
Shown is the radial velocity $v_r(r,z)$ for the dynamical time steps  $t= 40, 100, 1000, 2000$ (from top) 
indicating the asymmetric evolution of the disk and the disk wind.
}
 \label{bicase2_innerdisk1}
 \end{figure}

\subsection{A local prescription of magnetic diffusivity}
In the present paper we investigate an asymmetric evolution of the disk-jet structure in both hemispheres.
Therefore, we need to apply a new, truly bipolar setup in which the leading parameters for the 
diffusivity profile are not defined with respect to the equatorial plane anymore.
It is clear that for an asymmetric disk evolution, a hypothetical disk mid-plane will not coincide with the 
equatorial plane.
Therefore, in order to consider the spatial evolution of the asymmetric disk structure, it is essential 
to apply a {\em local} prescription of diffusivity.

Among other options, one possibility is to relate the magnetic diffusivity to the local pressure or density. 
One may assume that the turbulent Alfv\'enic pressure is proportional to the thermal pressure 
as discussed e.g. by \citet{1997ApJ...482..712O}.
In particular, \citet{2002A&A...395.1045F} suggested that if magnetic diffusivity is interrelated to the 
turbulent {\em Alfv\'enic pressure}, the magnetic diffusivity profile will follow a power law 
$\eta_{\rm m} \sim \rho^{1/3}$. 
In this casse the diffusivity is proportional to the local sound speed $\eta_{\rm m} \propto c_{\rm S}$.
We may generalize the power law 
\begin{equation}
   \eta_{\rm p}= \alpha_{\rm m ,2} f_2(r,z) \equiv \alpha_{\rm m ,2} \rho^{\Gamma}.
\label{eq:magdiff_cemel}
\end{equation}
%
%
The profile with $\Gamma \simeq 1/3$ results in a relatively broad vertical diffusivity profile, 
wider than the previously used exponential profile Eq.\ref{eq:magdiff_global}.
It has been shown that this will impact the jet acceleration and collimation \citep{2002A&A...395.1045F},
maybe more than it effects the launching mechanism itself.
For comparison we have also applied different power laws, such as $\Gamma = 2/3$, or even steeper profiles.
The drawback of the simple profile \ref{eq:magdiff_cemel} is that the diffusivity {\em decreases} along the 
disk is decreasing with radius.
The outer disk material is therefore strongly coupled to the magnetic field, leading to an super efficient
angular momentum removal, rapid accretion, and, thus, a short life-time of the outer disk.
On the other hand, in case of MRI driven turbulence, the MRI activity is expected to cease for large radii, 
where "dead zones" for active accretion may exist 
\citep{1996ApJ...457..355G, 2011MNRAS.416..361B, 2013ApJ...764...66S}.

We have favored another option for the magnetic diffusivity which allows for a disk diffusivity 
{\em increasing} with radius,
\begin{equation}
\eta_{\rm p}(r,z) = \alpha_{\rm m,3}\, f_3(r,z)
            \equiv  \alpha_{\rm m,3}\, \tilde{H}_{\rm L}(r,z) \frac{1}{1 + \frac{\tilde{H}_{\rm L}(r,z)}{\tilde{H}_0}. },
\label{eq:magdiff_local}
\end{equation}
Thus, we apply a density-weighted "local disk scale height" $\tilde{H}_{\rm L}(r,z)$ mainly following 
the local sound speed in the gas flow,
\begin{equation}
\tilde{H}_{\rm L}(r,z) = \rho(r,z)^{\sigma_{\rho}} \sqrt{ \gamma \frac{P(r,z)}{\rho(r,z)} } \frac{1}{v_{\rm Kep}(r)} r^{\sigma_{r}},
\end{equation}
and a quenching term in order to avoid numerically problematic diffusivities.
We typically choose $\tilde{H}_0 = 0.5$, $\sigma_{\rho} = 3/2$, $\sigma_{r} = 3/2$,or
                    $\tilde{H}_0 = 0.1$, $\sigma_{\rho} = 3/2$, $\sigma_{r} = 5/2$ (simulation run cb14).

Essentially, both prescriptions for the magnetic diffusivity Eqs.~\ref{eq:magdiff_cemel} and
\ref{eq:magdiff_local} allow for a smooth transition from accretion to ejection, and, also
to follow the changes in the local disk structure.
As the outflow density decreases along the streamlines, also the outflow diffusivity 
decreases. 
For low densities towards the asymptotic outflow, the ideal MHD will be approached.
A physical motivation can be the following.
Turbulently diffusive disk material is lifted from the disk into the disk, and is further 
accelerated along the outflow while, however, the turbulent motions decay.
In paper I we have estimated the scale height where the turbulence will be damped to be several 
disk thermal scale heights (see our discussion above citing \citealt{2010MNRAS.405...41G}).

\section{Test case - launching symmetric bipolar jets}
We first present jet launching simulations resulting in symmetric bipolar jets.
The first example is the evolution of jets following a magnetic diffusivity description
fixed in time and space (case sb1).
This case serves as reference simulation for this paper, and also allows for a comparison
to the one-hemispheric simulations of paper I.
This inflow-outflow evolution is shown in Fig.~\ref{fig:bipo1_fix_case1}.

As we see, the outflow evolves perfectly symmetric in both hemispheres for almost 3000
dynamical time steps, until the outer disk starts to deviate from symmetry due to
numerical effects.
However, even for these late time steps, the inner disk, which is the main jet launching area,
is still highly symmetric.

In particular this could be seen in the time-evolution of the mass fluxes.
Figure \ref{fig:bipo1_fix_case1_mass} shows the accretion rate and the ejection rates
into the two hemispheres.
The ejection rates are integrated from $r=2$ to $r=20$ at $z = \pm 3\,H(r)$, while the
accretion rate is integrated from $z=-3H$ till $z=3H$ at $r=10$.
It is not surprising that due to the symmetric diffusivity profile which remains fixed in time 
together with symmetric initial and boundary conditions, we find a symmetric evolution of both
the disk and the outflows.
The symmetric bipolar jet structure we obtain does mainly serve as a test case for the numerical
setup, in particular the sink boundary conditions.
While the symmetric evolution is expected on physical grounds, even a little numerical failure in 
the setup would lead to asymmetry quickly.

In comparison to our one-hemispheric simulation in paper I (reference run case1), the mass fluxes 
we measure now are quite similar.
As accretion rate for the bipolar simulation we measure $M_{\rm acc}(t=2000) \simeq 0.03$, while for
the previous reference case1 it was $M_{\rm acc}(t=2000) \simeq 0.015$ 
The outflow rates for the bipolar simulation are $M_{\rm ejec}(t=2000)\simeq 0.01$ in each direction,
which agrees with the $M_{\rm ejec}(t=2000) = 0.007$ in paper I nicely. 
However, due to the smaller grid extension, the disk mass reservoir is smaller, leading
to a faster decay of the disk mass (see the outer disk structure at $t=3000$).
Therefore, disk accretion and mass ejection decay faster as well.


\section{Bipolar jets from asymmetric disks}
Here we present simulations in which we have disturbed the internal hemispheric disk symmetry.
We have applied two options for disturbing the disk symmetry - either 
by prescribing a {\em global} asymmetric initial state, 
or by injecting a {\em localized} overpressure (an explosion at certain time).
For both alternatives we apply a symmetric magnetic diffusivity prescription Eq.~\ref{eq:magdiff_global}, 
constant in time.

\subsection{An asymmetric disk scale height}
Option I is to prescribe an initial disk structure with a global pressure asymmetry in both disk hemispheres.
In our model, we achieve this by applying a different thermal disk scale height for the initial disk in each 
hemisphere.
In simulation sb2 we have applied $\epsilon = H/r = 0.1$ for the upper hemisphere and $\epsilon  = H/r = 0.15$ 
for the lower hemisphere.
Consequently, we have a density and a pressure jump across the equatorial plane, $\Delta P /P = 0.2$.

The disk turns into an asymmetric evolution right from the beginning, evolving into a warped 
structure\footnote{Note, however, the axisymmetric setup} along the midplane (Fig.~\ref{fig:bipo1_fix_case2}).
A series of warps is visible along the disk with warp amplitudes of a few local disk scale heights.
After about 1000 dynamical time steps the warp amplitudes start to decrease - first along the inner 
disk, while the outer disk still in a warped state.

The disk asymmetry is reflected in the jet evolution.
Along with the initial asymmetric disk evolution, the jets launched from the inner disk are asymmetric
as well. 
This is clearly visible in the poloidal magnetic field structure (Fig.~\ref{fig:bipo1_fix_case2}), 
but also in the mass fluxes we measure.
We also note a different time scale in the jet propagation. 
The upper jet reaches the grid boundary earlier than the lower jet which is delayed by about $\Delta t = 10\%$ 
(note that this obviously depends on the grid size, and happens rather early at $t < 50$).

Figure \ref{fig:bicase2_massfluxes} shows the time evolution of the mass fluxes.
Comparing both jet fluxes we find that at early stages, $t<700$, the lower 
outflow  carries about 80\% of the mass load of the upper outflow.
From $t\simeq 700-1500$ an asymmetric inflow-outflow system is established.
The differences in mass flux are now about 10\%.
After $t\simeq 1500$, the variation in the outflow rates decreases, and the inner disk has 
established a symmetric structure.

The same behavior is also visible in the velocity distribution.
Figure \ref{bicase2_innerdisk1} shows the complex velocity field of the inner disk.
Accretion starts first in the innermost disk regions. 
The disk asymmetry is well reflected in the $v_r$-distribution.
However, after $t=1000$ the system turns to a symmetric geometry.
We also see the highest accretion velocities in the upper disk layers.
One may indicate this as {\em layered accretion}, however, due to the higher density
in the lower disk layers, the radial mass flux is more almost distributed over the
disk height.

We believe that the disk evolution into a symmetric state is given by two reasons.
Firstly by the restoring force of the symmetric gravitational potential, and 
secondly by the symmetry and the time-independent prescription of the magnetic diffusivity.
The latter is interesting since it is a second-order effect only, as the magnetic diffusivity 
does not provide a force term in the MHD equations which could directly re-configure the disk 
structure.
The diffusion time scale $\tau_{\eta} = \Delta Z^2 /\eta \simeq H^2/\eta  \simeq $ 
is about 100 dynamical time scales  for $\eta \simeq 0.01,  H \simeq 1$ at radius $r =10$.

So far we have concentrated on the inflow-outflow dynamics close to the jet launching region.
We now consider the evolution further downstream the outflows.
This is interesting as it is this part of the outflow which is in principle accessible by 
the observations.
Figure \ref{fig:bicase2_massfluxes} (bottom) shows the mass fluxes of jet and counter-jet far
from the source, integrated from $r = 2$ to $r = 50$ at $z = 75$, and $z = -75$, respectively.
We find that the mass flux asymmetry of about 5\% at the launching region propagates 
to the asymptotic region, where we find a similar mass flux difference.
The time lag between launching region and asymptotic domain of about
$\Delta t = 300 = 1200 -900$ can be explained by the propagation period of the 
accelerating outflow of velocities $v_z \simeq 0.2-0.8$ propagating a distance $\Delta z = 75$.
The maximum jet velocity is achieved along a narrow cone between $r=10$ and $r=20$ (at $z=100$).
The outflow velocity of the lower jet is about $1.2$, while for the upper jet it is $1.1$
times the Keplerian speed at the inner disk radius.
The bulk of the mass flux of the outflow is, however, located between $r=15$ and $r=30$, and is
a factor 4 higher than for the fast flow and moves with a speed $v_z \simeq 0.4$.


\begin{figure}
\centering
\includegraphics[width=6cm]{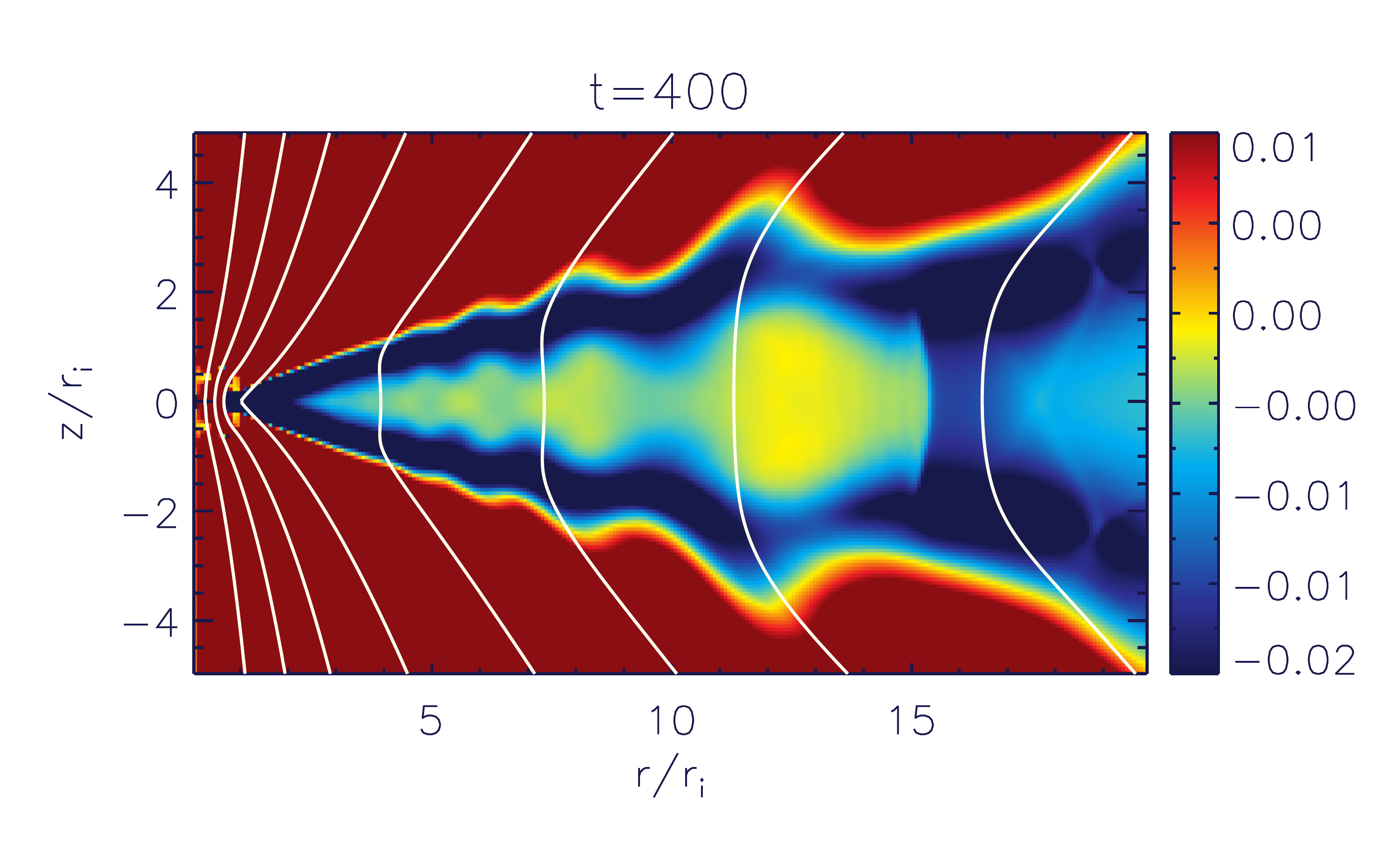}
\includegraphics[width=6cm]{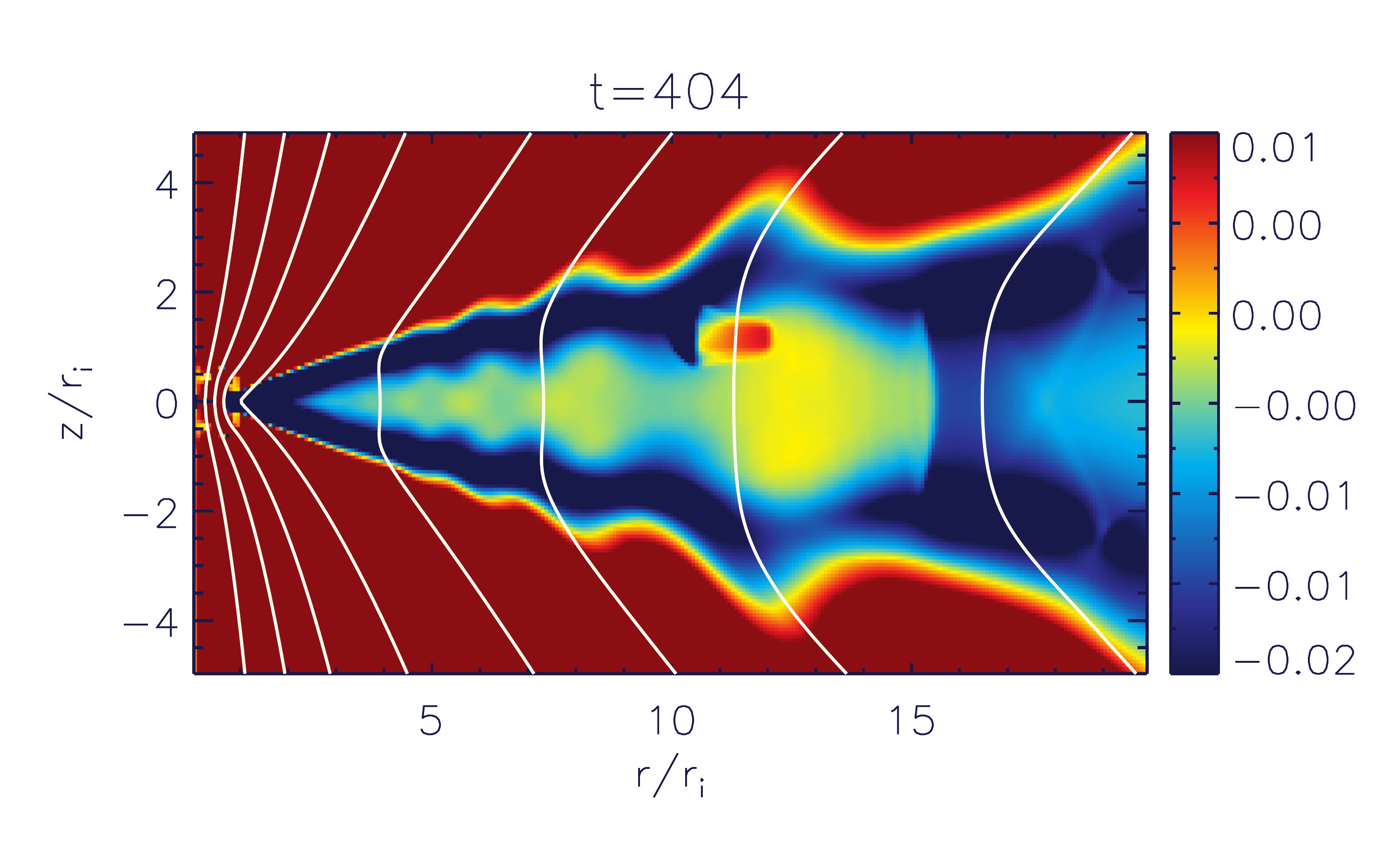}
\includegraphics[width=6cm]{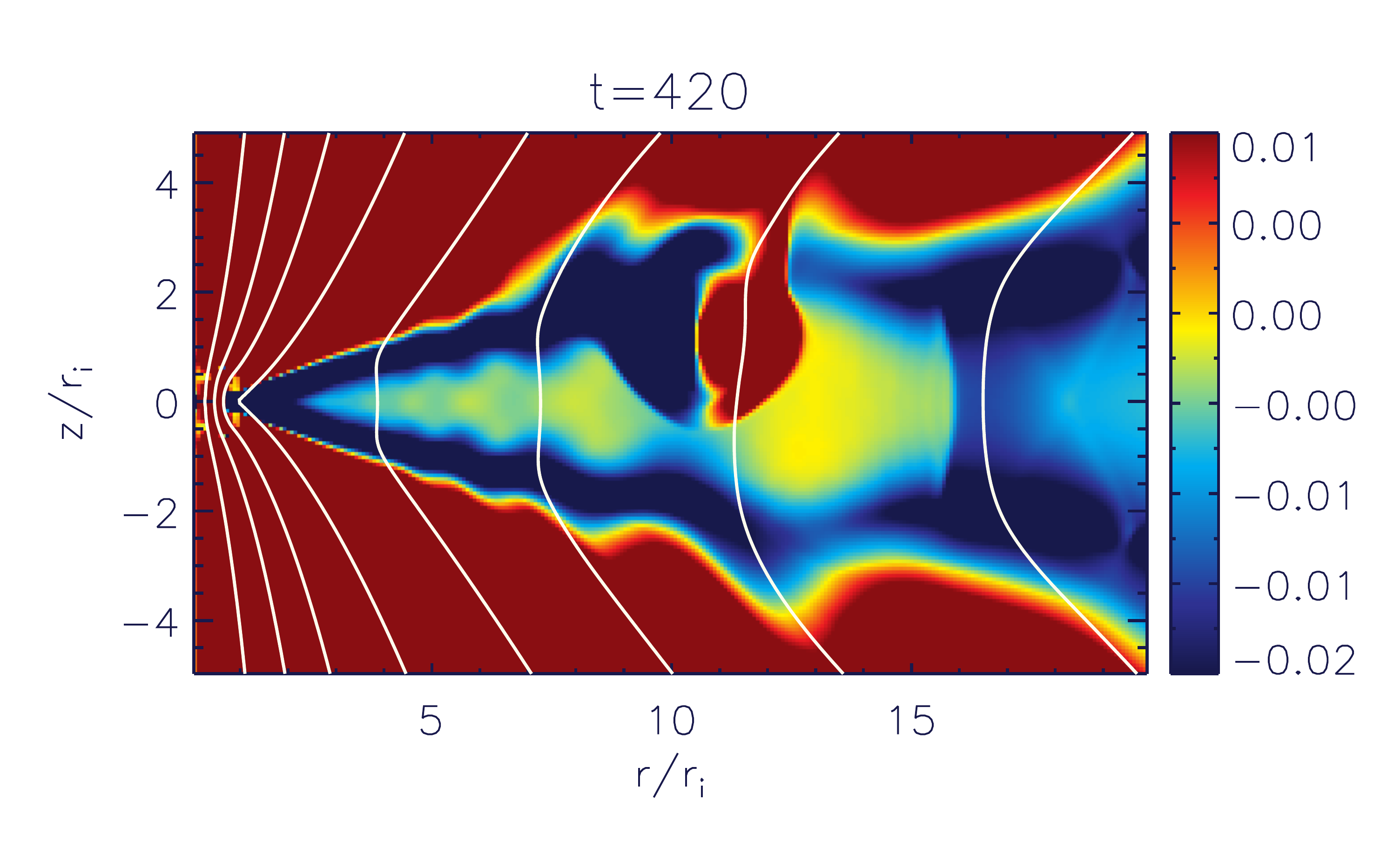}
\includegraphics[width=6cm]{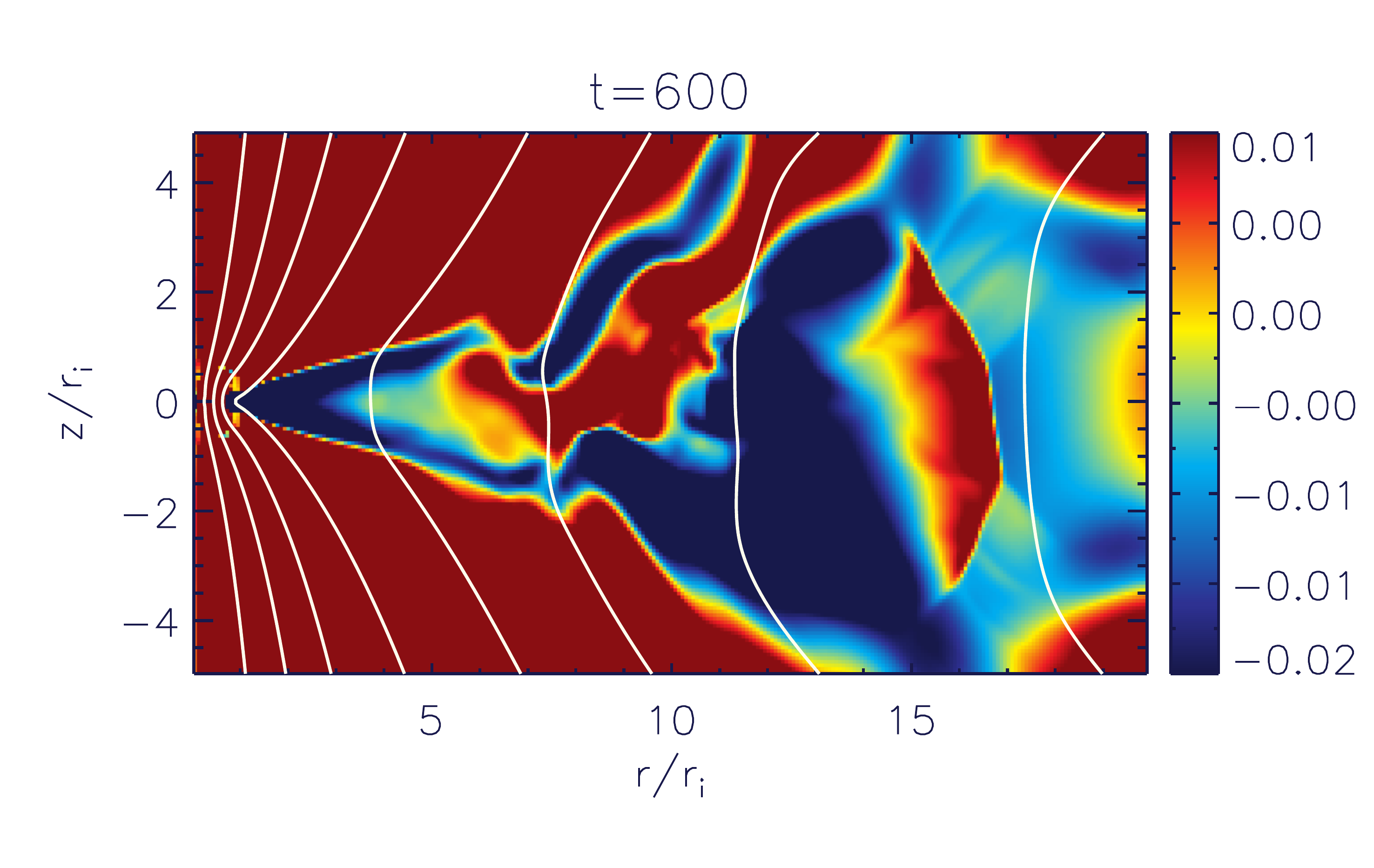}
\includegraphics[width=6cm]{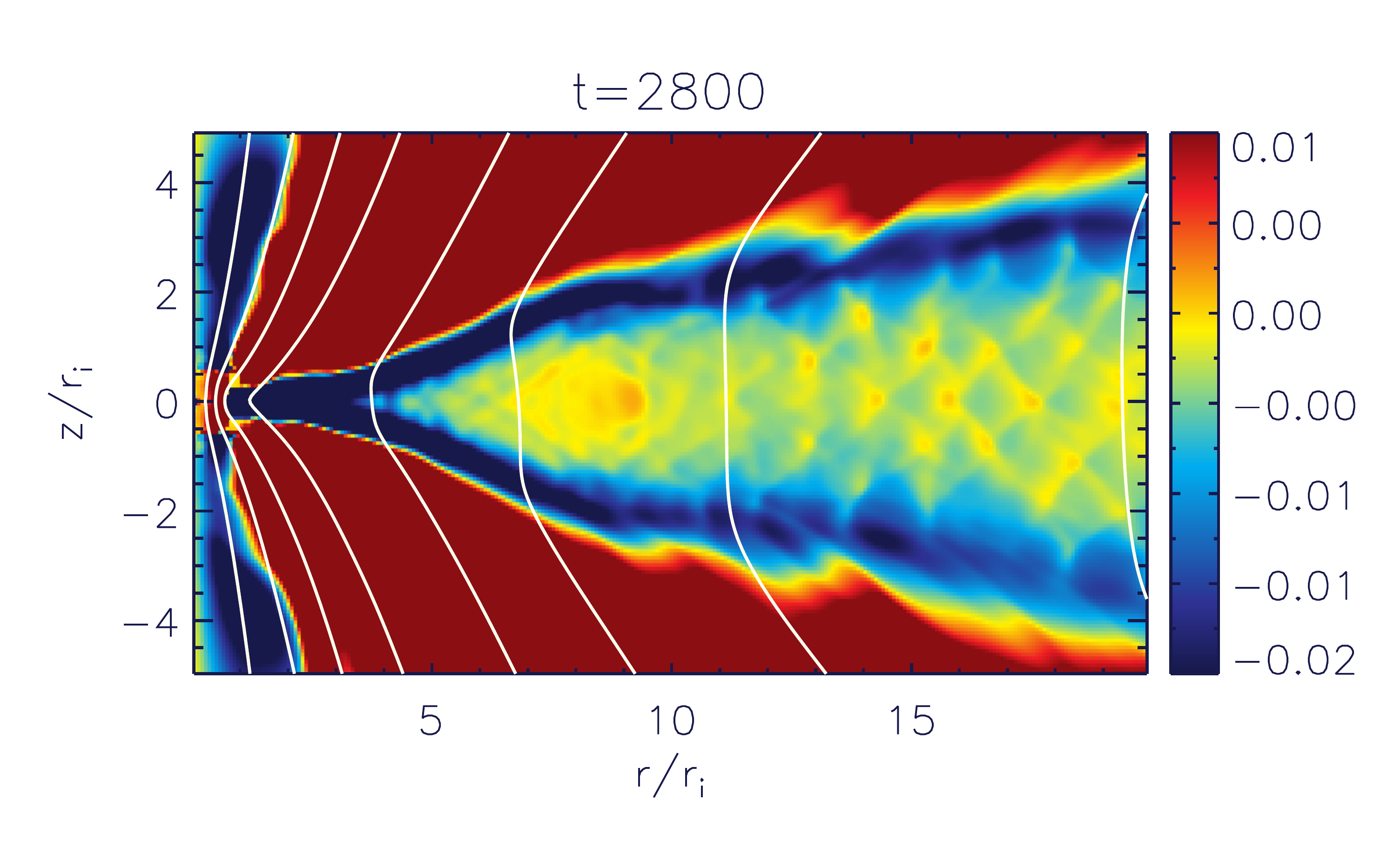}
\caption{Time evolution of the inner disk accretion for simulation sb3,
applying the diffusivity distribution fixed in time Eq.~\ref{eq:magdiff_global}.
An overpressure is added at $t=400$ in the upper disk hemisphere at radius $r = 12$,
lasting for few rotations.
Shown is the evolution of the mass density (color) 
for the dynamical times $t = 400, 404, 420, 600, 2800$.
}
\label{fig:bicase3-v1}
\end{figure}

\begin{figure*}
\centering
\includegraphics[width=3.5cm]{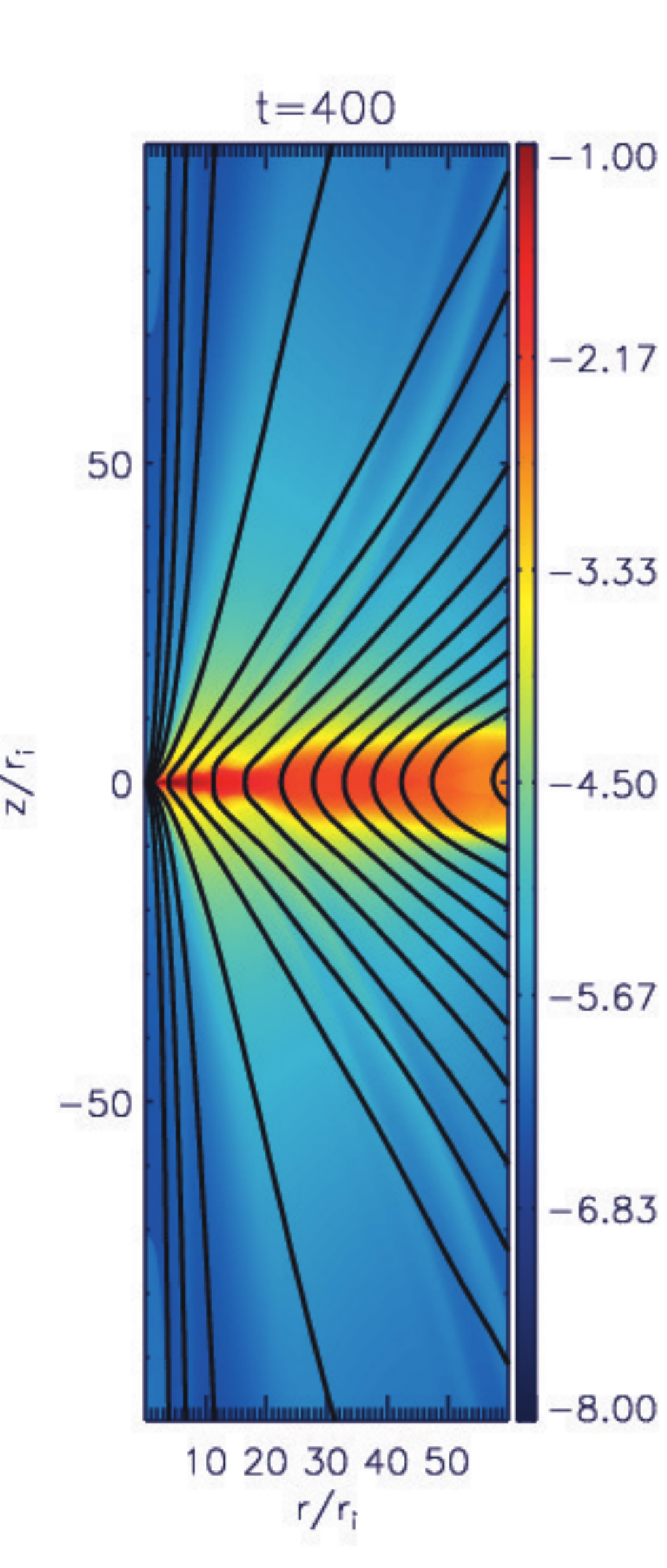}
\includegraphics[width=3.5cm]{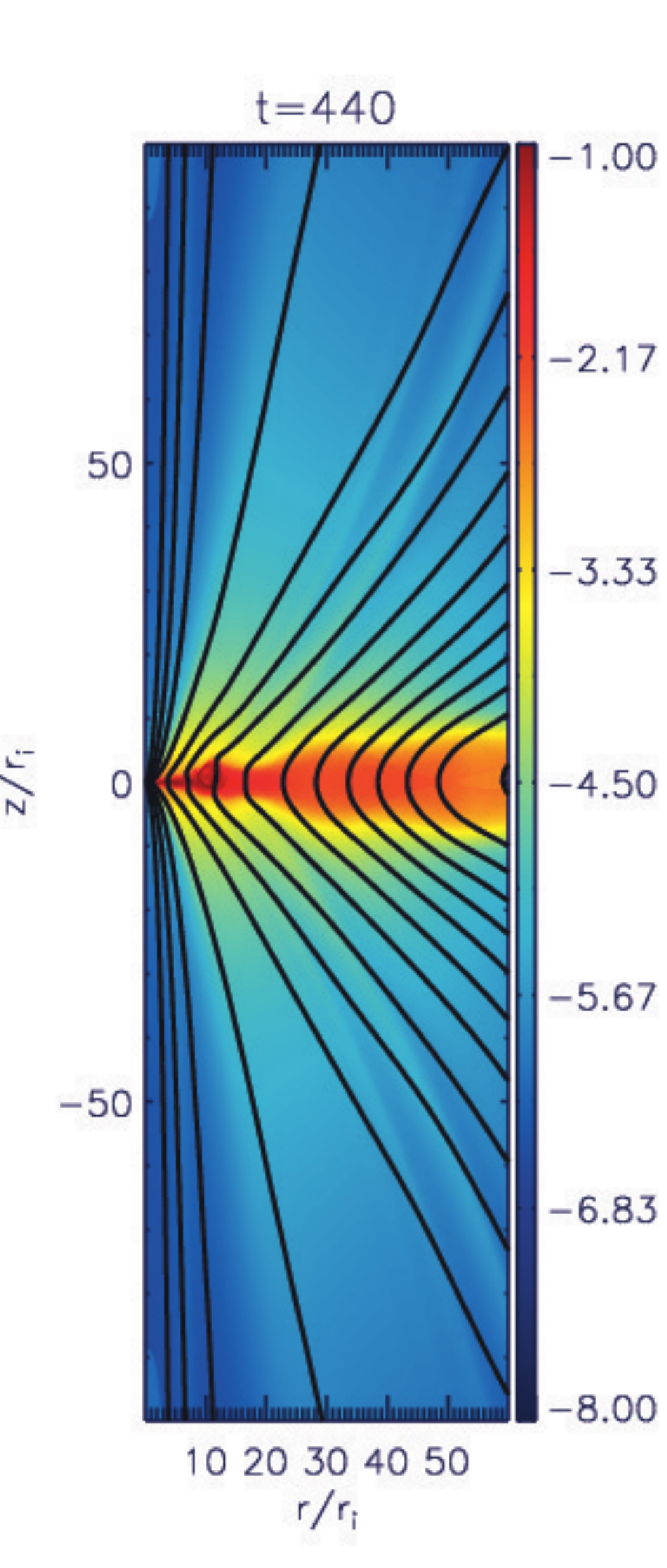}
\includegraphics[width=3.5cm]{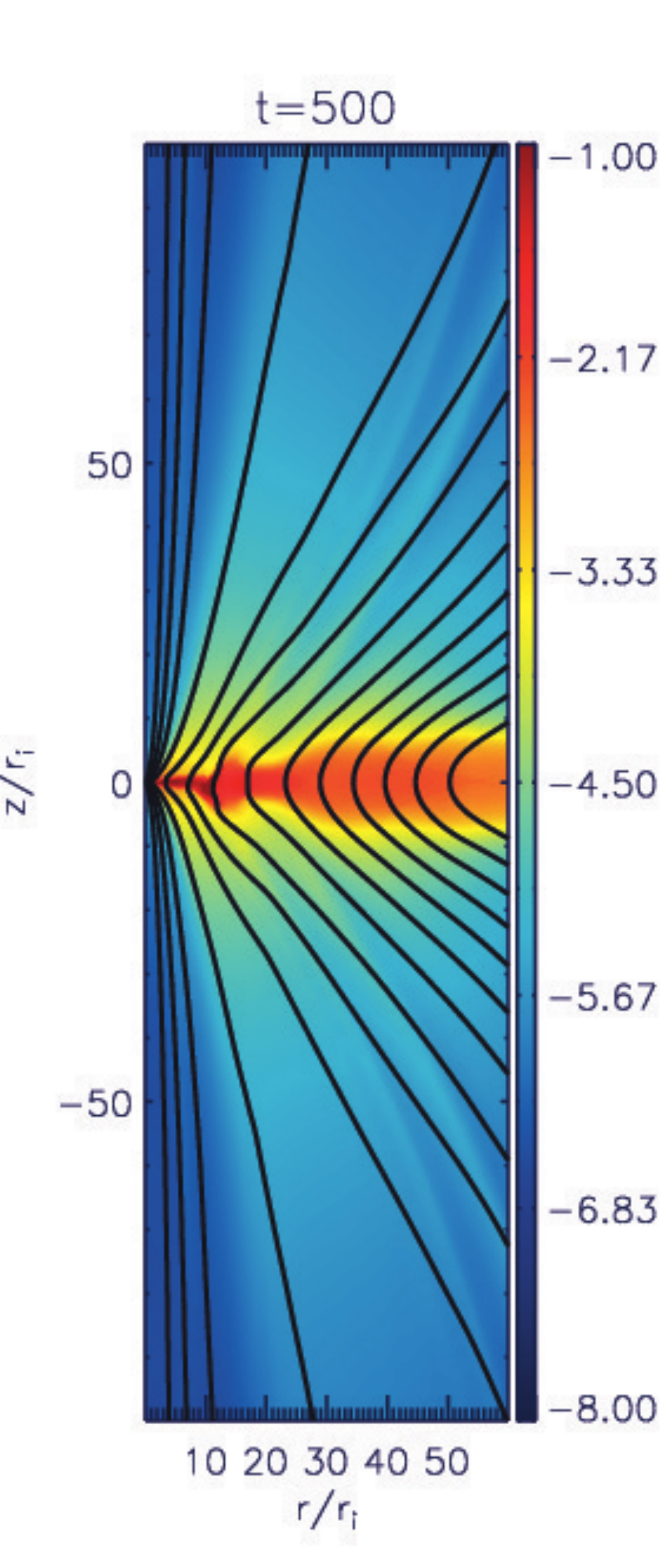}
\includegraphics[width=3.5cm]{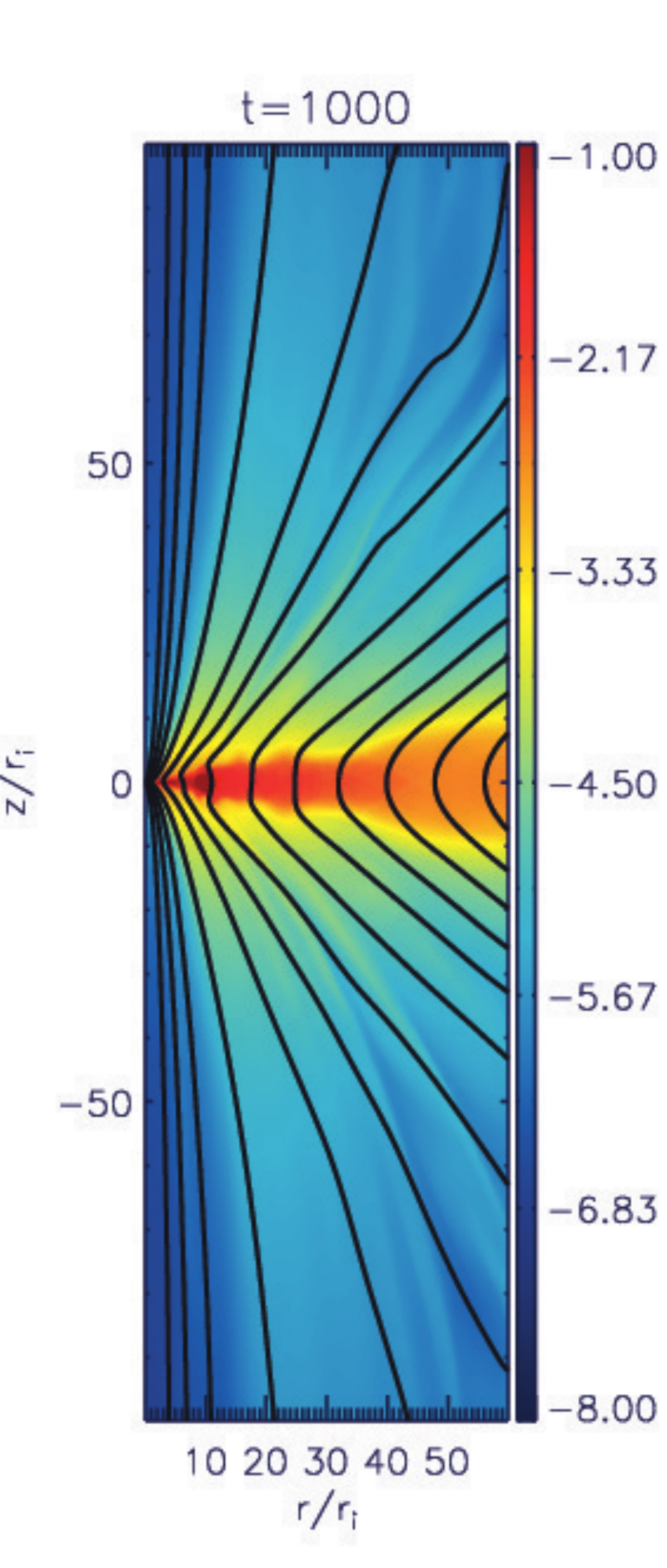}
\includegraphics[width=3.5cm]{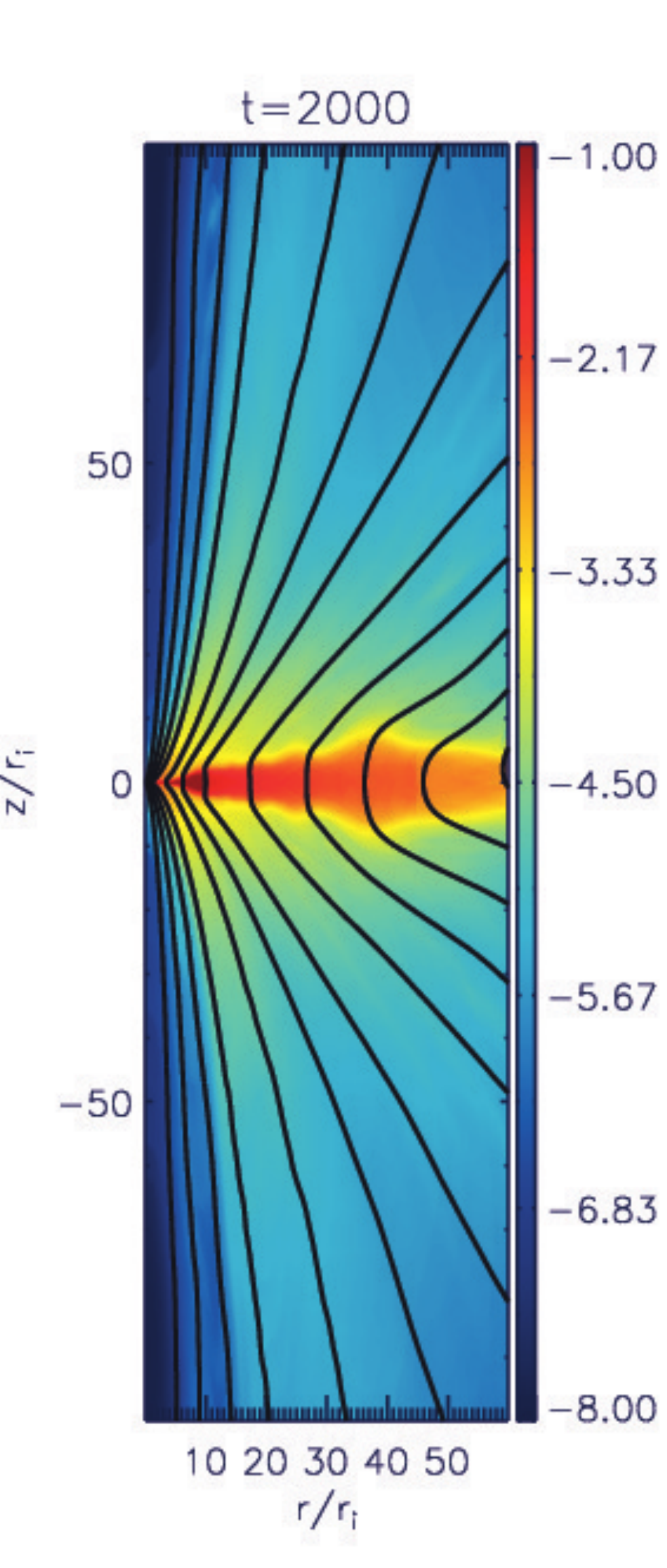}
\caption{Time evolution of the bipolar jet-disk structure for a simulation sb3 with 
the fixed-in-time diffusivity distribution in time, Eq.~\ref{eq:magdiff_global}, but with a 
localized overpressure added at $t=400$ for $\Delta t = 20$. 
See Fig.~\ref{fig:bicase3-v1} for a higher resolution of the inner disk area.
We show the evolution for time $t = 0, 100, 1000, 2000, 3000$ of the mass density (color) 
and the poloidal magnetic field (lines), i.e. contours of the magnetic flux, with  
flux levels $\Psi = 0.01, 0.03, 0.06, 0.1, 0.15, 0.2$, $0.26, 0.35, 0.45, 0.55$, 
$0.65, 0.75, 0.85, 0.95$, $1.1, 1.3, 1.5, 1.7$.
}
\label{fig:bicase3-rho-mag}
\end{figure*}

\begin{figure*}
\centering
 \includegraphics[width=5.9cm]{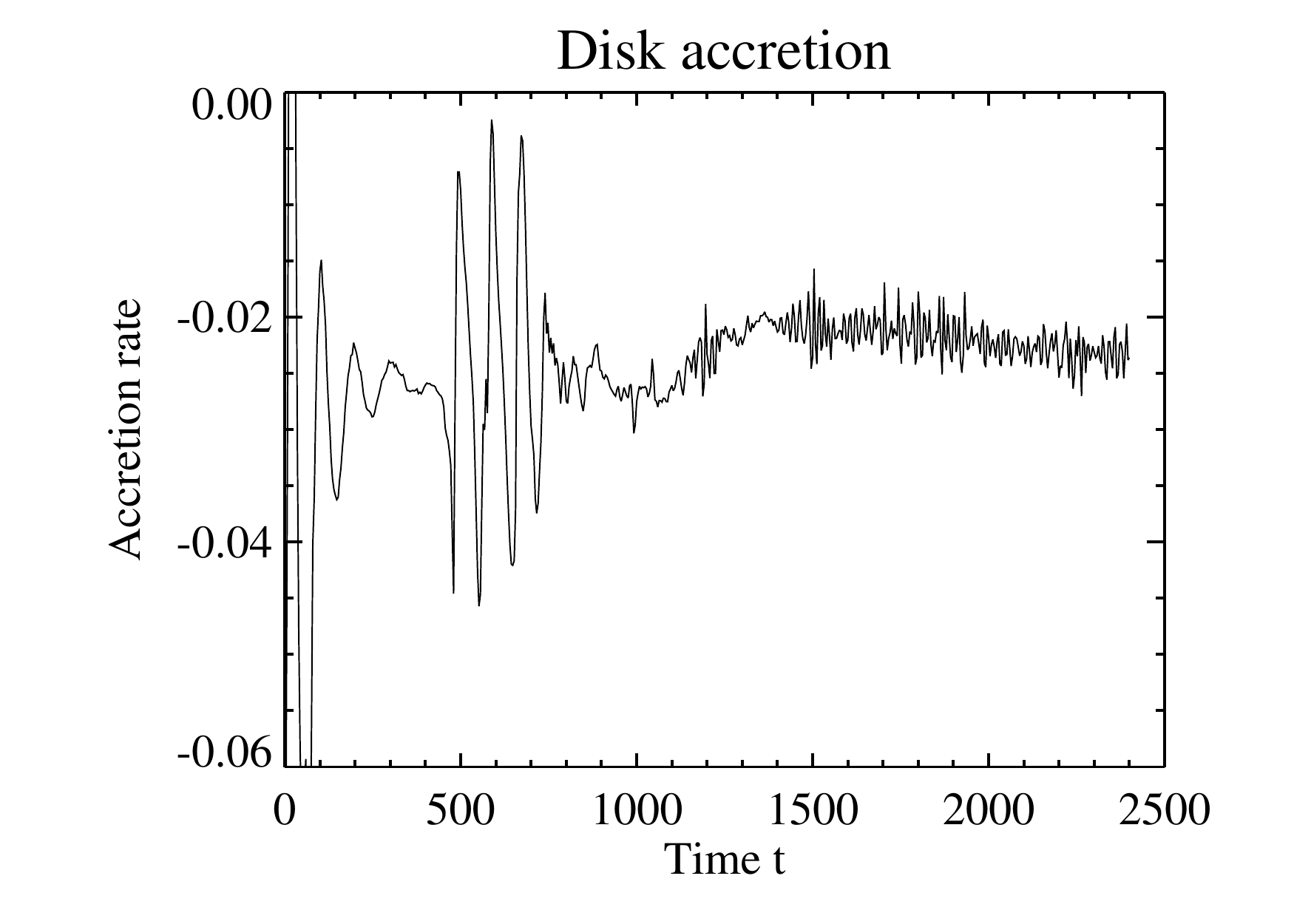}
 \includegraphics[width=5.9cm]{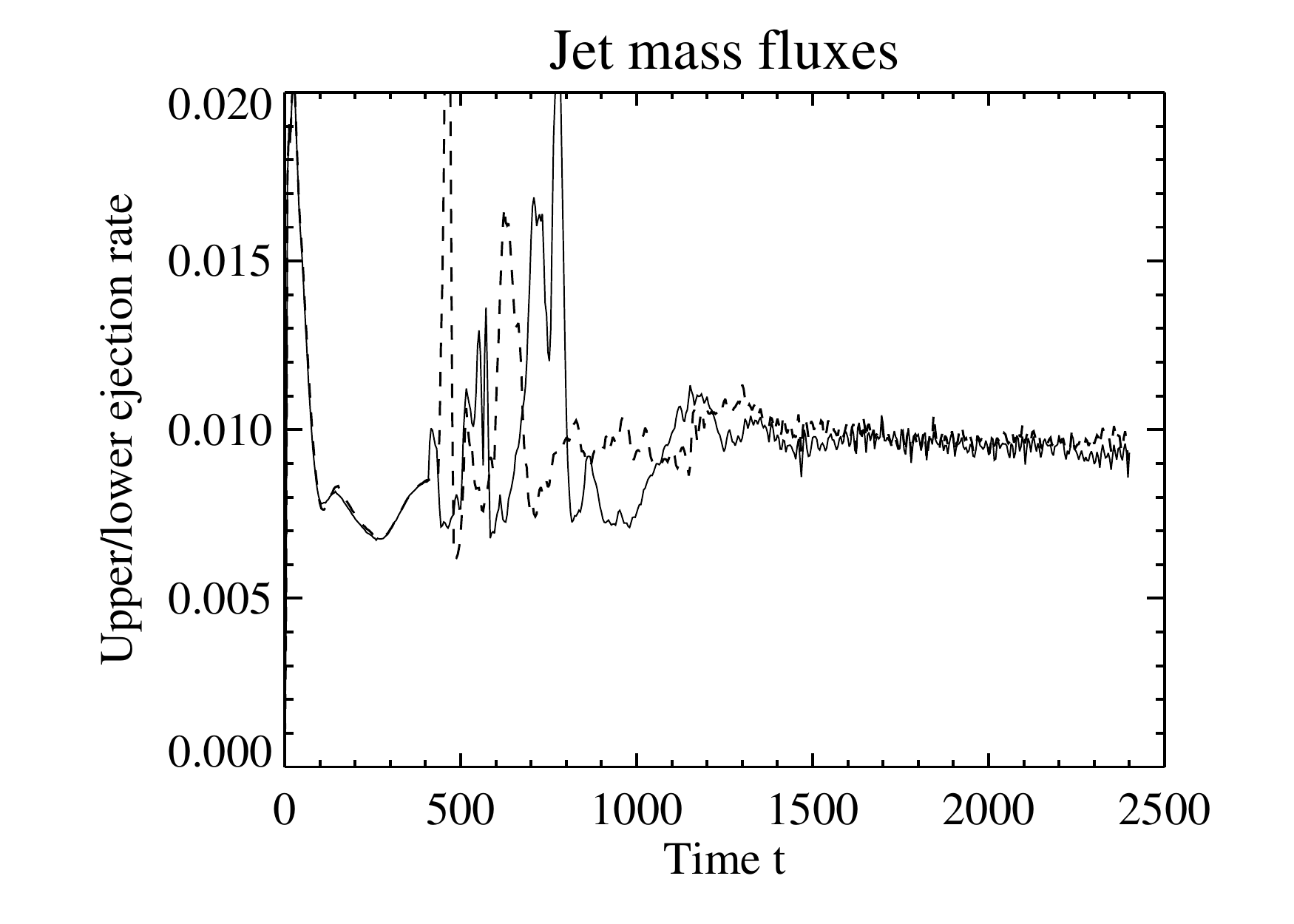}
 \includegraphics[width=5.9cm]{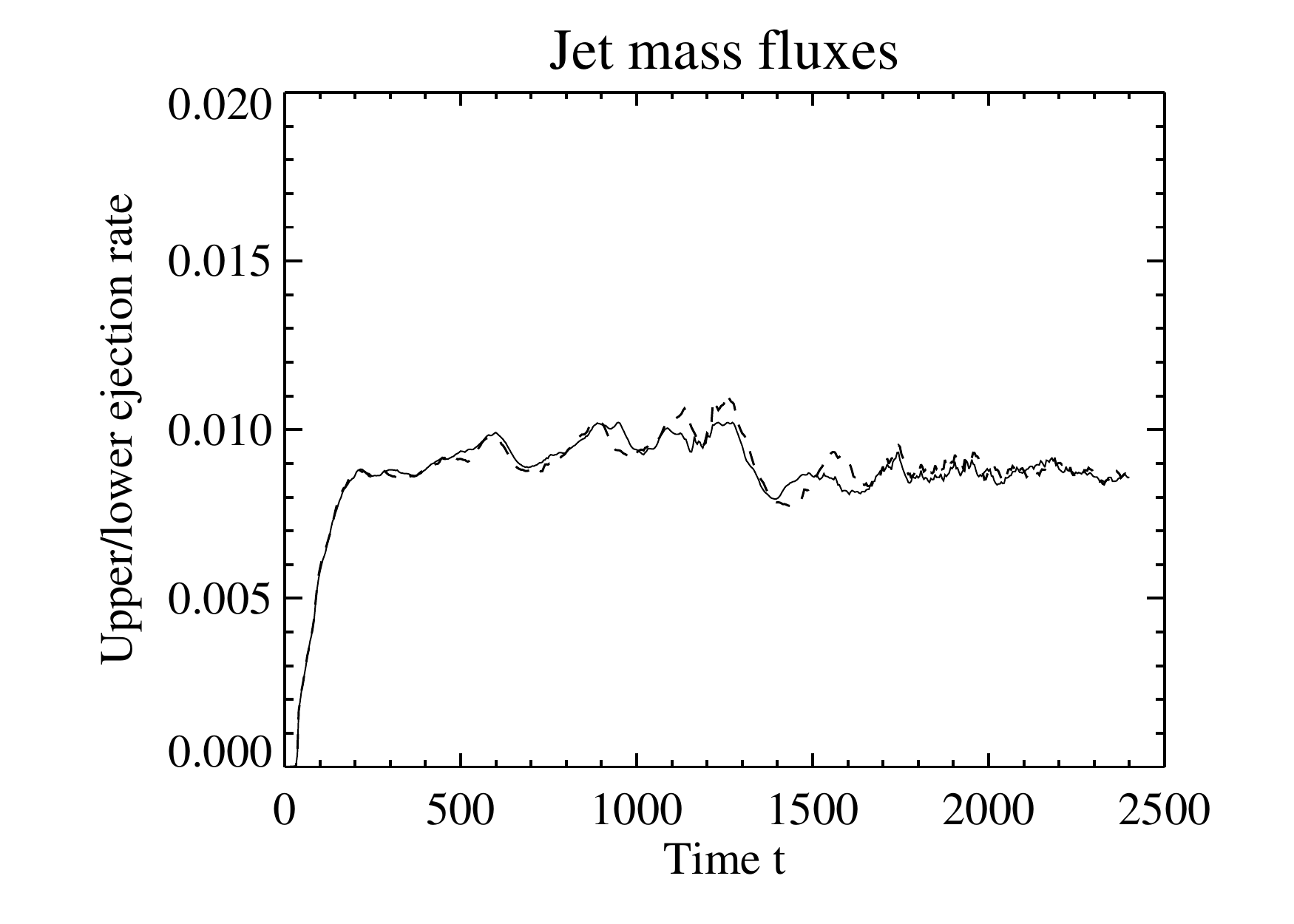}
\caption{Time evolution of the mass fluxes for simulation sb3.
Shown is evolution of accretion rate and the mass ejection rates from the upper (solid) and lower (dashed)
disk surfaces (all in code units). 
Ejection rates (middle) are measured in the control volumes with $r_1=1.5$ and $r_2 = 10.0$, while the accretion
rate (top) is integrated at $r=5$.
For comparison we show the asymptotic (vertical) mass fluxes through the jet, integrated from $r_1=2.0$ to $r_2 = 50.0$
at $z= -75, 75$.}
\label{fig:bicase3_massfluxes}
\end{figure*}

\subsection{A localized disk asymmetry in one disk hemisphere}
We start simulation sb3 with a symmetric initial disk structure. 
As for reference case sb1 the outflows evolve into a symmetric jet-counter jet 
structure (Fig.~\ref{fig:bicase3-rho-mag}).
However, at $t=400$, when a quasi steady state of the inner inflow-outflow structure is reached, we 
disturb the symmetry of the disk structure by inserting a localized over pressure in the upper disk 
hemisphere, probably similar to an explosion (see Fig.~\ref{fig:bicase3-v1}).
This injection is localized within a box of size ($\Delta r \times \Delta z) = (1.5 \times 0.4)$
located at $(r,z) = (11.25, 1.2)$, and is switched on for $\Delta t = 20$, corresponding to 0.1 
of an orbital period at this radius.
The injected material has on average a 20 times higher density and a 2000 times higher pressure
compared to the ambient disk material.


The injected material disturbs the disk symmetry as it expands across the disk.
The disturbance is slowly advected into the jet launching region.
However, we observe that the disk asymmetry decays faster than it is advected along
the disk into the jet launching area at small disk radii.
This is easy to understand. 
The expansion is roughly with sound speed, while the advection happens a sub-sonic 
speed $v_r \simeq \epsilon v_{\rm Kep}$.
Figure \ref{fig:bicase3-v1} shows the disk accretion velocity evolution.
The expansion following the "explosion" first penetrates the whole disk in vertical 
direction, before the over-density is advected slowly inwards. 
Thus, the symmetry of the inner disk is not really affected by the explosion in the 
upper hemisphere at $r=12$.
The dominant outflow from the inner disk is only weakly affected.
When the size of the launching area reaches the explosion site, the asymmetry has 
almost disappeared.

Up to $t=1000$ the mass fluxes show an asymmetry on a 10\%-level. 
Also a slight structural asymmetry is visible (see the $6^{\rm th}$ flux surface contour
in Fig.~\ref{fig:bicase3-rho-mag})
which has propagated from the disk surface to a distance $z \simeq 70$ at this time.

That part of the outflow which originates from the initially asymmetric part of the 
disk (where the blob was injected), starts indeed asymmetric.
As the mass flux launched from larger disk radii is small compared to the jet launched 
from the inner part, the total jet mass fluxes into the two hemispheres differ only 
marginally.
Figure \ref{fig:bicase3_massfluxes} show the time evolution of the mass fluxes.
The ejection rates are integrated along the disk surface between $r=1.5$ and $r=10$
thus inside the radius where the explosion happens.
Still the explosion disturbs the launching site such that we see a 5-10\% effect in the
outflow mass fluxes.
This effect is somewhat delayed from the explosion time since the disturbance needs
time to be advected inwards.
The asymptotic outflow rates, integrated from $r=2$ to $r=50$ at $z=\pm 75$, show
only marginal differences, however.
Note that the flux surface passing through $(r,z) = (50,75)$ anchors at $r=5$ in the
disk, and thus in a weekly disturbed region.

In summary, although the initially symmetric disk is clearly disturbed by the asymmetric
explosion, the asymmetry decays faster than it propagates to the inner jet launching
site. 
Thus, the asymptotic, collimated jet, which originates in the inner disk area is only
marginally asymmetric, moreover as the main mass flux is launched along the innermost
field lines.

\begin{figure*}
\centering
 \includegraphics[width=5.5cm]{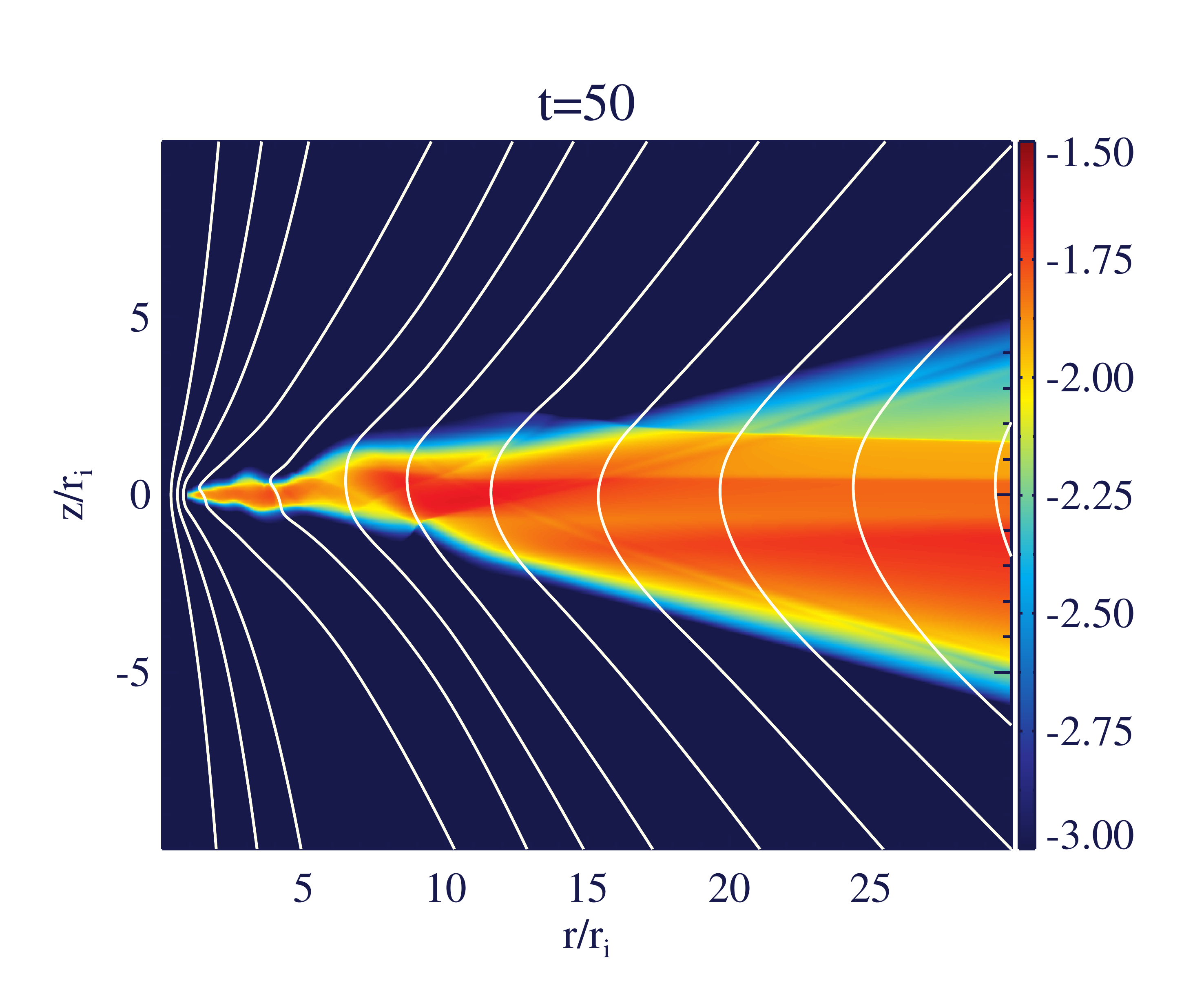}
 \includegraphics[width=5.5cm]{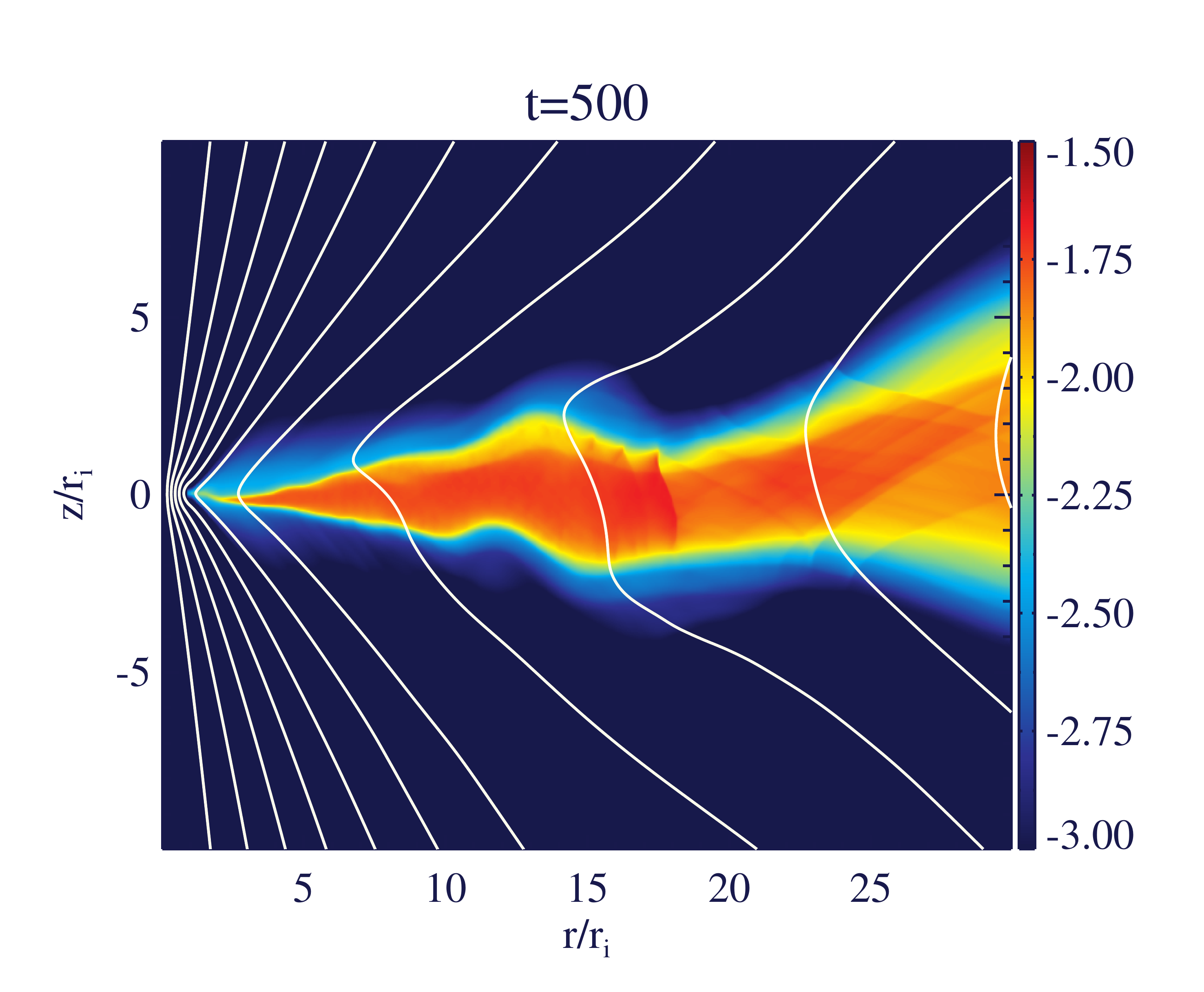}
 \includegraphics[width=5.5cm]{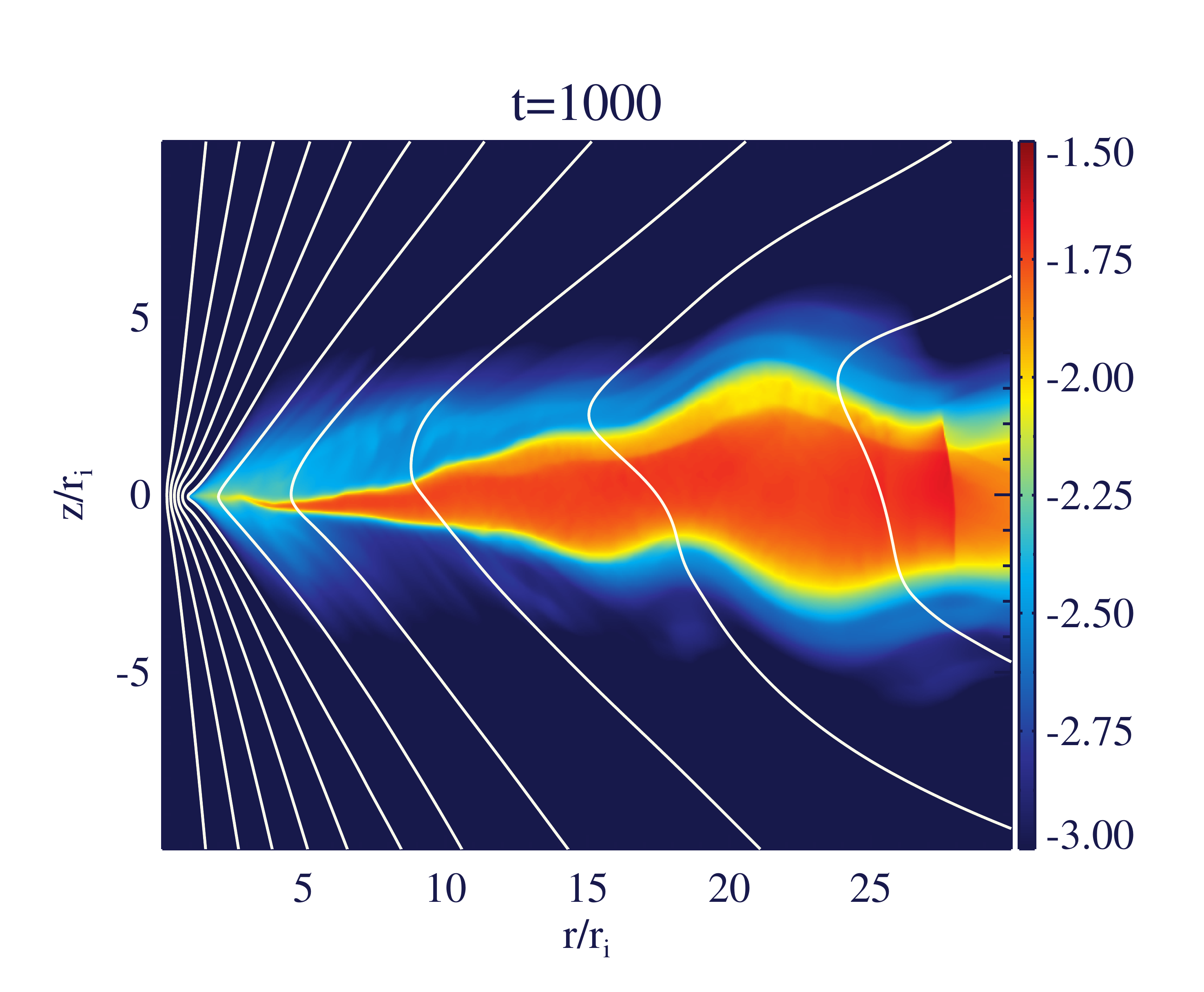}
\caption{Time evolution of the inner disk structure for simulation run cb14. 
Here the diffusivity profile is prescribed as a local Eq.~\ref{eq:magdiff_local}.
Shown is the evolution of magnetic diffusivity (color) and the poloidal magnetic field (contours of poloidal magnetic flux $\Psi(r,z)$)
for the dynamical times steps $t = 50, 500, 1000$ (from left).
}
\label{fig:cf10-eta}
\end{figure*}

\begin{figure}
\centering
 \includegraphics[width=8.0cm]{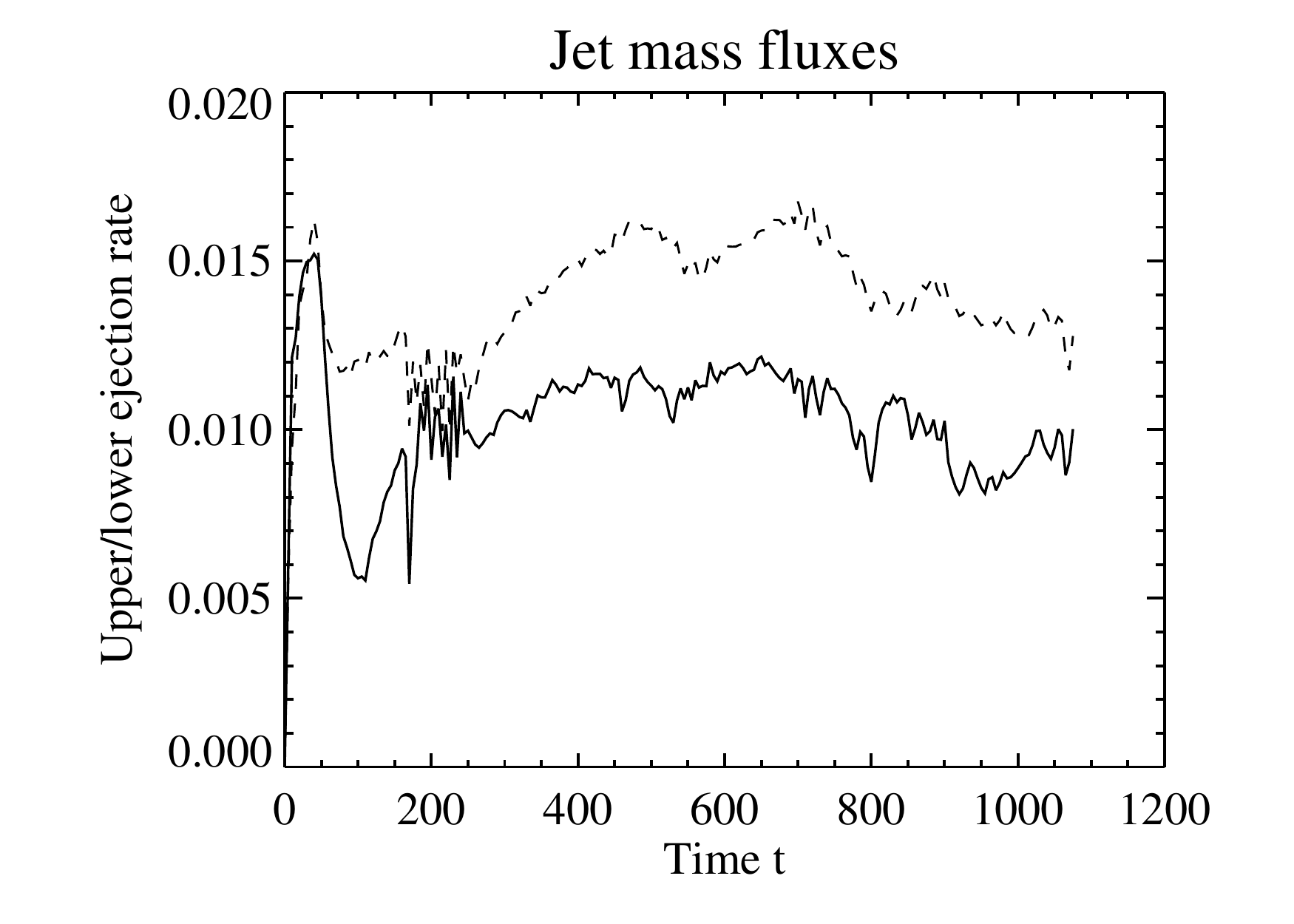}
 \includegraphics[width=8.0cm]{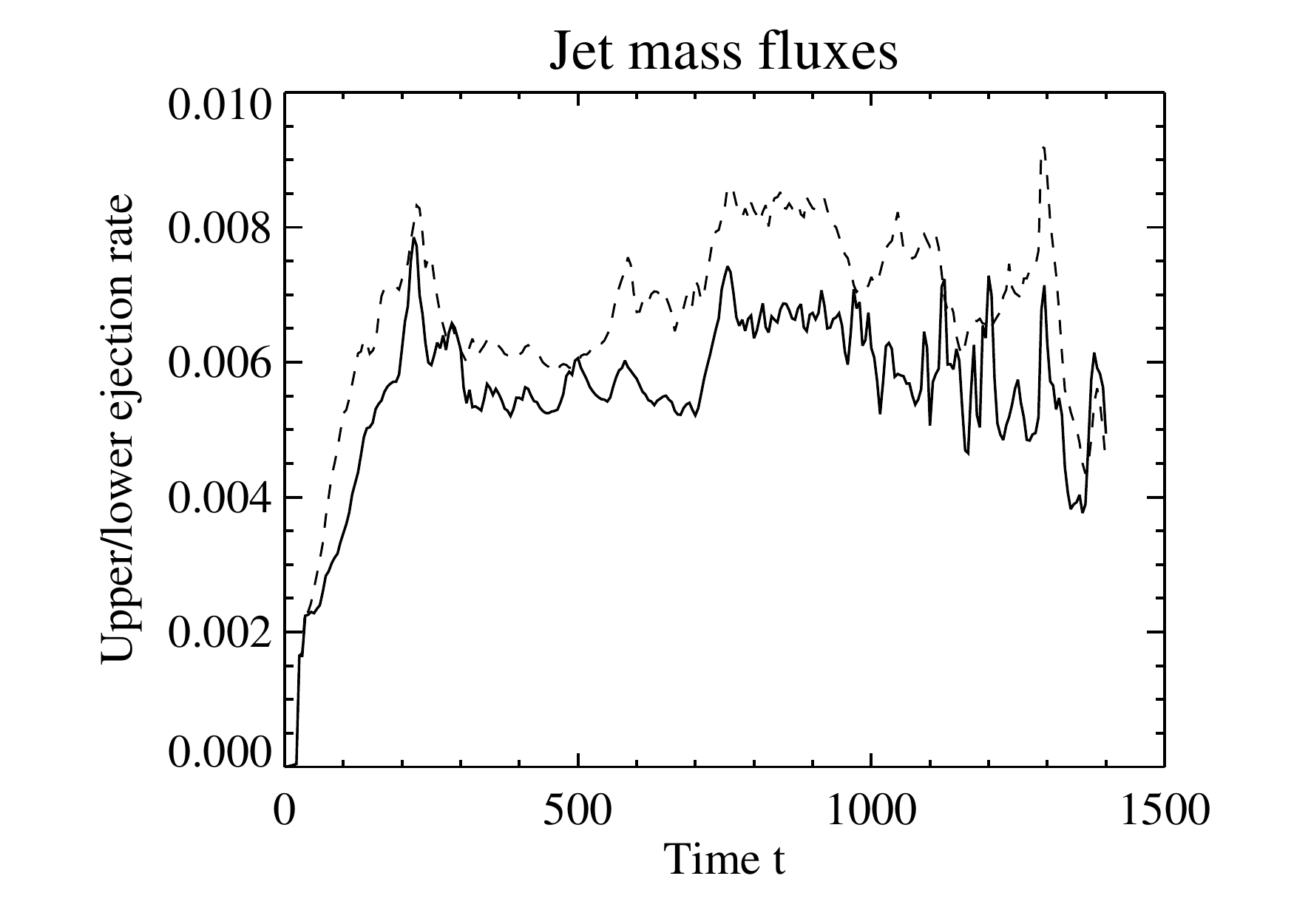}
\caption{Time evolution of the jet mass flux rates for simulation cb14. 
Shown is also the launching mass flux measured at 3 disk scale heights (top) and
the asymptotic fluxes (bottom) integrated at $z= \pm 50$ from $r= 2$ to $r=40$.
}
\label{fig:cf10-rates}
\end{figure}

\section{A local magnetic diffusivity model and bipolar jet launching}
In order to allow the disk and jet to follow a truly asymmetric evolution, without 
the restoring effects of a symmetric, global magnetic diffusivity distribution, 
we have applied a {\em local} description of diffusivity which follows the local 
evolution of disk and outflow (see Tab.~\ref{tbl:bicases}).

The results of this section are somewhat preliminary, as a physically self-consistent parametrisation for a local 
magnetic (turbulent) diffusivity is not available.
In order to investigate the main features of the local approach, we have applied two diffusivity distributions. 
One follows Eq.~\ref{eq:magdiff_cemel},
the other one refers to Eq.~\ref{eq:magdiff_local}.
The main difficulty one faces with a local description is a feedback mechanism such that low densities lead to
a lower diffusivity, thus stronger matter-field coupling, thus more efficient angular momentum 
removal and a faster accretion, which leads to even lower densities.

Moreover, the local prescription Eq.~\ref{eq:magdiff_cemel} gives a diffusivity profile decreasing with radius, 
contrary to the usual choice in the literature, Eq.~\ref{eq:magdiff_global}, which is also realized in 
Eq.~\ref{eq:magdiff_local}.

We find that the feedback mechanism is most efficient in simulations with a strong coupling between diffusivity
and density, such as a diffusivity profile Eq.~\ref{eq:magdiff_cemel}.
In the extreme cases we see that the feedback mechanism may lead to a temporary gap opening over the innermost 
disk radii.
As a result, the accretion in this area is strongly episodic, and, subsequently also the mass ejection.
We will present a detailed investigation of this feedback mechanism in a forthcoming paper.

Applying a more sophisticated prescription of the magnetic diffusivity following Eq.~\ref{eq:magdiff_local},
the over-all accretion-ejection evolution is more similar to the picture established by the simulations
using a global diffusivity profile.
The fundamental difference is, however, the longer lasting and more persistent asymmetry in the disk
and the outflows.

Figure \ref{fig:cf10-eta} shows the time evolution of magnetic diffusivity and the magnetic
field for a simulation with a more sophisticated local setup of magnetic diffusivity following
Eq.~\ref{eq:magdiff_local}. 
We clearly see how the magnetic diffusivity follows the structure of the disk and the outflow.
The disk "{warping}" is seen diffusivity distribution as well. 
Mass loading and matter-field coupling depends on the local disk properties 
(defined by Eq.~\ref{eq:magdiff_local}).
Since the diffusivity profile is broader in vertical direction, mass loading and angular momentum
removal is more efficient, establishing higher accretion and outflow rates - in agreement with our
results in paper I.

The outflow mass fluxes develop a clearly asymmetric structure.
Figure \ref{fig:cf10-rates} show the mass fluxes integrated from $r=1.5$ to $r = 10$,
along a surface $z=0.3 r$ parallel to the initial disk surface.
Jet and counter jet injection mass flow rates differ substantially - by about 30\% over at least 
1000 time steps.
The asymptotic jet mass fluxes integrated from $r=2$ to $r = 40$,
along a surface $z = \pm 50$  are somewhat lower (due to the fact that some of the material
injected into the outflow leaves the grid in radial direction), and still differ on a 15\%-level.



\section{Impact of the environment}
This section is devoted to the question whether jet asymmetries are due to external of internal 
interaction.
It seems clear that the conditions in the ambient gas the jet is penetrating, will certainly 
affect the expansion and the collimation on the large scales.

Here we consider a model setup, in which the bipolar outflow is launched intrinsically symmetric
from a (symmetric) disk into an asymmetric star-disk corona.
We do this by prescribing a different initial coronal density (and thus pressure), simply choosing 
$\delta_{\rm up} = 10^{-4}$ and $\delta_{\rm up} = 10^{-3}$.

\begin{figure*}
\centering
\includegraphics[width=3.5cm]{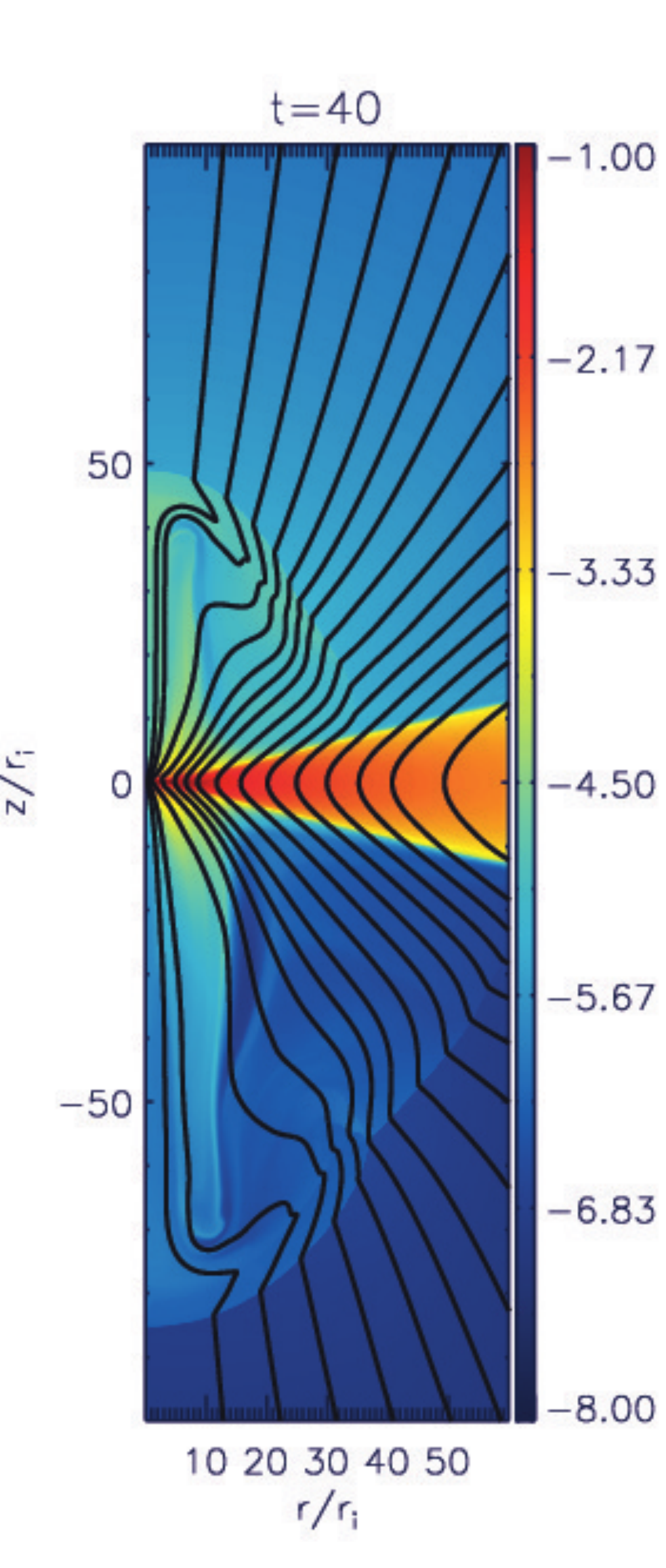}
\includegraphics[width=3.5cm]{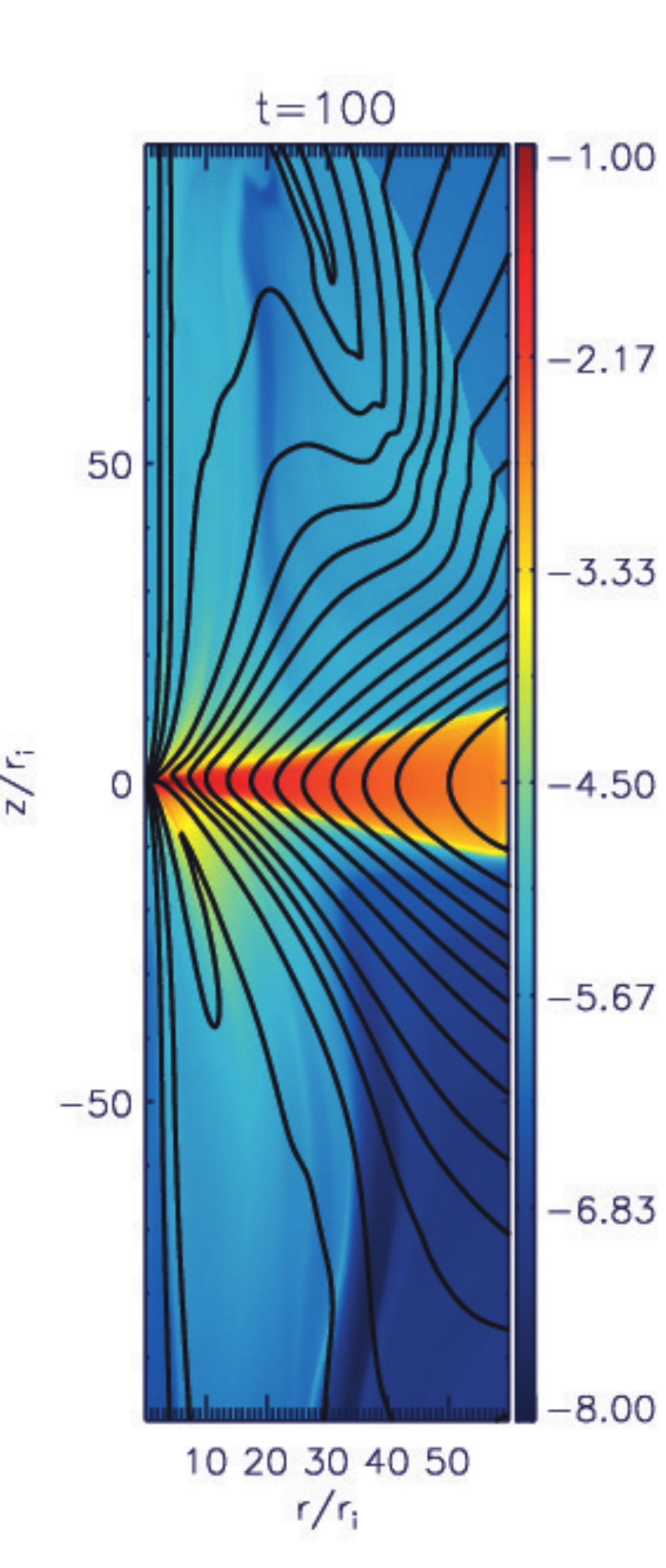}
\includegraphics[width=3.5cm]{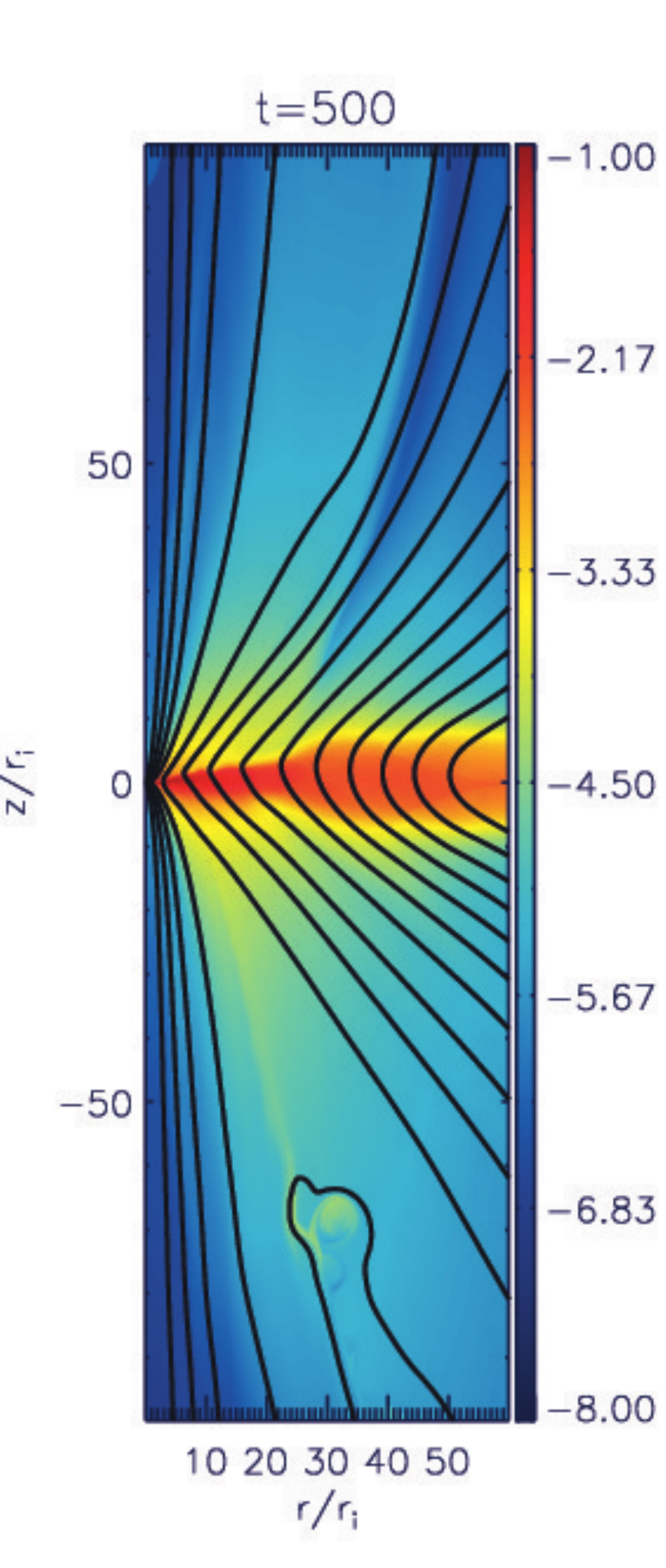}
\includegraphics[width=3.5cm]{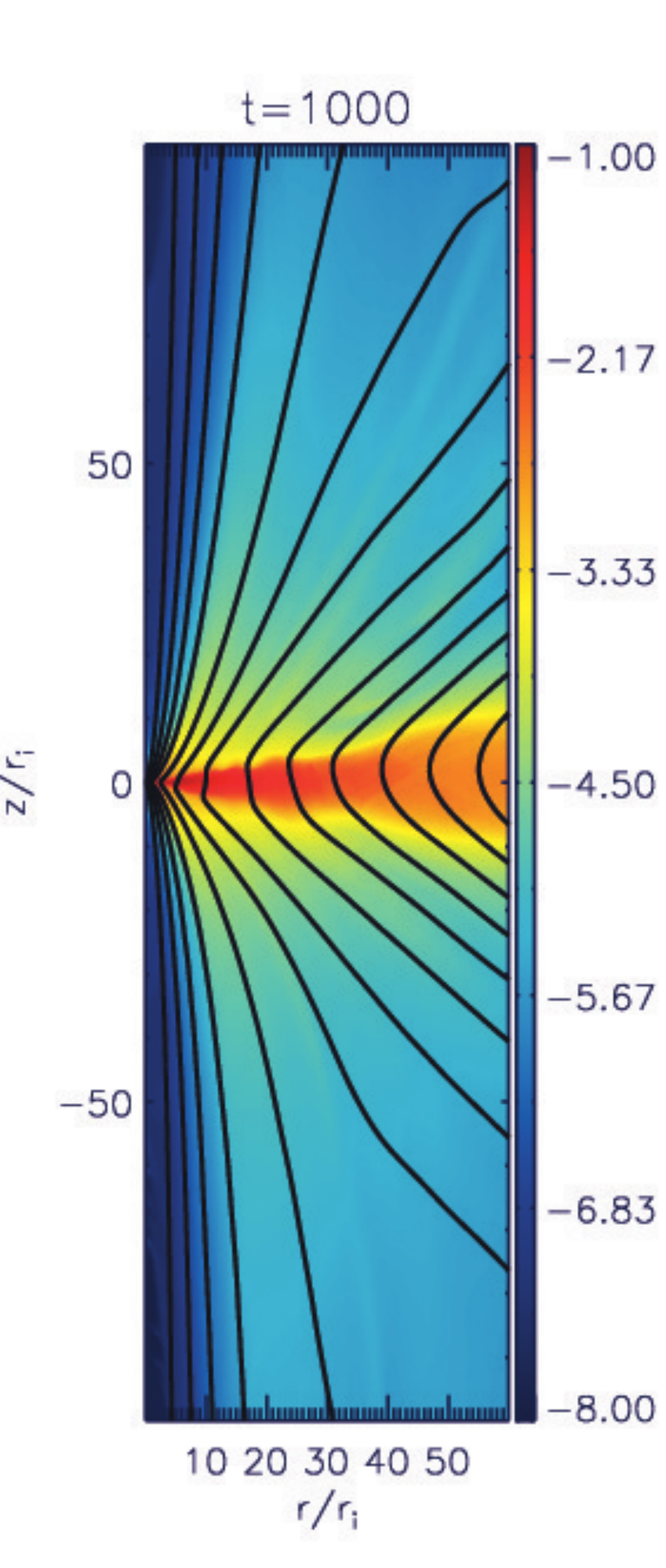}
\includegraphics[width=3.5cm]{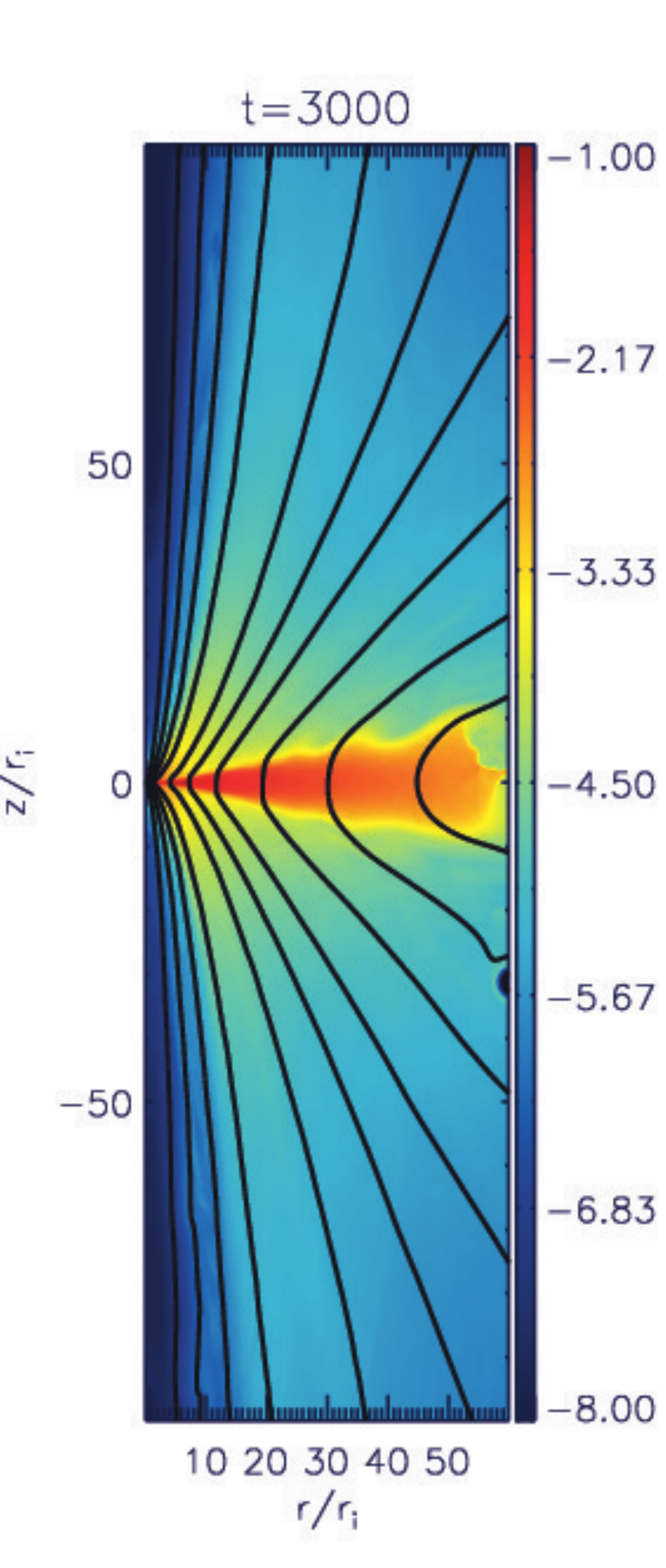}
\caption{Time evolution of the bipolar jet-disk structure for simulation sb4. 
Here the outflow is launched in hemispheres with different density contrast.
Shown is the evolution of mass density (color) and the poloidal magnetic field (contours of poloidal magnetic flux $\Psi(r,z)$)
for the dynamical times steps $t = 0, 100, 500, 1000, 3000$.
}
\label{fig:bipo1-fix-case5}
\end{figure*}

Figure \ref{fig:bipo1-fix-case5} shows the time evolution of density and magnetic field for
such a setup.

\begin{figure*}
\centering
 \includegraphics[width=5.9cm]{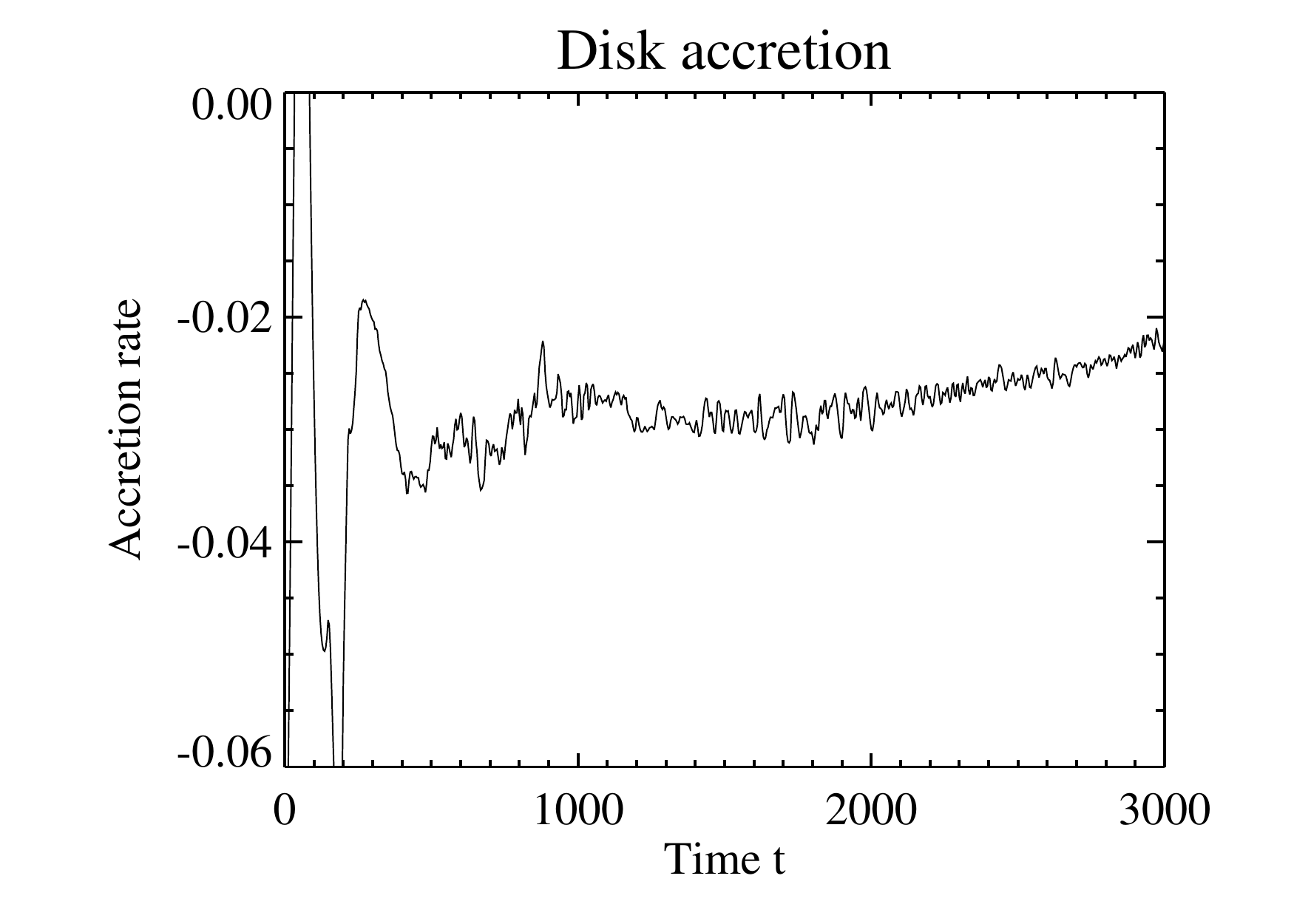}
 \includegraphics[width=5.9cm]{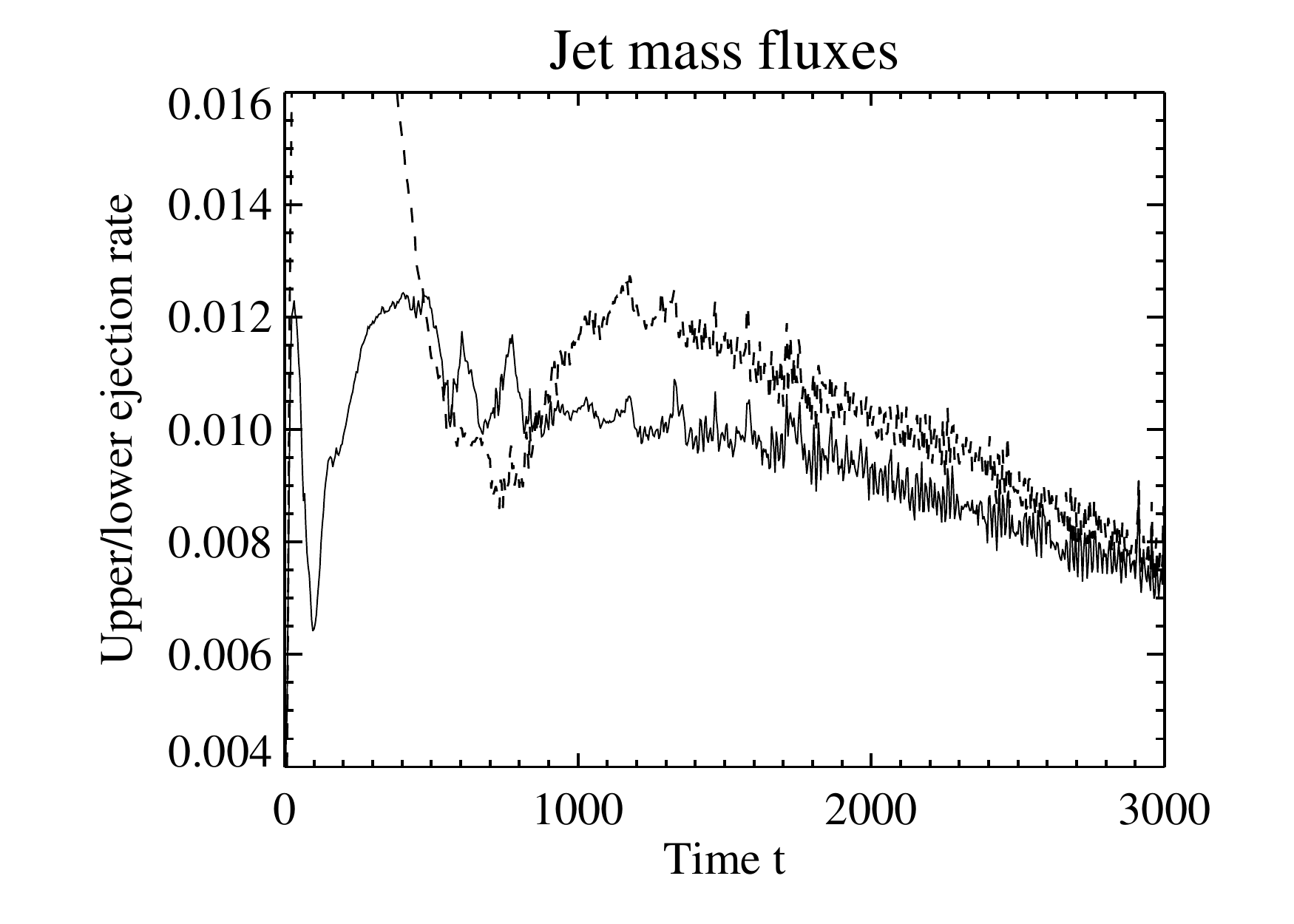}
 \includegraphics[width=5.9cm]{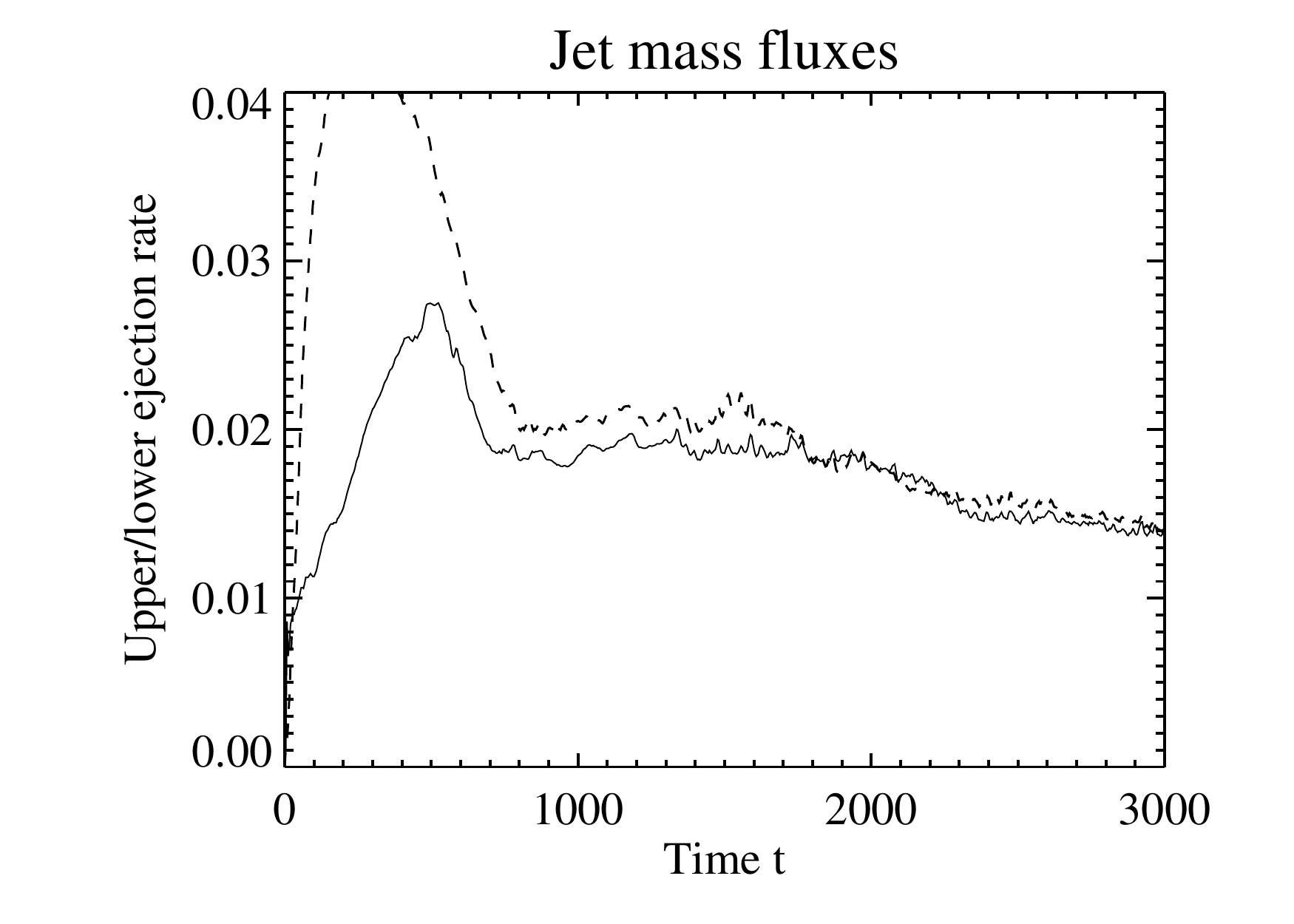}
\caption{Time evolution of the mass fluxes for simulation sb4.
Shown is evolution of accretion rate and the mass ejection rates from the upper (solid) and lower (dashed)
disk surfaces (all in code units). 
Ejection rates (middle) are measured in the control volumes with $r_1=1.5$ and $r_2 = 10.0$, while the accretion
rate (top) is integrated at $r=5$.
For comparison we show the asymptotic (vertical) mass fluxes through the jet, integrated from 
$r_1 = 1.5$ to $r_2 = 40.0$
at $z= -80, 80$.}
\label{fig:bicase5_massfluxes}
\end{figure*}

As expected the initial outflow is highly asymmetric, more than the outflows disturbed by 
intrinsic disk asymmetries.
However, the asymmetry is clearly transient.
As soon as the initial coronal material is swept out of the computational domain,
the accretion-ejection system returns to hemispheric symmetry.
This is well visible in the mass flux evolution (Fig.~\ref{fig:bicase5_massfluxes}), which
shows a drastic asymmetry during the first hundreds of time steps.
Again it is interesting to see various time lags in the flow evolution.
Firstly, the time lag between the launching time of the initial asymmetry and the arrival time 
of asymmetric features in the asymptotic region.
Secondly, the time lag between the time step when the disk-outflow system has returned to symmetry, 
and the time when the outflow symmetry has reached the asymptotic jet.

Note that in these simulations the disk structure itself stays rather symmetric, in difference to 
the simulations discussed above, where we see the disk evolving into a warped structure.
However, we can see back reaction of the asymmetric outflow onto the disk structure in this approach.
The two asymmetric jets drive a different electric current system in both hemispheres.
Both the current systems are connected within the disk - subsequent MHD forces therefore slightly
distort the symmetric hydrodynamic disk structure.
The deviation from symmetry is, however, not as strong as for the previous cases in which we
start with an initially asymmetry disk.

\section{Resolution study}
\begin{figure}
\centering
 \includegraphics[width=9.0cm]{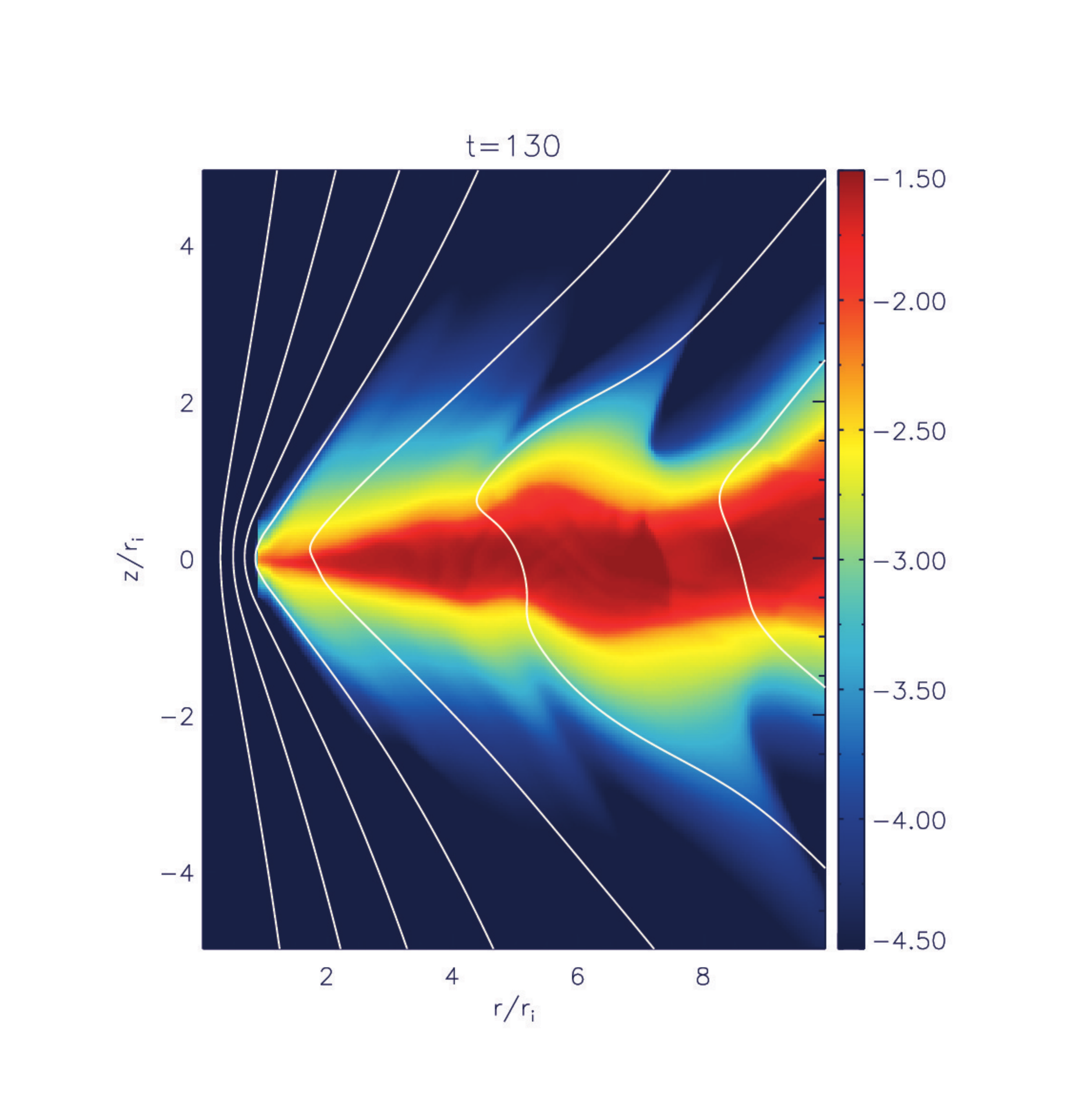}
 \includegraphics[width=8.0cm]{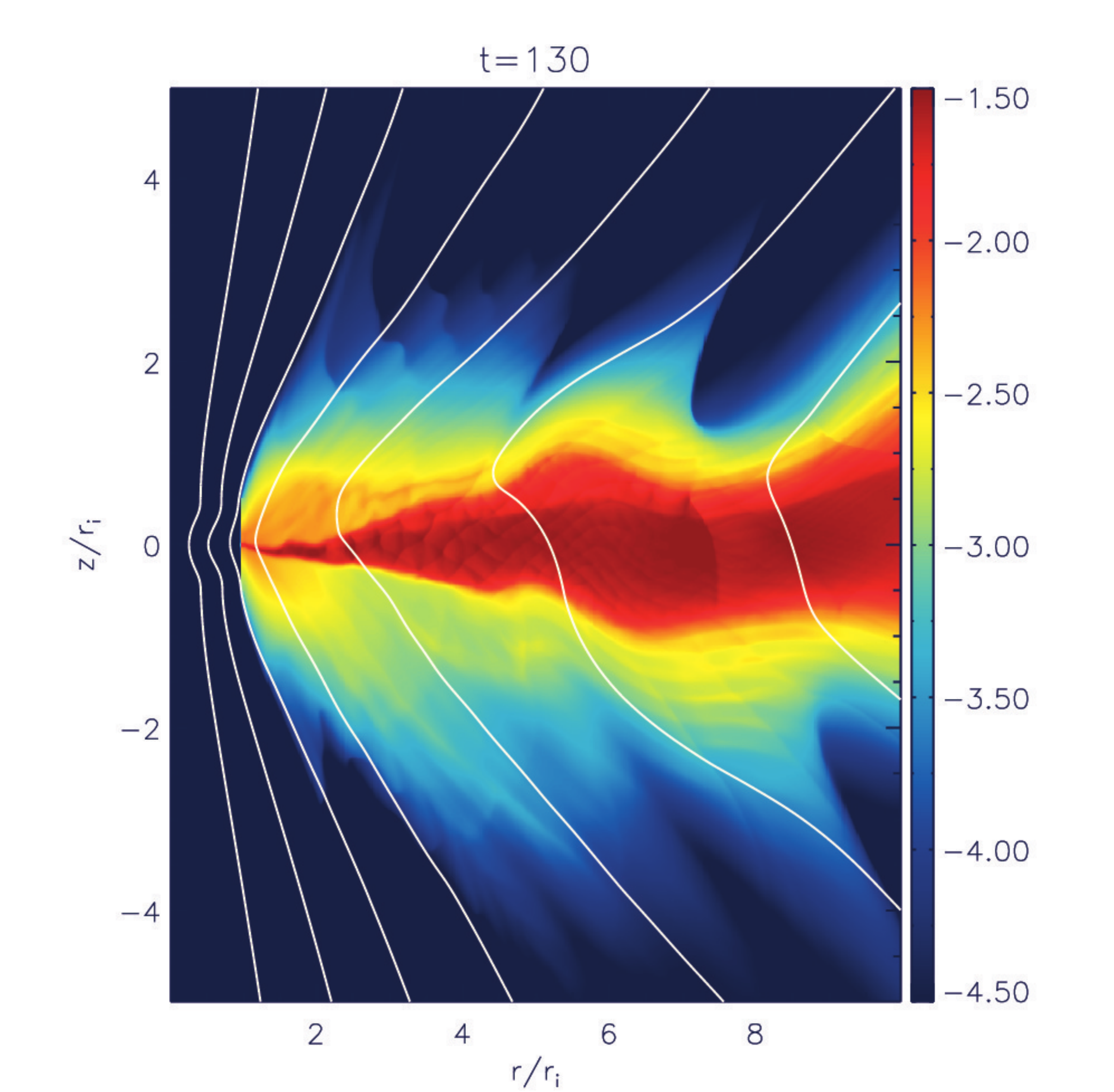}
\caption{Resolution study for our setup.
Subset of the innermost disk-jet area which is most critical to resolution issues.
Shown is the magnetic diffusivity distribution following Eq.~\ref{eq:magdiff_local}
for simulation run cb14 with (low resolution, top) with run cb16 (high resolution, bottom).
}
\label{fig:resolution}
\end{figure}

Finally we present example results of our resolution study in brief.
Figure \ref{fig:resolution} shows the inner part of the disk at the dynamical time step $t=130$
when the inner disk evolution is rather violent.
We compare simulation cb14 applying our standard resolution of
$\Delta r = 0.0417 $, $\Delta z = 0.0333$ in the disk, with simulation cb16,
applying a 3 times higher resolution, 
$\Delta r = 0.0165 $, $\Delta z = 0.00824$.
Although the higher resolution run shows somewhat more sub-structure, such as internal shocks or a more
peaked diffusivity distribution for $r < 3$, the main features of the disk dynamics are just the same.
In particular the disk height is similar, as well as the structure and opening angle of the
magnetic flux surfaces.
Also the radially structured features of the outflow close to the disk is very similar
in both hemispheres.
The overall mass fluxes measured in the simulations are similar as well, however, the
jet-counter jet asymmetry is somewhat more pronounced in the high-resolution run.

We note that high-resolution simulations are particularly difficult to manage in 
diffusive MHD since the CF time stepping condition $\Delta t_{\eta} \leq (\Delta x)^2 / \eta$.

\section{Conclusions}
We have presented results of MHD simulations investigating the launching of jets and outflows
from a magnetically diffusive disk in Keplerian rotation.
The time evolution of the accretion disk structure is self-consistently taken into account.
The simulations are performed in axisymmetry applying the MHD code PLUTO.

Based, on paper I which studied how magnetic diffusivity and magnetization affect the disk and 
outflow properties, such as the mass and angular momentum fluxes, jet collimation, or jet radius,
the present paper investigates the hemispheric symmetry of outflows.
We have setup a numerical scheme that can treat the outflows from an equatorial disk in bipolar
direction. 
The setup has been carefully checked against numerical artifacts triggering asymmetry 
in the disk-outflow evolution.

We disturb the disk-outflow bipolar asymmetry applying several approaches.
Bipolar symmetry is re-achieved on the long term by gravitational forces, in particular the component 
vertical to the disk, and (in some cases) a symmetric magnetic diffusivity prescription. 

In particular, we have obtained the following results.

(1) A test case with a symmetric setup gave a perfectly symmetric bipolar evolution of
disk and outflow for several thousand rotations, until numerical effects by the outer disk
boundary condition start to disturb the symmetry of the outermost disk. The measured mass fluxes
compare well with the one-hemispheric simulations in paper I. 

(2) We then applied various options to disturb the outflow bipolar symmetry. 
First we prescribed an initially asymmetric disk applying a different pressure scale 
height in both disk hemispheres, $\epsilon_{\rm up} = 0.1$,  $\epsilon_{\rm down} = 0.15$.
In general the disk structure evolves into a warped configuration, with warp amplitudes 
of a few initial disk scale heights. 
Electric currents are driven across the equatorial plane. 
The mass fluxes of the resulting outflows differ by about 10\% over less than 1000 time steps,
until a symmetric launching is achieved again. 
The outflow rates at larger distance from the source differ similarly, however, asymmetry
last longer, just because of the propagation time of the material launched from the disk.

(3) We then started the simulation with symmetric initial setup, but disturbing the
symmetric disk structure by a localized (in time and space) explosion at the time when a
symmetric outflow is already well established. 
As for (2), a asymmetric outflow arises after the injection, however, after a time delay
due to the propagation time of the asymmetric injection along the disk towards the main
jet launching area. On the long term a symmetric outflow is re-established.

(4) We then investigated the disk-jet evolution applying a {\em local} prescription of 
magnetic diffusivity $\eta = \eta(r,z,t)$ following the local disk evolution.
We find that - as discussed in paper I - the distribution of magnetic diffusivity affects
the disk evolution (and subsequently the outflow evolution) essentially, as it governs 
the coupling between matter and field, and, thus the angular momentum evolution,  
the disk accretion, the magneto-centrifugal acceleration, and also the mass loading of 
the outflow.
We first applied a simple power law diffusivity distribution as motivated by earlier 
papers. 
We observed a strong feedback such that for low densities an almost ideal MHD situation
leads to super-efficient accretion with weak outflows and strongly decaying disk mass.
These features deserve further attention which will be the focus of a forthcoming
paper.

(5) For a more sophisticated magnetic diffusivity distribution following the density-weighted
{\em local} sound speed, and an increasing diffusivity with radius, the accretion evolution
is more persistent.
We were able to follow the accretion-ejection process over many thousands of dynamical
time steps.
The most interesting result is that the bipolar asymmetry of jet and counter jet is
{\em long-lasting}.
We find that in this case the warped structure of the disk survive many dynamical time scales
also in the inner disk.
We interpret this as due to the lack of a symmetric diffusivity distribution. 
There are the same restoring forces of gravity as in the case of a symmetric diffusivity
profile.
However, the matter is more directly coupled to the distorted magnetic field which in principle 
opposes the return to symmetry.
In the end we observe a persistent difference in the jet-counter jet mass fluxes - up to 30\% and 
lasting longer than the simulation run time.

(6) In order to compare between internal and external effects we also investigated the launch
of a jet outflow from a symmetric disk into an asymmetric ambient gas distribution.
As probably expected the initial outflow is strongly asymmetric, with 20\% different mass fluxes
for jet and counter jet, however as soon as the outflow has penetrated the asymmetric corona
(i.e. when the outflow has left the grid, and the initial condition has swept out of the
domain, the outflow returns to symmetry (after about 2000 dynamical time steps).
In comparison to the simulations discussed about, the initial disk structure is symmetric
and does not evolve into a warped structure.

\acknowledgements
We thank the Andrea Mignone and the PLUTO team for the possibility to use their code.
S.S. acknowledges the hospitality by the Max Planck Institute for Astronomy.
The simulations were performed on the THEO cluster of the Max Planck Institute for Astronomy.
This work was financed by the SFB 881, subproject B4, of the German science foundation DFG,
and partly by a scholarship of the Ministry of Science, Research, and Technology of Iran.




\begin{thebibliography}{26}
\expandafter\ifx\csname natexlab\endcsname\relax\def\natexlab#1{#1}\fi

\bibitem[{{Bai} \& {Stone}(2013)}]{2013ApJ...767...30B}
{Bai}, X.-N. \& {Stone}, J.~M. 2013, \apj, 767, 30

\bibitem[{{Beckwith} {et~al.}(2011){Beckwith}, {Armitage}, \&
  {Simon}}]{2011MNRAS.416..361B}
{Beckwith}, K., {Armitage}, P.~J., \& {Simon}, J.~B. 2011, \mnras, 416, 361

\bibitem[{{Chagelishvili} {et~al.}(1996){Chagelishvili}, {Bodo}, \&
  {Trussoni}}]{1996A&A...306..329C}
{Chagelishvili}, G.~D., {Bodo}, G., \& {Trussoni}, E. 1996, \aap, 306, 329

\bibitem[{{Ellerbroek} {et~al.}(2013){Ellerbroek}, {Podio}, {Kaper}, {Sana},
  {Huppenkothen}, {de Koter}, \& {Monaco}}]{2013A&A...551A...5E}
{Ellerbroek}, L.~E., {Podio}, L., {Kaper}, L., {Sana}, H., {Huppenkothen}, D.,
  {de Koter}, A., \& {Monaco}, L. 2013, \aap, 551, A5

\bibitem[{{Fendt} \& {{\v C}emelji{\'c}}(2002)}]{2002A&A...395.1045F}
{Fendt}, C. \& {{\v C}emelji{\'c}}, M. 2002, \aap, 395, 1045

\bibitem[{{Fromang} {et~al.}(2012){Fromang}, {Latter}, {Lesur}, \&
  {Ogilvie}}]{2012arXiv1210.6664F}
{Fromang}, S., {Latter}, H.~N., {Lesur}, G., \& {Ogilvie}, G.~I. 2012, ArXiv
  e-prints

\bibitem[{{Gammie}(1996)}]{1996ApJ...457..355G}
{Gammie}, C.~F. 1996, \apj, 457, 355

\bibitem[{{Gressel}(2010)}]{2010MNRAS.405...41G}
{Gressel}, O. 2010, \mnras, 405, 41

\bibitem[{{Lesur} {et~al.}(2013){Lesur}, {Ferreira}, \&
  {Ogilvie}}]{2013A&A...550A..61L}
{Lesur}, G., {Ferreira}, J., \& {Ogilvie}, G.~I. 2013, \aap, 550, A61

\bibitem[{{Lovelace} {et~al.}(2010){Lovelace}, {Romanova}, {Ustyugova}, \&
  {Koldoba}}]{2010MNRAS.408.2083L}
{Lovelace}, R.~V.~E., {Romanova}, M.~M., {Ustyugova}, G.~V., \& {Koldoba},
  A.~V. 2010, \mnras, 408, 2083

\bibitem[{{Matsakos} {et~al.}(2012){Matsakos}, {Vlahakis}, {Tsinganos},
  {Karampelas}, {Sauty}, {Cayatte}, {Matt}, {Massaglia}, {Trussoni}, \&
  {Mignone}}]{2012A&A...545A..53M}
{Matsakos}, T., {Vlahakis}, N., {Tsinganos}, K., {Karampelas}, K., {Sauty}, C.,
  {Cayatte}, V., {Matt}, S.~P., {Massaglia}, S., {Trussoni}, E., \& {Mignone},
  A. 2012, \aap, 545, A53

\bibitem[{{Mignone} {et~al.}(2007){Mignone}, {Bodo}, {Massaglia}, {Matsakos},
  {Tesileanu}, {Zanni}, \& {Ferrari}}]{2007ApJS..170..228M}
{Mignone}, A., {Bodo}, G., {Massaglia}, S., {Matsakos}, T., {Tesileanu}, O.,
  {Zanni}, C., \& {Ferrari}, A. 2007, \apjs, 170, 228

\bibitem[{{Mignone} {et~al.}(2012){Mignone}, {Zanni}, {Tzeferacos}, {van
  Straalen}, {Colella}, \& {Bodo}}]{2012ApJS..198....7M}
{Mignone}, A., {Zanni}, C., {Tzeferacos}, P., {van Straalen}, B., {Colella},
  P., \& {Bodo}, G. 2012, \apjs, 198, 7

\bibitem[{{Mundt} {et~al.}(1990){Mundt}, {Buehrke}, {Solf}, {Ray}, \&
  {Raga}}]{1990A&A...232...37M}
{Mundt}, R., {Buehrke}, T., {Solf}, J., {Ray}, T.~P., \& {Raga}, A.~C. 1990,
  \aap, 232, 37

\bibitem[{{Murphy} {et~al.}(2010){Murphy}, {Ferreira}, \&
  {Zanni}}]{2010A&A...512A..82M}
{Murphy}, G.~C., {Ferreira}, J., \& {Zanni}, C. 2010, \aap, 512, A82+

\bibitem[{{Ouyed} \& {Pudritz}(1997)}]{1997ApJ...482..712O}
{Ouyed}, R. \& {Pudritz}, R.~E. 1997, \apj, 482, 712

\bibitem[{{Pessah} {et~al.}(2007){Pessah}, {Chan}, \&
  {Psaltis}}]{2007ApJ...668L..51P}
{Pessah}, M.~E., {Chan}, C.-k., \& {Psaltis}, D. 2007, \apjl, 668, L51

\bibitem[{{Porth} \& {Fendt}(2010)}]{2010ApJ...709.1100P}
{Porth}, O. \& {Fendt}, C. 2010, \apj, 709, 1100

\bibitem[{{Ray} {et~al.}(2007){Ray}, {Dougados}, {Bacciotti}, {Eisl{\"o}ffel},
  \& {Chrysostomou}}]{2007prpl.conf..231R}
{Ray}, T., {Dougados}, C., {Bacciotti}, F., {Eisl{\"o}ffel}, J., \&
  {Chrysostomou}, A. 2007, Protostars and Planets V, 231

\bibitem[{{Ray} {et~al.}(1996){Ray}, {Mundt}, {Dyson}, {Falle}, \&
  {Raga}}]{1996ApJ...468L.103R}
{Ray}, T.~P., {Mundt}, R., {Dyson}, J.~E., {Falle}, S.~A.~E.~G., \& {Raga},
  A.~C. 1996, \apjl, 468, L103

\bibitem[{{Sauty} {et~al.}(2012){Sauty}, {Cayatte}, {Lima}, {Matsakos}, \&
  {Tsinganos}}]{2012ApJ...759L...1S}
{Sauty}, C., {Cayatte}, V., {Lima}, J.~J.~G., {Matsakos}, T., \& {Tsinganos},
  K. 2012, \apjl, 759, L1

\bibitem[{{Sheikhnezami} {et~al.}(2012){Sheikhnezami}, {Fendt}, {Porth},
  {Vaidya}, \& {Ghanbari}}]{2012ApJ...757...65S}
{Sheikhnezami}, S., {Fendt}, C., {Porth}, O., {Vaidya}, B., \& {Ghanbari}, J.
  2012, \apj, 757, 65

\bibitem[{{Simon} {et~al.}(2013){Simon}, {Bai}, {Stone}, {Armitage}, \&
  {Beckwith}}]{2013ApJ...764...66S}
{Simon}, J.~B., {Bai}, X.-N., {Stone}, J.~M., {Armitage}, P.~J., \& {Beckwith},
  K. 2013, \apj, 764, 66

\bibitem[{{von Rekowski} {et~al.}(2003){von Rekowski}, {Brandenburg}, {Dobler},
  {Dobler}, \& {Shukurov}}]{2003A&A...398..825V}
{von Rekowski}, B., {Brandenburg}, A., {Dobler}, W., {Dobler}, W., \&
  {Shukurov}, A. 2003, \aap, 398, 825

\bibitem[{{Zanni} {et~al.}(2007){Zanni}, {Ferrari}, {Rosner}, {Bodo}, \&
  {Massaglia}}]{2007A&A...469..811Z}
{Zanni}, C., {Ferrari}, A., {Rosner}, R., {Bodo}, G., \& {Massaglia}, S. 2007,
  \aap, 469, 811

\bibitem[{{Zinnecker} {et~al.}(1998){Zinnecker}, {McCaughrean}, \&
  {Rayner}}]{1998Natur.394..862Z}
{Zinnecker}, H., {McCaughrean}, M.~J., \& {Rayner}, J.~T. 1998, \nat, 394, 862

\end{thebibliography}

\bibliographystyle{apj}

\end{document}